\documentclass[lettersize,journal]{IEEEtran}
\usepackage{cite}
\usepackage{hyperref}

\usepackage{amsmath,amsfonts}
\usepackage{color}

\usepackage{url}

\usepackage{graphicx}
\usepackage{float}
\usepackage[table,xcdraw]{xcolor}
\usepackage{multirow}
\usepackage{siunitx}
\usepackage{chemformula}
\usepackage[caption=false,font=normalsize,labelfont=sf,textfont=sf,subrefformat=parens,labelformat=parens]{subfig}
\usepackage{verbatim}
\usepackage{stfloats}
\usepackage{textcomp}
\usepackage{nomencl}
\usepackage[acronym]{glossaries}
\usepackage{tabularx}
\usepackage{color,soul}
\DeclareSIUnit\mile{mi}
\DeclareSIUnit\feet{ft}
\makeglossaries

\newacronym{3gpp}{3GPP}{3rd Generation Partnership Project}
\newacronym{5g}{5G}{5th Generation}
\newacronym{mmwave}{mmWave}{Millimeter-Wave}
\newacronym{ofdm}{OFDM}{Orthogonal Frequency Division Multiplexing}
\newacronym{6g}{6G}{6th Generation}
\newacronym{phy}{PHY}{Physical Layer}
\newacronym{ai}{AI}{Artificial Intelligence}
\newacronym{mimo}{MIMO}{Multiple-Input Multiple-Output}
\newacronym{ris}{RIS}{Reconfigurable Intelligent Surfaces}
\newacronym{nn}{NN}{Neural Network}
\newacronym{ml}{ML}{Machine Learning}
\newacronym{mcs}{MCS}{Modulation and Coding Scheme}
\newacronym{e2e}{E2E}{End-to-End}
\newacronym{dnn}{DNN}{Deep Neural Network}
\newacronym{mc}{MC}{Multicarrier}
\newacronym{noma}{NOMA}{Non-Orthogonal Multiple Access}
\newacronym{los}{LOS}{Line-of-Sight}
\newacronym{nlos}{NLOS}{Non-Line-of-Sight}
\newacronym{bpsk}{BPSK}{Binary Phase Shift Keying}
\newacronym{qpsk}{QPSK}{Quadrature Phase Shift Keying}
\newacronym{oqpsk}{OQPSK}{Offset Quadrature Phase Shift Keying}
\newacronym{awgn}{AWGN}{Additive White Gaussian Noise}
\newacronym{snr}{SNR}{Signal-to-Noise Ratio}
\newacronym{mse}{MSE}{Mean Squared Error}
\newacronym{qam}{QAM}{Quadrature Amplitude Modulation}
\newacronym{psk}{PSK}{Phase-Shift Keying}
\newacronym{bs}{BS}{Base Station}
\newacronym{mu}{MU}{Multi-User}
\newacronym{ber}{BER}{Bit Error Rate}
\newacronym{pdp}{PDP}{Power Delay Profile}
\newacronym{nasa}{NASA}{National Aeronautics and Space Administration}
\newacronym{dsn}{DSN}{Deep Space Network}
\newacronym{noct}{NOCT}{Network Operations Control Team}
\newacronym{jpl}{JPL}{Jet Propulsion Laboratory}
\newacronym{twta}{TWTA}{Traveling Wave Tube Amplifier}
\newacronym{mgs}{MGS}{Mars Global Surveyor}
\newacronym{dgmsk}{DGMSK}{Differential Gaussian Minimum Shift Keying}
\newacronym{mer}{MER}{Mars Exploration Rovers}
\newacronym{pcm}{PCM}{Pulse Code Modulation}
\newacronym{ccsds}{CCSDS}{Consultative Committee for Space Data Systems}
\newacronym{esa}{ESA}{European Space Agency}
\newacronym{fsa}{FSA}{Russian Federal Space Agency}
\newacronym{jaxa}{JAXA}{Japan Aerospace Exploration Agency}
\newacronym{cnsa}{CNSA}{China National Space Administration}
\newacronym{mro}{MRO}{Mars Reconnaissance Orbiter}
\newacronym{eirp}{EIRP}{Equivalent Isotropic Radiated Power}
\newacronym{msl}{MSL}{Mars Science Laboratory}
\newacronym{idsn}{IDSN}{Indian Deep Space Network}
\newacronym{edl}{EDL}{Entry-Decent-Landing}
\newacronym{imf}{IMF}{Interplanetary Magnetic Field}
\newacronym{sep}{SEP}{Sun–Earth–probe}
\newacronym{stk}{STK}{System Tool Kit}
\newacronym{wsn}{WSN}{Wireless Sensor Network}
\newacronym{per}{PER}{Packet Error Rate}
\newacronym{lte}{LTE}{Long-Term Evolution}
\newacronym{ue}{UE}{User Equipments}
\newacronym{dem}{DEM}{Digital Elevation Model}
\newacronym{wi-fi}{Wi-Fi}{Wireless Fidelity}
\newacronym{man}{MAN}{Metropolitan Area Network}
\newacronym{dtn}{DTN}{Delay Tolerant Networking}
\newacronym{sdr}{SDR}{Software-Defined Radio}
\newacronym{lcrd}{LCRD}{Laser Communications Relay Demonstration}
\newacronym{illuma-t}{ILLUMA-T}{Integrated LCRD Low-Earth Orbit User Modem and Amplifier Terminal}
\newacronym{psg}{PSG}{Planetary Spectrum Generator}
\newacronym{uav}{UAV}{Unmanned Aerial Vehicles}
\newacronym{nfp}{NFP}{Networked Flying Platforms}
\newacronym{haps}{HAPS}{High-Altitude Platform Station}
\newacronym{ap}{AP}{Access Point}
\newacronym{fso}{FSO}{Free-Space Optics}
\newacronym{leo}{LEO}{Low Earth orbit}
\newacronym{v-hetnet}{V-HetNet}{Vertical Heterogeneous Network}
\newacronym{tcs}{TCS}{Thermal Control System}
\newacronym{dsh}{DSH}{Deep Space Habitat}
\newacronym{satcom}{SatCom}{Satellite Communications}
\newacronym{geo}{GEO}{Geostationary Orbit}
\newacronym{isl}{ISL}{Inter-Satellite Links}
\newacronym{mspa}{MSPA}{Multiple Spacecraft Per Antenna}
\newacronym{mupa}{MUPA}{Multiple Uplinks Per Antenna}
\newacronym{w-lan}{W-LAN}{Wireless Local Area Network}
\newacronym{dsrt}{DSRT}{Deep Space Relay Terminal}
\newacronym{drl}{DRL}{Deep Reinforcement Learning}
\newacronym{dl}{DL}{Deep Learning}
\newacronym{mac}{MAC}{Media Access Control}
\newacronym{sdn}{SDN}{Software-Defined Networks}
\newacronym{son}{SON}{Self-organizing networks}
\newacronym{mrnss}{MRNSS}{Mars Regional Navigation Satellite System}
\newacronym{od}{OD}{Orbit Determination}
\newacronym{scan}{SCaN}{Space Communications and Navigation}
\newacronym{ioc}{IOC}{Initial Operating Capability}
\newacronym{pace}{PACE}{Plankton, Aerosol, Cloud, Ocean Ecosystem}
\newacronym{kplo}{KPLO}{Korea Pathfinder Lunar Orbiter}
\newacronym{cdma}{CDMA}{Code-Division Multiple Access}
\newacronym{pn}{PN}{Pseudo-Random Noise}
\newacronym{ofdma}{OFDMA}{Orthogonal Frequency-Division Multiple Access}
\newacronym{oma}{OMA}{Orthogonal Multiple Access}
\newacronym{fdma}{FDMA}{Frequency Division Multiple Access}
\newacronym{tdma}{TDMA}{Time Division Multiple Access}
\newacronym{mapp}{MAPP}{Mobile Autonomous Prospecting Platform}
\newacronym{3d}{3D}{Three-Dimensional}
\newacronym{cpm}{CPM}{Channel Path Map}
\newacronym{csm}{CSM}{Channel Shadowing Map}
\newacronym{cgm}{CGM}{Channel Gain Map}
\newacronym{csi}{CSI}{Channel State Information}
\newacronym{ckm}{CKM}{Channel Knowledge Map}
\newacronym{qos}{QoS}{Quality of Service}
\newacronym{rms}{RMS}{Root-mean-square}
\newacronym{fm}{FM}{Frequency Modulation}
\newacronym{isro}{ISRO}{Indian Space Research Organisation}
\newacronym{jsc}{JSC}{Johnson Space Center}
\newacronym{uhf}{UHF}{Ultra High Frequency}
\newacronym{mr}{MR}{Mars Relay}
\newacronym{iss}{ISS}{International Space Station}
\newacronym{ip}{IP}{Internet Protocol}
\newacronym{mbr}{MBR}{Mars Balloon Relay}
\newacronym{gso}{GSO}{Geosynchronous Orbit}
\newacronym{aso}{ASO}{Areo­synchronous Orbit}
\newacronym{rf}{RF}{Radio Frequency}
\newacronym{acm}{ACM}{Adaptive Coding and Modulation}
\newacronym{cgr}{CGR}{Contact Graph Routing}
\newacronym{icn}{ICN}{Information Centric Networking}
\newacronym{ltp}{LTP}{Licklider Transmission Protocol}
\newacronym{rs}{RS}{Reed-Solomon}
\newacronym{bp}{BP}{Bundle Protocol}
\newacronym{dttp}{DTTP}{Delay-Tolerant Transport Protocol}
\newacronym{msr}{MSR}{Moderate Source Routing}
\newacronym{lms}{LMS}{Land Mobile Satellite}
\newacronym{wdm}{WDM}{Wavelength Division Multiplexing}
\newacronym{wdma}{WDMA}{Wavelength Division Multiple Access}
\newacronym{mpc}{MPC}{Multi-Period-Clustering}
\newacronym{sspa}{SSPA}{Solid-State Power Amplifier }
\newacronym{sdst}{SDST}{Small Deep-Space Transponder}
\newacronym{edm}{EDM}{Entry Descent and Landing Demonstrator Module}
\newacronym{tgo}{TGO}{Trace Gas Orbiter}
\newacronym{hga}{HGA}{High-Gain Antenna}
\newacronym{mga}{MGA}{Medium-Gain Antenna}
\newacronym{lga}{LGA}{Low-Gain Antenna}
\newacronym{rsp}{RSP}{Rover and Surface Platform}
\newacronym{istrac}{ISTRAC}{ISRO Telemetry, Tracking and Command Network}
\newacronym{isse}{ISSE}{Beijing Institute of Spacecraft System Engineering}
\newacronym{rhcp}{RHCP}{Right Hand Circular Polarization}
\newacronym{vswr}{VSWR}{Voltage Standing Wave Ratio}

\begin{document}

	\title{Communications for the Planet Mars: \\ Past, Present, and Future}
	%
	%
	%
	
	\author{Enes Koktas~\IEEEmembership{Student Member,~IEEE} and Ertugrul Basar,~\IEEEmembership{Fellow,~IEEE} 
		\thanks{E. Koktas and E. Basar are with CoreLab, Department of Electrical and Electronics Engineering, Ko\c{c} University, Istanbul, Turkey (e-mail: ekoktas21@ku.edu.tr, ebasar@ku.edu.tr).}
	}

	\maketitle
	
	
	%
	\begin{abstract}
			Space exploration has been on the rise since the 1960s. Along with the other planets such as Mercury, Venus, Saturn, and Jupiter, Mars certainly plays a significant role in the history of space exploration and has the potential to be the first extraterrestrial planet to host human life. In this context, tremendous effort has been put into developing new technologies to photograph, measure, and analyze the red planet. As the amount of data collected from science instruments around and on Mars increased, the need for fast and reliable communication between Earth and space probes has emerged. However, communicating over deep space has always been a big challenge due to the propagation characteristics of radio waves. Nowadays, the collaboration of private companies like SpaceX with space agencies to make Mars colonization a reality, introduces even more challenges, such as providing high data rate, low latency, energy-efficient, reliable, and mobility-resistant communication infrastructures in the Martian environment. Propagation medium and wireless channel characteristics of Mars should be extensively studied to achieve these goals. This survey article presents a comprehensive overview of the Mars missions and channel modeling studies of the near-Earth, interstellar, and near-planet links. Studies featuring three-dimensional (\acrshort{3d}) channel modeling simulations on the Martian surface are also reviewed. We have also presented our own computer simulations considering various scenarios based on realistic 3D Martian terrains using the Wireless Insite software. Path loss exponent, power delay profile, and root-mean-square delay spread for these scenarios are calculated and tabularized in this study. Furthermore, future insights on emerging communication technologies for Mars are given.   
	\end{abstract}
	

	\printglossary[style=tree, type=\acronymtype, title=Nomenclature, nogroupskip, nonumberlist]

	\section{Introduction}
	%
	%
	%
	%
	\IEEEPARstart{T}{he} red planet: the ultimate dream of humankind.
	Mars has captured humankind's attention since it was spotted glowing in the night sky. Early on, the planet's ruddy color distinguished it from its neighbors, each of which was unique in its manner, but none of them traced a reddish route across Earth's skies. Then, telescopes first showed diverse features, patterns, and landforms of the Martian surface that researchers initially mistook for evidence of a vibrant Martian civilization. As of today, it is known that Mars is devoid of artificial structures. However, it has also been discovered that the poisonous and dry world we see now may have once been just as livable as Earth $ 3.5 $ billion years ago. 
	Humans have been exploring the planet Mars since the 1960s in an effort to learn more about how the planet formed and whether it has ever supported extraterrestrial life. Only uncrewed spacecraft have visited the red planet thus far, but that may soon change. In order to expand the exploration of the Martian surface, numerous additional missions will be launched by the National Aeronautics and Space Administration (\acrshort{nasa}) to reach the goal of having the first people set foot on Mars in the 2030s \cite{drake_2021_Why_we_exp}.
	
	Mars is a suitable exploration destination since it is the second-closest planet to Earth, has a climate and atmosphere comparable to that of Earth, provides a significant potential for scientific study, and has been mapped and studied for decades \cite{west_2020_five_reasons}.
	In this regard, many successful and unsuccessful missions have been performed since the 1960s. Space agencies worldwide are continuously conducting research and development on new spacecraft that would hopefully open new opportunities to better understand the nature of the planet Mars. On the other hand, journeys of spacecraft to Mars are not easy. They require extremely detailed mission planning and integration of highly complex engineering disciplines. One of the most important disciplines is electrical engineering, specifically telecommunications, because the essential duty of these missions is to collect and transfer as much data as possible. Also, it is crucial to send command data to spacecraft from the control center.
	
	This article presents a comprehensive overview of Mars communications in light of past experiences and future aspects. First, a summary of early and present missions is provided by discussing their objectives, antenna features, frequency bands, communication standards, and relay capabilities. Comparisons of the Mars missions regarding technical features and communication parameters are highlighted. Some of the exemplary tables given in this article are adapted from relevant sources to compare certain parameters; however, we have tried to provide some useful insights for readers in that regard. In the next step, general statistical parameters of Mars and their comparisons with the Earth's are provided. The effects of the Martian atmosphere, including ionosphere, troposphere, cloud, fog, and various gases, along with dust, on radio propagation are discussed. Theoretical channel modeling studies of near-Earth, deep space, and near-planet links are reviewed. The effect of solar scintillation is examined. Additionally, 3D channel modeling computer simulations from the literature utilizing various software such as System Tool Kit (\acrshort{stk}), HertzMapper, OMNET++, SIMULINK, and MATLAB are summarized. One of the aims of this study is to make an original contribution to the literature on Martian channel modeling. In light of this goal, we have performed our own computer simulations using the Wireless Insite software. These new computer simulations consider ray-tracing-based channel modeling for various communication scenarios envisioned on different Martian terrains, which include the landing sites of the Curiosity and Perseverance rovers. Path loss exponent, power delay profile (\acrshort{pdp}), root-mean-square (\acrshort{rms}) delay spread, and shadowing parameters are calculated for each scenario. Provided results can be used to study potential Mars communication standards and design new waveforms for the Mars environment. Lastly, this survey article also provides very interesting new insights and perspectives regarding the use of emerging technologies like reconfigurable intelligent surfaces (\acrshort{ris}s) for the first time for Mars communications, as well as multiple access schemes, cognitive radio, terahertz (\si{\tera\hertz}) communication, software-defined radio (\acrshort{sdr}), and environment-aware communication. 
	
	\cite{KonsgenForster_2021} also provides a brief overview of the current state of space communication in national agencies, standardization organizations, and active research in the area, targeting readers who are new to the subject. Additionally, this study provides a brief summary of historical evolution before summarizing the authors' views on potential challenges for space communication in the future.
	Different from \cite{KonsgenForster_2021}, which is one of the earliest surveys in the literature, our article provides a comprehensive overview of all the successful missions from the past and present, demonstrates some valuable insights on the physical problems, gives technical details, and provides novel channel modeling computer simulations.
	
	The rest of the paper is summarized as follows: Section \ref{sec:historical_overview} provides a brief historical overview of Mars missions and their communication capabilities. Section \ref{sec:mars_atmosphere} introduces Martian atmospheric effects on radio propagation. Channel modeling studies are given in Section \ref{sec:channel_modeling}. Section \ref{sec:emerging_tech} discusses the emerging technologies for Mars communication. Finally, Section \ref{sec:Conclusions} concludes the paper.
	\section{A Historical Overview} \label{sec:historical_overview}
	This section provides a brief historical overview of Mars missions and analyzes them from a communication engineering perspective. 
	
	Before getting into the Mars missions, NASA's Deep Space Network (\acrshort{dsn}), a global satellite network that has given life to all these missions over the years, needs to be mentioned. The DSN is essentially a group of giant antennas ranging in diameter from $ 34 $ to \SI{70}{\meter}. There are three DSN facilities around the world having several $ 34 $-\si{\meter} antennas and one $ 70 $-\si{\meter} antenna. One DSN facility is in Goldstone, California's Mojave Desert, another in Madrid, Spain, and the third is outside Canberra, Australia. These locations are positioned around \ang{120} apart from each other to provide 24-hour coverage for all missions beyond Earth's orbit. In order to prevent signal interference from surrounding regions, each facility is placed in a semi-mountainous, bowl-shaped territory. Since Earth rotates on its axis at a rate of $ 0.004 $ degrees per second, the antennas should be able to navigate at an extremely precise rate of near thousandths of a degree per second to stay pointed at the spaceship. DSN's daily operators are $ 34 $-\si{\meter} antennas which can handle most of the links demanded by NASA alone or in groups of two or three. However, if the distance is very large for several cooperating $ 34 $-\si{\meter} antennas, the DSN officials working on Network Operations Control Team (\acrshort{noct}) at Jet Propulsion Laboratory's (\acrshort{jpl}) Deep Space Operations Center employ their $ 70 $-\si{\meter} antenna. The $ 70 $-\si{\meter} antennas are also employed when a spacecraft is unable to connect with its high-gain antenna (\acrshort{hga}) due to a predetermined important event, such as orbit insertion. For example, a $ 70 $-\si{\meter} antenna was utilized to return Apollo 13 to Earth successfully. Additionally, Neil Armstrong's well-known "That is one small step for a man. One giant leap for mankind" message is received by this antenna. The DSN is still the most sophisticated and sensitive telecommunications system that exists today \cite{krywko_2019_deep_space_dial_up}, \cite{Mukherjee_2013_Interplanetary_Internet}. Further details on the DSN antennas are given in \cite{taylor_deep_space}.
	\subsection{Early Missions}	
	The first successful mission to Mars was the Mariner 4 mission, carried out by NASA in 1964. The Mariner 4 spacecraft was the first to send photos from Mars. During the mission, $ 5.2 $ million bits of data were received thanks to its parabolic HGA with a diameter of \SI{116.8}{\centi\meter} and omnidirectional low-gain antenna (\acrshort{lga}) with a length of \SI{223.5}{\centi\meter}. The receivers and transmitters worked on the S-band because the signals in this band easily penetrate through Earth's atmosphere \cite{reiff_sloan_james_Mariner4}.
	
	Flights to Mars until 1971 were "flyby" flights. In these flights, spacecraft did not orbit the planet; they passed as close to it as possible, collected the data, and then sent it back to Earth. The Mariner 9, sent in 1971, was the first spacecraft to orbit Mars. In this way, Mariner was able to take nine thousand photos and obtain much information about the atmosphere of Mars. It had a \SI{1.44}{\meter} tall omnidirectional LGA and a parabolic HGA. It also had a medium-gain cone antenna. The \SI{168}{\meter} long $ 8 $-track tape it had could record $ 180 $ million bits at $ 132 $ $ \mathrm{kbit}/\mathrm{s} $. The receivers and transmitters worked on the S-band as in Mariner 4 \cite{Schneiderman_Mariner9}.
	
	The first successful missions that an orbiter and a lander were sent together were Viking 1 and Viking 2, carried out in 1975. Viking 1 and Viking 2 landers stayed operational for six and four years, respectively. The landers conducted various experiments and collected soil samples. On the other side, Viking 1 and Viking 2 orbiters took tens of thousands of photos, contributing to the creation of a global map of Mars.
	In the orbiter, communication was carried out with a \SI{20}{\watt} transmitter in the S-band, with a frequency of \SI{2.3}{\giga\hertz} and two traveling wave tube amplifiers (\acrshort{twta}). X-band  (\SI{8.4}{\giga\hertz}) downlink has also been added for communication experiments. Uplink was in S-band with a carrier frequency of \SI{2.1}{\giga\hertz}. The orbiter had a $ 2 $-axis steerable high-gain dish antenna with a diameter of approximately \SI{1.5}{\meter}. There was also a fixed LGA. Two recorders could store $ 1280 $ $ \mathrm{Mbits} $, and there was a \SI{381}{\mega\hertz} relay radio on the spacecraft \cite{Strobel_Vikings_Orbiter}.	
	In the lander, communication was carried out in the S-band with a \SI{20}{\watt} transmitter, as in the orbiter. It had a $ 2 $-axis steerable high-gain dish antenna. There was also a fixed LGA. Both of these antennas could communicate directly with Earth. The antenna in the ultra high frequency (\acrshort{uhf}) band (\SI{381}{\mega\hertz}) served as a one-way relay to the orbiter using the \SI{30}{\watt} relay radio. The data was recorded on a $ 40 $ $ \mathrm{Mbit} $ recorder \cite{Strobel_Vikings_Lander}.
	
	\subsection{Missions in 1990s-2000s}
	A major turning point for Mars exploration after the Viking missions was the Mars Global Surveyor (\acrshort{mgs}) mission in 1996. This mission lasted nine years, being the longest mission ever. It also took the highest resolution photos possible at those times and helped to perform other tasks. When it was first launched, it used the LGA until it got far enough away from Earth to use its \SI{1.5}{\meter} diameter HGA \cite{nasa_MGS_facts}.
	The MGS used a \SI{1.5}{\meter} diameter high-gain dish antenna to communicate with Earth. The speed at which data is transmitted from Earth to orbiter on uplink was $ 500 $ $ \mathrm{bps} $. The orbiter used radio signals in the X-band (\SI{8.4}{\giga\hertz}) for all transmissions. Surveyor had only \SI{25}{\watt} power for transmission. Therefore, the DSN was the only way to obtain signals on Earth. Thanks to the DSN, up to $ 85.3 $ $ \mathrm{kbps} $ could be delivered to Earth on the downlink \cite{nasa_keeping_in_touch}. MGS's average data return throughput for each rover is estimated to be around $ 49 $ $ \mathrm{Mbits} $ per Martian day ($ \mathrm{sol}) $ \cite{taylor_2016_mars_exp_rovers}.
	Mars Relay (\acrshort{mr}), one of the six primary devices on the Surveyor, acted as a relay between the landers on the Martian surface and Earth. In this way, landers did not need large-scale, high-mass dish antennas \cite{nasa_Mars_relay}.
	Relay’s transmit frequency to stations was \SI{437.1}{\mega\hertz} and receive frequencies from stations are \SI{401.5}{\mega\hertz} and \SI{405.6}{\mega\hertz}. At the time, due to the closeness of these frequencies to the amateur radio bands, Hams (amateur radio operators) had the opportunity to assist NASA by keeping an eye out for the MR beacon signal at \SI{437.1}{\mega\hertz} \cite{NASA_1996_MGS_relay}. It could achieve $ 8 $ to $ 128 $ $ \mathrm{kbps} $ on the RF link \cite{nasa_MR_info}.
	
	The first successful rover was made in 1996, 20 years after the Viking mission. The Mars Pathfinder \& Sojourner mission was planned for a month, but the data transfer lasted about two months. 
	Similar to the Viking lander, the Pathfinder had a low-gain and steerable HGA. It communicated with Earth in the X-band, which means it used the frequency of \SI{8.4}{\giga\hertz} for downlink and \SI{7.2}{\giga\hertz} for uplink \cite{Kahan_mars_precession_rate}.
	The steerable HGA on the lander allowed communication with the $ 70 $-\si{\meter} DSN antenna at a rate of $ 5.5 $ $ \mathrm{kbps} $. The solar panels on the lander could generate enough energy to transmit for two-four hours in every $ 24.6 $ hours (one Martian day) \cite{Golombek_mars_pathfinder}.
	The radio modem inside the Sojourner rover operated on \SI{459.7}{\mega\hertz}. Its channel bandwidth was \SI{25}{\kilo\hertz}. It used a version of frequency modulation (\acrshort{fm}); differential Gaussian minimum shift keying (\acrshort{dgmsk}). Its transmit power was \SI{100}{\milli\watt}, and the maximum data rate was $ 9.6 $ $ \mathrm{kbps} $ \cite{stride_microrover_radios_antennas}.

	After Sojourner, Opportunity and Spirit twin rovers were sent to take the rover missions to the next level. Spirit continued its mission for ten years, traveling \SI{7.73}{\kilo\meter}, and Opportunity for fourteen years, traveling \SI{45.16}{\kilo\meter}. The main goal of the Mars exploration rovers (\acrshort{mer}) mission was to deploy two mobile research laboratories on the Martian surface for remote geology studies, including the analysis of a variety of rocks and soils that could include hints about previous water content. With the objective of adjusting and verifying orbital spectroscopic remote sensing, the mission aimed to carry out novel discoveries of Mars geology, consisting of the first microscopic level examinations of rock samples and comprehensive analysis of the surface conditions. The mission intended to accomplish these goals while bringing the general public to the excitement and wonder of space exploration \cite{Taylor2016MarsER}.
	During missions on the surface, rovers used LGAs to receive commands at a minimum speed of $ 7.8125 $ $ \mathrm{bps} $ and transmitted data at a minimum speed of $ 10 $ $ \mathrm{bps} $. The speed of the steerable HGA depended on the distance from Mars to Earth. In short distances, its command receiving speed was up to $ 2 $ $ \mathrm{kbps} $, and the data transmitting speed up to $ 28.8 $ $ \mathrm{kbps} $. 
	At the physical layer (\acrshort{phy}) of MERs, some important topics need to be considered. Firstly, they consume \SI{6}{\watt} and \SI{12}{\watt} on receiving and transmitting, respectively. They had one forward (orbiter to rover) frequency of \SI{437.1}{\mega\hertz}. They had one return (rover to orbiter) frequency of $ 401.585625 $ $ \mathrm{MHz} $ to Odyssey and Mars Express and another frequency of $ 401.528711 $ $ \mathrm{MHz} $ to MGS. They used a version of pulse code modulation (\acrshort{pcm}) with a residual carrier and a modulation index of $ 1.05 $ \si{\radian}. Supported data rates were $ 8 $, $ 32 $, $ 128 $, and $ 256 $ $ \mathrm{kbps} $ for both forward and return links. No encoding scheme was used in the forward link. On the other hand, convolutional encoding with $ 1/2 $ rate and $ 7 $ constraint length was used in the return link \cite{Taylor_MER_telecom}. Detailed information about the rover is given in \cite{Taylor2016MarsER}.

	Odyssey orbiter was launched on April 7, 2001. It was developed to create a detailed map of chemical elements and minerals, study the radiation levels for human missions and search for proof that water actually exists on Mars \cite{Every_miss_odyssey}. Since it became operational, it has also been used as a relay to send data from surface rovers and landers. 
	An organization called The Consultative Committee for Space Data Systems (\acrshort{ccsds}), which has member agencies like NASA, European Space Agency (\acrshort{esa}), The Russian Federal Space Agency (\acrshort{fsa}), Japan Aerospace Exploration Agency (\acrshort{jaxa}), and China National Space Administration (\acrshort{cnsa}) established the Proximity-1 protocol, which is a bi-directional space link layer protocol for use of space missions \cite{proximity1}. This protocol provides interoperability between spacecraft like landers, rovers, and orbiters of different space agencies. Mars Odyssey was the first orbiter compatible with the CCSDS Proximity-1 protocol. 
	Odyssey carries redundant Cincinnati Electronics-505 (CE-505) UHF transceivers as its relay payload. It uses \SI{437.1}{\mega\hertz} for the orbiter-to-lander link and $ 401.585625 $ $\mathrm{MHz}$ for the lander-to-orbiter link. It supports data rates of $ 8 $, $ 32 $, $ 128 $, and $ 256 $ $ \mathrm{kbps} $. Odyssey's average data return throughput for each rover is estimated to be around $ 56 $ $ \mathrm{Mbits} $ per $ \mathrm{sol} $ \cite{taylor_2016_mars_exp_rovers}. To communicate with Earth, it uses a \SI{1.3}{\meter} diameter HGA, a medium-gain antenna (\acrshort{mga}), mounted on the HGA, and a LGA in the X-band supplied with \SI{15}{\watt} power amplifiers \cite{Edwards_relay_comm}, \cite{NASA_Odyssey_descanso}.

	Although most successful missions to Mars belong to NASA, the discovery of Mars has been a topic of global interest. Researchers and space agencies from many countries have recently been working on new Mars missions. One is the Mars Express mission, launched by ESA in 2003.
	Depending on the distance between Mars and Earth, the time it takes to send data to the Mars Express orbiter ranges from $ 4 $ to \SI{25}{\minute}. The data rate at the closest distance is $ 228 $ $ \mathrm{kbps} $, while at the furthest distance, it is $ 57 $ $ \mathrm{kbps} $. It has a dish antenna with a diameter of \SI{1.65}{\meter} and two LGAs used when the spacecraft is close to Earth \cite{esa_mars_express}.	
	The UHF radio Mars Express is carrying is called Melacom \cite{Schmit2003}. Melacom uses CCSDS Promximity-1 protocol as Odyssey and The Mars Reconnaissance Orbiter (\acrshort{mro}). It supports data rates from $ 2 $ to $ 128 $ $ \mathrm{kbps} $, including only powers of two \cite{Edwards_relay_comm}.
	Signals are transmitted to Earth using \SI{7.1}{\giga\hertz} frequency in the X-band and to the orbiter using \SI{2.1}{\giga\hertz} frequency in the S-band. The spacecraft has $ 12 $ $ \mathrm{Gbit} $ of solid-state mass memory \cite{esa_mars_express_telecoms}. Detailed information about the mission is given in in \cite{esa_earth_observation_portal}.

	Another spacecraft sent to study the geological and climatic structure of Mars is MRO, which launched on August 12, 2005. It also acts as a communications relay. MRO has $ 160 $ $ \mathrm{Gbit} $ of solid-state memory and can reach a data rate of $ 6 $ $ \mathrm{Mbps} $, but this is a maximum of $ 3.5 $ $ \mathrm{Mbps} $ in practice. The communication module operates on the X-band at a frequency of about \SI{8}{\giga\hertz}. One-way communication takes about \SI{21}{\minute} when Mars is furthest from Earth. 
	\begin{table*}[!t]
		\caption{Comparison of Relay Characteristics of Mars Orbiters \cite{Edwards2004AMT}.}
		\label{table:comparison_of_mars_relay}
		\centering
		\renewcommand{\arraystretch}{1.4}
		\begin{tabular}{lcccc}
			\rowcolor[HTML]{D6D6D6} 
			& \textbf{\begin{tabular}[c]{@{}c@{}}Mars Global \vspace{-1mm}\\ Surveyor\end{tabular}} & \textbf{\begin{tabular}[c]{@{}c@{}}Mars \vspace{-1mm}\\ Odyssey\end{tabular}} & \textbf{\begin{tabular}[c]{@{}c@{}}Mars \vspace{-1mm}\\ Express\end{tabular}} & \textbf{\begin{tabular}[c]{@{}c@{}}Mars Reconnaissance \vspace{-1mm}\\ Orbiter\end{tabular}} \\
			\rowcolor[HTML]{EFEFEF} 
			\textbf{Agency}        & NASA                                                                     & NASA                                                            & ESA                                                             & NASA                                                                           \\
			\rowcolor[HTML]{D6D6D6} 
			\textbf{Launch Date}        & Nov. 8, 1996                                                             & April 7, 2001                                                   & June 2, 2003                                                    & Mar., 2006                                                                      \\
			\rowcolor[HTML]{EFEFEF} 
			\textbf{UHF Radio}     & Mars Relay (CNES)                                                        & CE-505                                                          & Melacom                                                         & Electra                                                                        \\
			\rowcolor[HTML]{D6D6D6} 
			\textbf{Link Protocol} & Mars Balloon Relay (\acrshort{mbr})                                                 & CCSDS Proximity-1                                               & CCSDS Proximity-1                                               & CCSDS Proximity-1                                                              \\
			\multicolumn{5}{l}{\cellcolor[HTML]{EFEFEF}\textbf{Forward Link}}                                                                                                                                                                                                                                                                              \\
			\rowcolor[HTML]{D6D6D6} 
			\quad- Frequency   & 437.1 MHz                                                                & 437.1 MHz                                                       & 437.1 MHz                                                       & 435-450 MHz                                                                    \\
			\rowcolor[HTML]{EFEFEF} 
			\quad- Data Rates  & n/a (MBR tones only)                                                     & 8 kbps                                                          & 2, 8 kbps                                                       & 1, 2, 4, ... , 1024 kbps                                                       \\
			\multicolumn{5}{l}{\textbf{\cellcolor[HTML]{D6D6D6}Return Link}}                                                                                                                                                                                                                                                                               \\
			\rowcolor[HTML]{EFEFEF} 
			\quad- Frequency   & 401.528711 MHz                                                           & 401.585625 MHz                                                  & 401.585625 MHz                                                  & 390-405 MHz                                                                    \\
			\rowcolor[HTML]{D6D6D6} 
			\quad- Data Rates  & 8, 128 kbps                                                              & 8, 32, 128, 256 kbps                                            & 2, 4, ... , 128 kbps                                            & 1, 2, 4, ... , 1024 kbps                                                       \\
			\end{tabular}
	\end{table*}
	At this distance, the MRO transmits to the \SI{34}{\meter} antenna at a rate of $ 600 $ $ \mathrm{kbps} $. At Mars's closest approach to Earth, the MRO sends data at about $ 2.6 $ $ \mathrm{Mbps} $ to the \SI{34}{\meter} antenna and about $ 3.5 $ $ \mathrm{Mbps} $ to the \SI{70}{\meter} antenna \cite{nasa_MRO_telecom}. MRO's average data return throughput for each rover is estimated to be around $ 49 \mathrm{Mb/sol} $ \cite{taylor_2016_mars_exp_rovers}.	
	MRO has a \SI{100}{\watt} TWTA and a \SI{3}{\meter} HGA for the Earth link, which can be considered a highly-capable link and a significant improvement over the Odyssey and MGS's $ 15 $ and \SI{25}{\watt} power amplifiers, respectively. In comparison to MGS and Odyssey, the corresponding X-band equivalent isotropic radiated power (\acrshort{eirp}) of $ 96 $ $ \mathrm{dBm} $ indicates an improvement of $ 13 $–\SI{15}{\decibel} \cite{EDWARDS2007_Key_Tel_Tech}. Apart from this, it has an Electra proximity payload for the relay link \cite{Edwards_relay_comm}. A comparison of relay characteristics and capabilities of these orbiters and the Mars Express is given in Table \ref{table:comparison_of_mars_relay}. The relay technologies of the orbiters are similar to each other except for different data rates. So, it can be claimed that there is room for growth in orbiter relay technologies. Key telecommunication system parameters of MGS, Odyssey, and MRO are compared in Table \ref{table:key_telecommunication_system}. MRO has remarkable dominance over MGS and Odyssey, pushing the limits of physical constraints such as size, power, and mass. Also, the data rates of these orbiters in specific frequency bands are given in Table \ref{table:characteristics_of_Comm_assets}. MRO gives the maximum data rate at Ka-band, but MGS gives it at the UHF band because data rates of corresponding frequency bands vary with antenna diameters, gains, and transmit and receive powers.
	A comparison over a simple communication scenario can be given to understand more clearly how fast the MRO is. For instance, it would require $ 460 $ times the universe's age for a standard mobile phone photo to get from Jupiter to Earth using Explorer 1 radio equipment. For the same operation, the early 1960s Pioneers 2 and 4 would require more than $ 633,000 $ years. Mariner 9 would need $ 55 $ $ \mathrm{hours} $ to complete. However, with MRO, the transfer could be done in three minutes \cite{krywko_2019_deep_space_dial_up}. Further details on MRO can be found in \cite{taylor_2016_MRO}.

	In the pursuit of finding water on Mars, Phoenix was designed to land farther north than any previous mission to determine whether this area was ever hospitable to life or not by studying soil and ice \cite{nasa_Mars_phoenix}.
	Phoenix decreases the cost and mass by not using any X-band telecommunication system. It uses only UHF relay telecommunications instead. It uses a CE-505 UHF transceiver same as MERs. Because it has no X-band direct-to-Earth link, it uses UHF relay links, especially with MRO and Odyssey. Both MRO and Odyssey offer more than $ 50 $ $ \mathrm{Mb/sol} $ from a single pass and between  $ 100 $-$ 150 $ $ \mathrm{Mb/sol} $ for two passes per sol \cite{Edwards_relay_comm}.

	Curiosity, also known as Mars Science Laboratory (\acrshort{msl}), is one of the rovers sent to look for evidence that microbial life forms previously existed on Mars. The main goal of the MSL mission is to deploy a mobile science laboratory on the Martian surface to analyze the geology of the landing area, examine planetary factors that affect habitability, and explore a wide spectrum of surface radiation. MSL is studying the radiation environment and researching the surface conditions while making discoveries about the geology of Mars utilizing enhanced micro-imaging and spectrometry \cite{Taylor2009MSL}.
	Curiosity communicates with orbiters in the UHF band and DSN in the X-band. The data rate directly between Earth varies between $ 500 $ $ \mathrm{bps} $ and $ 32 $ $ \mathrm{kbps} $. The data rate between Curiosity and MRO is automatically adjusted and can reach a maximum of $ 2 $ $ \mathrm{Mbps} $. The data rate between Curiosity and Odyssey can be selected as $ 128 $ $ \mathrm{kbps} $ or $ 256 $ $\mathrm{kbps}$. The rover communicates with each orbiter for around eight minutes per $ \mathrm{sol} $. $ 100 $ to $ 250 $ megabits of data can be transferred to an orbiter at that period. For the same $ 250 $ megabits to be transmitted directly to Earth, it may require up to twenty hours. Although Earth may be visible for a significantly extended period of time, the rover can only broadcast directly to Earth for a few hours per day because of power restrictions or conflicts with other scheduled tasks \cite{nasa_curiosity_comm}. 
	\begin{table*}[!t]
		\caption{Comparison of Communication Param eters of Mars Orbiters \cite{NASA_Odyssey_descanso}, \cite{NASA_MGS_descanso}, \cite{MGS_charact}, \cite{NASA_MRO_descanso}.}
		\label{table:key_telecommunication_system}
		\resizebox{\textwidth}{!}{%
			\renewcommand{\arraystretch}{1.5}
			\begin{tabular}{|
					>{\columncolor[HTML]{EEF5FF}}l |
					>{\columncolor[HTML]{FFF9F2}}c 
					>{\columncolor[HTML]{FFF9F2}}c 
					>{\columncolor[HTML]{FFF9F2}}c |
					>{\columncolor[HTML]{EFFFEB}}c 
					>{\columncolor[HTML]{EFFFEB}}c 
					>{\columncolor[HTML]{EFFFEB}}c |
					>{\columncolor[HTML]{E9FEFF}}c 
					>{\columncolor[HTML]{E9FEFF}}c 
					>{\columncolor[HTML]{E9FEFF}}c |}
				\hline
				\cellcolor[HTML]{DCEBFF}\textbf{Parameter \textbackslash Orbiter} &
				\multicolumn{3}{c|}{\cellcolor[HTML]{FFE1BE}\textbf{MGS}} &
				\multicolumn{3}{c|}{\cellcolor[HTML]{CCFFCB}\textbf{Odyssey}} &
				\multicolumn{3}{c|}{\cellcolor[HTML]{C1FDFF}\textbf{MRO}} \\ \hline
				\cellcolor[HTML]{EEF5FF} &
				\multicolumn{1}{c|}{\cellcolor[HTML]{FFF9F2}Uplink} &
				\multicolumn{1}{c|}{\cellcolor[HTML]{FFF9F2}X band} &
				7.2 GHz &
				\multicolumn{1}{c|}{\cellcolor[HTML]{EFFFEB}Uplink} &
				\multicolumn{1}{c|}{\cellcolor[HTML]{EFFFEB}X band} &
				7.2 GHz &
				\multicolumn{1}{c|}{\cellcolor[HTML]{E9FEFF}Uplink} &
				\multicolumn{1}{c|}{\cellcolor[HTML]{E9FEFF}X band} &
				7.183 GHz \\ \cline{2-10} 
				\cellcolor[HTML]{EEF5FF} &
				\multicolumn{1}{c|}{\cellcolor[HTML]{FFF9F2}} &
				\multicolumn{1}{c|}{\cellcolor[HTML]{FFF9F2}X band} &
				8.4 GHz &
				\multicolumn{1}{c|}{\cellcolor[HTML]{EFFFEB}} &
				\multicolumn{1}{c|}{\cellcolor[HTML]{EFFFEB}} &
				\cellcolor[HTML]{EFFFEB} &
				\multicolumn{1}{c|}{\cellcolor[HTML]{E9FEFF}} &
				\multicolumn{1}{c|}{\cellcolor[HTML]{E9FEFF}X band} &
				8.4 GHz \\ \cline{3-4} \cline{9-10} 
				\cellcolor[HTML]{EEF5FF} &
				\multicolumn{1}{c|}{\multirow{-2}{*}{\cellcolor[HTML]{FFF9F2}Downlink}} &
				\multicolumn{1}{c|}{\cellcolor[HTML]{FFF9F2}Ka band} &
				32 GHz &
				\multicolumn{1}{c|}{\multirow{-2}{*}{\cellcolor[HTML]{EFFFEB}Downlink}} &
				\multicolumn{1}{c|}{\multirow{-2}{*}{\cellcolor[HTML]{EFFFEB}X band}} &
				\multirow{-2}{*}{\cellcolor[HTML]{EFFFEB}8.4 GHz} &
				\multicolumn{1}{c|}{\multirow{-2}{*}{\cellcolor[HTML]{E9FEFF}Downlink}} &
				\multicolumn{1}{c|}{\cellcolor[HTML]{E9FEFF}Ka band} &
				32 GHz \\ \cline{2-10} 
				\multirow{-4}{*}{\cellcolor[HTML]{EEF5FF}\textit{Frequency Bands}} &
				\multicolumn{1}{c|}{\cellcolor[HTML]{FFF9F2}\begin{tabular}[c]{@{}c@{}}Uplink \& \vspace{-1mm}\\ Downlink\end{tabular}} &
				\multicolumn{1}{c|}{\cellcolor[HTML]{FFF9F2}UHF} &
				\begin{tabular}[c]{@{}c@{}}401 MHz - \vspace{-1mm}\\ 437.1 MHz\end{tabular} &
				\multicolumn{1}{c|}{\cellcolor[HTML]{EFFFEB}\begin{tabular}[c]{@{}c@{}}Uplink \& \vspace{-1mm}\\ Downlink\end{tabular}} &
				\multicolumn{1}{c|}{\cellcolor[HTML]{EFFFEB}UHF} &
				\begin{tabular}[c]{@{}c@{}}401 MHz - \vspace{-1mm}\\ 437.1 MHz\end{tabular} &
				\multicolumn{1}{c|}{\cellcolor[HTML]{E9FEFF}\begin{tabular}[c]{@{}c@{}}Uplink \& \\ Downlink\end{tabular}} &
				\multicolumn{1}{c|}{\cellcolor[HTML]{E9FEFF}UHF} &
				\begin{tabular}[c]{@{}c@{}}401 MHz - \vspace{-1mm}\\ 437.1 MHz\end{tabular} \\ \hline
				\cellcolor[HTML]{DCEBFF} &
				\multicolumn{3}{c|}{\cellcolor[HTML]{FFE1BE}4 $\times$ LGA (2Rx, 2Tx)} &
				\multicolumn{3}{c|}{\cellcolor[HTML]{CCFFCB}1 $\times$ LGA (Rx only), 1 x MGA (Tx only)} &
				\multicolumn{3}{c|}{\cellcolor[HTML]{C1FDFF}2 $\times$ LGA (Tx and Rx)} \\ \cline{2-10} 
				\cellcolor[HTML]{DCEBFF} &
				\multicolumn{3}{c|}{\cellcolor[HTML]{FFE1BE}1 $\times$ HGA (Tx and Rx) (1.5 m diameter)} &
				\multicolumn{3}{c|}{\cellcolor[HTML]{CCFFCB}1 $\times$ HGA (Tx and Rx) (1.3 m diameter)} &
				\multicolumn{3}{c|}{\cellcolor[HTML]{C1FDFF}1 $\times$ HGA (Tx and Rx) (3 m diameter)} \\ \cline{2-10} 
				\multirow{-3}{*}{\cellcolor[HTML]{DCEBFF}\textit{Antenna Quantity}} &
				\multicolumn{3}{c|}{\cellcolor[HTML]{FFE1BE}1 $\times$ UHF (for surface operators)} &
				\multicolumn{3}{c|}{\cellcolor[HTML]{CCFFCB}1 $\times$ UHF (for surface operators)} &
				\multicolumn{3}{c|}{\cellcolor[HTML]{C1FDFF}1 $\times$ UHF (for surface operators)} \\ \hline
				\cellcolor[HTML]{EEF5FF} &
				\multicolumn{1}{c|}{\cellcolor[HTML]{FFF9F2}} &
				\multicolumn{1}{c|}{\cellcolor[HTML]{FFF9F2}7.2 GHz} &
				37.4 dBi &
				\multicolumn{1}{c|}{\cellcolor[HTML]{EFFFEB}} &
				\multicolumn{1}{c|}{\cellcolor[HTML]{EFFFEB}7.2 GHz} &
				36.6 dBi &
				\multicolumn{1}{c|}{\cellcolor[HTML]{E9FEFF}} &
				\multicolumn{1}{c|}{\cellcolor[HTML]{E9FEFF}7.2 GHz} &
				45.2 dBi \\ \cline{3-4} \cline{6-7} \cline{9-10} 
				\cellcolor[HTML]{EEF5FF} &
				\multicolumn{1}{c|}{\cellcolor[HTML]{FFF9F2}} &
				\multicolumn{1}{c|}{\cellcolor[HTML]{FFF9F2}8.4 GHz} &
				39 dBi &
				\multicolumn{1}{c|}{\multirow{-2}{*}{\cellcolor[HTML]{EFFFEB}HGA}} &
				\multicolumn{1}{c|}{\cellcolor[HTML]{EFFFEB}8.4 GHz} &
				38.3 dBi &
				\multicolumn{1}{c|}{\cellcolor[HTML]{E9FEFF}} &
				\multicolumn{1}{c|}{\cellcolor[HTML]{E9FEFF}8.4 GHz} &
				46.7 dBi \\ \cline{3-7} \cline{9-10} 
				\cellcolor[HTML]{EEF5FF} &
				\multicolumn{1}{c|}{\multirow{-3}{*}{\cellcolor[HTML]{FFF9F2}HGA}} &
				\multicolumn{1}{c|}{\cellcolor[HTML]{FFF9F2}32 GHz} &
				49 dBi &
				\multicolumn{1}{c|}{\cellcolor[HTML]{EFFFEB}MGA} &
				\multicolumn{1}{c|}{\cellcolor[HTML]{EFFFEB}8.4 GHz} &
				16.5 dBi &
				\multicolumn{1}{c|}{\multirow{-3}{*}{\cellcolor[HTML]{E9FEFF}HGA}} &
				\multicolumn{1}{c|}{\cellcolor[HTML]{E9FEFF}32 GHz} &
				56.4 dBi \\ \cline{2-10} 
				\multirow{-4}{*}{\cellcolor[HTML]{EEF5FF}\textit{Antenna Gain}} &
				\multicolumn{1}{c|}{\cellcolor[HTML]{FFF9F2}LGA} &
				\multicolumn{1}{c|}{\cellcolor[HTML]{FFF9F2}8.4 GHz} &
				6.5 dBi &
				\multicolumn{1}{c|}{\cellcolor[HTML]{EFFFEB}LGA} &
				\multicolumn{1}{c|}{\cellcolor[HTML]{EFFFEB}7.2 GHz} &
				7 +/- 0.4 dBi &
				\multicolumn{1}{c|}{\cellcolor[HTML]{E9FEFF}LGA} &
				\multicolumn{1}{c|}{\cellcolor[HTML]{E9FEFF}8.4 GHz} &
				8.8 dBi \\ \hline
				\cellcolor[HTML]{DCEBFF} &
				\multicolumn{1}{c|}{\cellcolor[HTML]{FFE1BE}} &
				\multicolumn{1}{c|}{\cellcolor[HTML]{FFE1BE}X band} &
				\cellcolor[HTML]{FFE1BE}±0.8 deg &
				\multicolumn{1}{c|}{\cellcolor[HTML]{CCFFCB}HGA} &
				\multicolumn{2}{c|}{\cellcolor[HTML]{CCFFCB}1.9 deg} &
				\multicolumn{1}{c|}{\cellcolor[HTML]{C1FDFF}} &
				\multicolumn{1}{c|}{\cellcolor[HTML]{C1FDFF}X band} &
				\cellcolor[HTML]{C1FDFF}0.69 deg \\ \cline{3-7} \cline{9-10} 
				\cellcolor[HTML]{DCEBFF} &
				\multicolumn{1}{c|}{\multirow{-2}{*}{\cellcolor[HTML]{FFE1BE}HGA}} &
				\multicolumn{1}{c|}{\cellcolor[HTML]{FFE1BE}Ka band} &
				\cellcolor[HTML]{FFE1BE}±0.2 deg &
				\multicolumn{1}{c|}{\cellcolor[HTML]{CCFFCB}MGA} &
				\multicolumn{2}{c|}{\cellcolor[HTML]{CCFFCB}28 deg} &
				\multicolumn{1}{c|}{\cellcolor[HTML]{C1FDFF}} &
				\multicolumn{1}{c|}{\cellcolor[HTML]{C1FDFF}} &
				\cellcolor[HTML]{C1FDFF} \\ \cline{2-7}
				\multirow{-3}{*}{\cellcolor[HTML]{DCEBFF}\textit{Antenna Half-Power Beamwidth}} &
				\multicolumn{1}{c|}{\cellcolor[HTML]{FFE1BE}LGA} &
				\multicolumn{1}{c|}{\cellcolor[HTML]{FFE1BE}X band} &
				\cellcolor[HTML]{FFE1BE}±40 deg &
				\multicolumn{1}{c|}{\cellcolor[HTML]{CCFFCB}LGA} &
				\multicolumn{2}{c|}{\cellcolor[HTML]{CCFFCB}82 deg} &
				\multicolumn{1}{c|}{\multirow{-3}{*}{\cellcolor[HTML]{C1FDFF}HGA}} &
				\multicolumn{1}{c|}{\multirow{-2}{*}{\cellcolor[HTML]{C1FDFF}Ka band}} &
				\multirow{-2}{*}{\cellcolor[HTML]{C1FDFF}0.18 deg} \\ \hline
				\textit{Telemetry Coding} &
				\multicolumn{3}{c|}{\cellcolor[HTML]{FFF9F2}convolutional (rate 1/2, constraint length 7)} &
				\multicolumn{3}{c|}{\cellcolor[HTML]{EFFFEB}convolutional (rate 1/2, constraint length 7)} &
				\multicolumn{3}{c|}{\cellcolor[HTML]{E9FEFF}\begin{tabular}[c]{@{}c@{}}convolutional (rate 1/2, constraint length 7), \vspace{-1mm}\\ turbo code, Reed–Solomon\end{tabular}} \\ \hline
				\cellcolor[HTML]{DCEBFF} &
				\multicolumn{3}{c|}{\cellcolor[HTML]{FFE1BE}\begin{tabular}[c]{@{}c@{}}1 $\times$ solid-state power amplifier \vspace{-1mm}\\ (\acrshort{sspa}) (Ka Band) (1W)\end{tabular}} &
				\multicolumn{3}{c|}{\cellcolor[HTML]{CCFFCB}\begin{tabular}[c]{@{}c@{}}2 $\times$ small deep-space transponder \vspace{-1mm}\\ (\acrshort{sdst}) (15 W)\end{tabular}} &
				\multicolumn{3}{c|}{\cellcolor[HTML]{C1FDFF}2 $\times$ SDST} \\ \cline{2-10} 
				\multirow{-2}{*}{\cellcolor[HTML]{DCEBFF}\textit{Power Amplifiers}} &
				\multicolumn{3}{c|}{\cellcolor[HTML]{FFE1BE}2 $\times$ X band TWTA (25 W)} &
				\multicolumn{3}{c|}{\cellcolor[HTML]{CCFFCB}2 $\times$ SSPA (15 W)} &
				\multicolumn{3}{c|}{\cellcolor[HTML]{C1FDFF}\begin{tabular}[c]{@{}c@{}}2 $\times$ X band TWTA (100 W), \vspace{-1mm}\\ 1 $\times$ Ka band TWTA (35 W)\end{tabular}} \\ \hline
				\cellcolor[HTML]{EEF5FF} &
				\multicolumn{1}{c|}{\cellcolor[HTML]{FFF9F2}} &
				\multicolumn{2}{c|}{\cellcolor[HTML]{FFF9F2}} &
				\multicolumn{1}{c|}{\cellcolor[HTML]{EFFFEB}X band} &
				\multicolumn{2}{c|}{\cellcolor[HTML]{EFFFEB}\begin{tabular}[c]{@{}c@{}}7.8125 bps (emergency) - \vspace{-1mm}\\ 500 bps (125 bps in normal)\end{tabular}} &
				\multicolumn{3}{c|}{\cellcolor[HTML]{E9FEFF}\begin{tabular}[c]{@{}c@{}}7.8125, 15.625, 31.25, 62.5, 125, 250, \vspace{-1mm}\\ 500, 1000, 2000 bps\end{tabular}} \\ \cline{5-10} 
				\multirow{-2}{*}{\cellcolor[HTML]{EEF5FF}\textit{\vspace{-1mm}Command Data Rate}} &
				\multicolumn{1}{c|}{\multirow{-2}{*}{\cellcolor[HTML]{FFF9F2}\vspace{-1mm}X band}} &
				\multicolumn{2}{c|}{\multirow{-2}{*}{\cellcolor[HTML]{FFF9F2}\begin{tabular}[c]{@{}c@{}}\vspace{-1mm}7.8125 bps (emergency) - \\ 500 bps (125 bps in normal)\end{tabular}}} &
				\multicolumn{1}{c|}{\cellcolor[HTML]{EFFFEB}UHF} &
				\multicolumn{2}{c|}{\cellcolor[HTML]{EFFFEB}8, 32, 128, and 256 kbps} &
				\multicolumn{1}{c|}{\cellcolor[HTML]{E9FEFF}UHF} &
				\multicolumn{2}{c|}{\cellcolor[HTML]{E9FEFF}\begin{tabular}[c]{@{}c@{}}1, 2, 4, 8, 16, 32, 64, 128, 256, \vspace{-1mm}\\ 512, 1024, 2048 kbps.\end{tabular}} \\ \hline
				\cellcolor[HTML]{DCEBFF}\textit{Downlink Data Rate} &
				\multicolumn{3}{c|}{\cellcolor[HTML]{FFE1BE}21.3 kbps - 85.33 kbps} &
				\multicolumn{3}{c|}{\cellcolor[HTML]{CCFFCB}3.6 kbps - 14.2 kbps} &
				\multicolumn{3}{c|}{\cellcolor[HTML]{C1FDFF}500 kbps - 6 Mbps} \\ \hline
			\end{tabular}%
		}
	\end{table*}
	
	MAVEN was launched in November 2013 and reached Mars in September of the following year. The primary goal of Maven is to map electric currents around Mars, which are essential to atmospheric loss. MAVEN's mission takes place in the upper atmosphere and ionosphere, recording their instantaneous state and the amount of gas lost to the atmosphere \cite{Every_miss_maven}. MAVEN also transmits data from rovers and landers to Earth with the help of the Electra communication package, which operates in the X-band. It has a HGA with a diameter of \SI{2}{\meter}. MAVEN stops collecting data every $ 3.5 $ day for $ 5 $ hours and communicates with the world. MAVEN can hold up to $ 32 $ $ \mathrm{Gbit} $ of data in its memory between these sessions. Maven's tracking is provided by its LGA \cite{nasa_maven_press_kit_2013}.
	
	\begin{table}[!t]
		\caption{Data Rates of Mars Orbiters on Various Frequency Bands \cite{Miranda_antenna_tech}.}
		\label{table:characteristics_of_Comm_assets}
		\centering
		\resizebox{\columnwidth}{!}{
			\renewcommand{\arraystretch}{1.5}
			\begin{tabular}{|l|l|c|c|}
				\hline
				\textbf{\begin{tabular}[c]{@{}l@{}}Communications \\ Asset\end{tabular}} & \multicolumn{1}{c|}{\textbf{Frequency Bands}} & \textbf{Data Rates} & \textbf{Purpose} \\ \hline
				\multirow{3}{*}{MRO}                                                     & X-band                                        & 300 kbps            & Tx/Rx to Earth   \\ \cline{2-4} 
				& UHF                                           & 100 kbps - 1 Mbps   & Tx/Rx to Mars    \\ \cline{2-4} 
				& Ka-band                                       & 5 Mbps       & Tx to Earth      \\ \hline
				\multirow{3}{*}{MGS}                                                     & X-band                                        & 20 kbps             & Tx/Rx to Earth   \\ \cline{2-4} 
				& UHF                                           & 128 kbps            & Tx/Rx to Mars    \\ \cline{2-4} 
				& Ka-band                                       & 85 kbps (max)       & Tx to Earth      \\ \hline
				\multirow{3}{*}{Mars Express}                                            & X-band                                        & 230 kbps            & Tx to Earth      \\ \cline{2-4} 
				& S-band                                        & up to 2 kbps        & Rx from Earth    \\ \cline{2-4} 
				& UHF                                           & 128 kbps            & Tx/Rx to Mars    \\ \hline
				\multirow{2}{*}{Mars Odyssey}                                            & X-band                                        & 128 kbps            & Tx/Rx to Earth   \\ \cline{2-4} 
				& UHF                                           & 128 kbps            & Tx/Rx to Mars    \\ \hline
			\end{tabular}
		}
	\end{table}
	Mangalyaan, which stands for spacecraft in the Indian language, is an orbiter developed by the Indiana Space Research Organization (\acrshort{isro}) and placed in the orbit of Mars in 2014. Communications were accomplished through two \SI{230}{\watt} TWTAs and two coherent transponders. It has LGA, MGA, and HGA. The HGA is a $ 2.2 $ diameter dish one. It uses S-band to transfer telemetry and command data via LGA. Data can be transferred without turbo coding at the rates of $ 5/10/20/40 $ $ \mathrm{kbps} $ according to choice of use. It uses delta differential one-way ranging (Delta-DOR or $\Delta$DOR) for better orbit determination (\acrshort{od}). On the ground side, spacecraft communicate with Earth via multiple networks such as NASA’s DSN, ISRO Telemetry, Tracking and Command Network's (\acrshort{istrac}) Ground Network, and Indian Deep Space Network (\acrshort{idsn}). IDSN consists of \SI{18}{\meter} and \SI{32}{\meter} diameter antennas with \SI{20}{\kilo\watt} power, located in Bangalore \cite{shamsudheen_mars_orbiter_mission}, \cite{Venkatesan_2013}.

	The  ExoMars is the first spacecraft built in collaboration with two different space agencies. These are the ESA and Roscosmos. This mission aims to examine methane and other atmospheric gases that are less than \SI{1}{\percent} of the atmosphere and to draw conclusions that could be evidence of possible biological activity.
	It was launched on 14 March 2016.
	The Trace Gas Orbiter (\acrshort{tgo}), and the entry descent and landing demonstrator module (\acrshort{edm}) were parts of the mission. While the EDM has a payload for in situ measurements during descent and on the surface of Mars, the TGO contains scientific instruments for remote observations. TGO is equipped with two Electra UHF transceivers from NASA that are identical in operation to the Electra transceiver on the MAVEN to communicate with rovers and landers. Each Electra transceiver aboard TGO is attached to its own UHF quadrifilar helix antenna, which has a $ 3 $ $\mathrm{dB}$ beamwidth of  ±$40$ $\mathrm{deg}$ and an on-boresight gain of $ 6 $ $ dBic $.
	The TGO features a \SI{65}{\watt} system with \SI{2.2}{\meter} diameter HGA and three LGAs that operate in the X-band to communicate with Earth \cite{esa_Exomars}.
	On the other hand, the EDM's UHF transceiver operates at \SI{437.1}{\mega\hertz} on the forward link and \SI{401.59}{\mega\hertz} on the return link. The EDM transceiver has a nominal \SI{4.8}{\watt} transmitter power and provides data rates of $ 8 $-$ 64 $ $\mathrm{kbps}$ and $ 8 $-$ 1024 $ $\mathrm{kbps}$ on the forward and return links, respectively, with optional ($ 7,1⁄2 $) convolutional coding. A Backshell LGA was used for the EDM to transmit the data during the initial phase of entry-decent-landing (\acrshort{edl}) \cite{Charles_Relay_2017}. Unfortunately, after the parachute deployment, the unit inside the EDM that measures the lander's rotation rate had a miscalculation due to unexpected high rotation rates, which then led the computer to calculate that the lander was below ground level. As a result, the backshell and parachute were released earlier than planned, the thrusters were fired for just three seconds as opposed to thirty, and the on-ground equipment was turned on as if Schiaparelli had touched down. The actual impact speed of the module was calculated to be $ 540 $ $\mathrm{km/h}$ because it was falling freely from a height of approximately $ 3.7 $ \si{\kilo\meter} \cite{esa_crush_2017}. Learning from these experiences, ESA and Roscosmos are planning to launch the ExoMars 2022 or the rover and surface platform (\acrshort{rsp}) mission, which will include a European ExoMars rover and a Russian surface platform. The main goal is to set down the rover at a location with a good chance of discovering organic material that has been well preserved, especially from the planet's early history \cite{exomars_2022_2019}.

	\subsection{New Missions}
	The InSight lander was launched in May 2018 and landed in November of the same year. Its goal is to understand more about the layers that make up the Martian subsurface so scientists might be able to make comparisons with what we know about other planets and Earth. Because the soil at the landing location was different than expected, one of InSight's key research equipment, a heat-flow probe, failed to perform. InSight is now listening for marsquakes to learn more about what lies under the Martian surface \cite{Every_miss_Insight}.
	Critical data about InSight's state was delivered through radio waves at \SI{401.586}{\mega\hertz} to two briefcase-size CubeSats named WALL-E and EVE, which relayed the data at $ 8 $ $ \mathrm{Kbps} $ back to massive $ 70 $-\si{\meter} antennas on Earth. The CubeSats were released on the same launcher as InSight and traveled together through the journey to Mars to monitor the EDL event and report back data as soon as possible. Other orbiters, such as the MRO, were not in position and could not initially communicate with the lander in real-time. If the CubeSats failed for whatever reason, the MRO was prepared to act. Each device functioned as a node in an Internet-like system, allowing data packages to be routed through several terminals built with various types of technology \cite{krywko_2019_deep_space_dial_up}. 
	As with Pathfinder and Opportunity, signals are sent from the DSN antenna to InSight at \SI{7.2}{\giga\hertz} frequency and from the spacecraft to the DSN antennas at \SI{8.4}{\giga\hertz} frequency. There is a turnaround ratio of $ 880/749 $ between the received and transmitted frequencies, so the signals do not interfere \cite{KAHAN_Mars_precession}.
	X-band antennas on Insight were used for communication during the spacecraft's journey from Earth to Mars. There were one MGA in the travel process and two LGAs, one as a receiver and one as a transmitter. The main method for transmitting data from the lander to Earth after landing is to transmit data from the lander's helical antenna to an available orbiter in the UHF band and relay it to Earth \cite{nasa_Insight_press_kit}.
	
	Chinese lander, rover, and orbiter mission Tianwen 1 was launched on July 23, 2020. The mission's scientific goals are to profile the martian ionosphere, temperature, and environment, describe the soil and its water-ice content, define the geography and geology of Mars, and constrain the gravity and magnetic fields. A high-gain dish antenna is installed on side of the orbiter bus, which is a short hexagonal cylinder with two solar panel wings. A HGA and a very low frequency antenna are used for communication.	The Zhurong rover weighs in at around \SI{240}{\kilo\gram}. It includes six wheels and a set of solar panels that can be folded out to supply electricity. At the rover's rear is a pole with a dish antenna attached \cite{tianwen_1_nasa}. The X band HGA is developed by the Beijing Institute of Spacecraft System Engineering (\acrshort{isse}). The HGA has right hand circular polarization (\acrshort{rhcp}) as with all other HGAs developed for spacecraft before Zhurong. The antenna enables communications between Mars and the Earth in both the uplink and downlink directions, as well as between Mars' surface and the orbiters. HGA's reception band is within $ 7145 $-$ 7190 $ $\mathrm{MHz}$ and transmission band is within $ 8400 $-$ 8450 $ $\mathrm{MHz}$. Edge of the coverage gain of the HGA including radio cable loss is greater than $ 21 $ $\mathrm{dBi}$ for both reception and transmission band. Detailed information about voltage standing wave ratio (\acrshort{vswr}) and radiation parameters are provided in \cite{Zhijia_X_2021}.

	Despite the challenges of the global COVID-19 pandemic, another rover by NASA, Perseverence, launched on July 30, 2020. It landed at Jezero crater on February 18, 2021, the site of an ancient lake and river delta. Perseverance stayed there for almost two weeks, completing checkups, calibrating subsystems, and gathering imagery and data. The rover then set out to find an appropriate area for the Mars Helicopter's (Ingenuity) flight tests. When the rover arrived at a proper test area, it deployed Ingenuity to the ground to conduct a number of test flights.
	After three successful flights, the helicopter concluded its technological demo. On April 19, 2021, Ingenuity lifted off, ascended to roughly \SI{3}{\meter} (\SI{10}{\feet}) above the surface, floated briefly in the air, performed a rotation, and landed afterward. It was a historic moment: the first powered, controlled flight in Mars's incredibly thin atmosphere and, indeed, the first such flight on any planet beyond Earth. Following that, the helicopter accomplished several experimental flights of increasing distance and altitude \cite{Mars_helicopter}. The rover is looking for proof of life that existed in the warm and wet times of Mars by analyzing the microbial fossils on rocks. It is also searching for organics, carbon-containing chemicals that are the fundamental elements of life on Earth. As it travels, Perseverance collects soil and rock samples, which it stores in tubes for future NASA and ESA missions to retrieve. Perseverance is a compact car-sized, one-ton Mars rover with six wheels. Perseverance can function in dust storms that prevent spacecraft from receiving sunlight thanks to having nuclear-powered architecture, the same as Curiosity \cite{Every_miss_perseverance}.

	The Perseverance rover has three antennas: a UHF antenna, an X-band HGA, and an X-band LGA. With the help of orbiters, it uses the UHF antenna, which operates at a frequency around \SI{400}{\mega\hertz}, when communicating with Earth. When the orbiter is close to the rover, the transmission speed between them can reach as high as $ 2 $ $ \mathrm{Mbps} $. Orbiters then transmit data to Earth using large dish antennas. The high-gain X-band antenna is steerable so that it can focus signals in a specific direction. This way, the rover does not have to rotate its body to communicate with Earth. The main task of the HGA is to transmit data to Earth using the frequency range of  $ 7 $-$ 8 $ \si{\giga\hertz}. It is hexagonal in shape and has a diameter of \SI{0.3}{\meter}. By using this antenna, data can be transferred at a speed of at least $ 160/500 $ $ \mathrm{bps} $ to \SI{34}{\meter} dish antennas and at least $ 800/3000 $ $ \mathrm{bps} $ to \SI{70}{\meter} dish antennas. The low-gain X-band antenna is omnidirectional, which means it transmits equally in all directions and can receive signals from all directions. It uses a frequency range of $ 7 $-$ 8 $ \si{\giga\hertz}, just like its HGA. By using this antenna, data can be transferred at a speed of at least $ 10 $ $ \mathrm{bps} $ with \SI{34}{\meter} dish antennas and at least $ 30 $ $ \mathrm{bps} $ with \SI{70}{\meter} dish antennas \cite{nasa_perseverance}.
	\begin{figure}[!t]
		\centering
		\includegraphics[scale = 0.50]{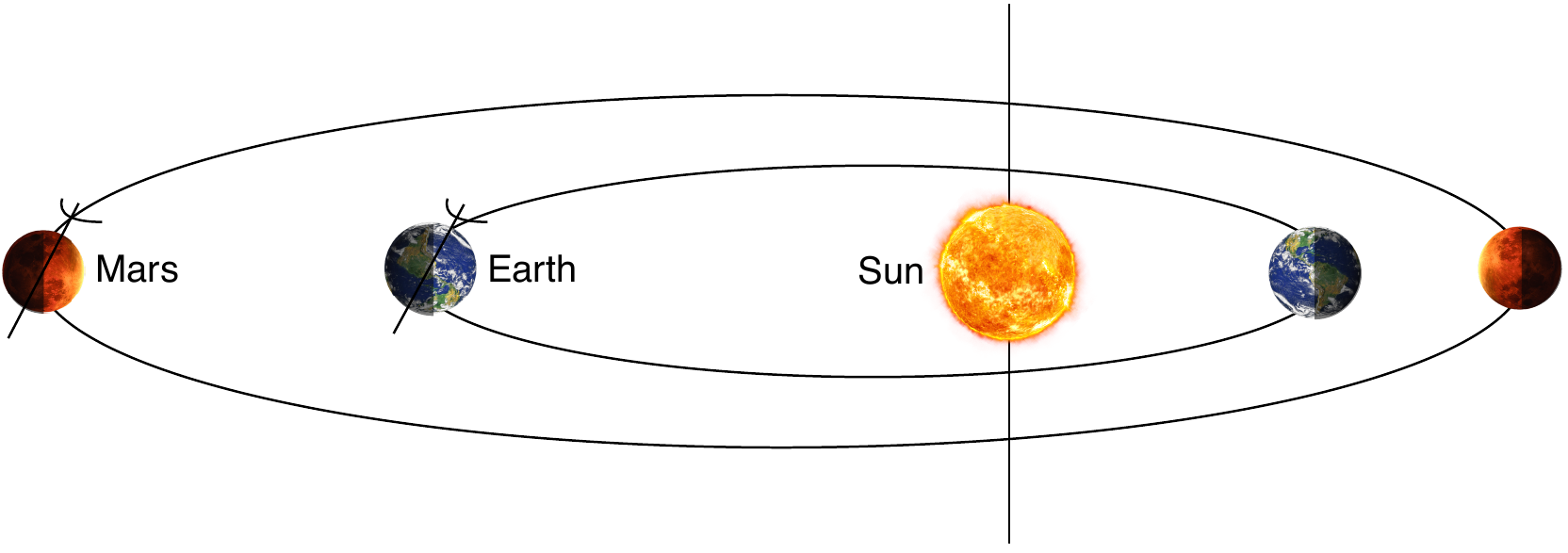}
		\caption{Martian and Earth orbits around the Sun.}
		\label{fig:orbit_of_mars}
	\end{figure}

	We conclude this section by providing data rate predictions for future missions. Data rates at planetary links are needed to be considerably increased to take the Mars exploration to another level. Solutions that could offer data rates of $ 10 $ $ \mathrm{Mbps} $ to $ 250 $ $ \mathrm{Mbps} $ in the return link (from Mars to Earth) and $ 3 $ $ \mathrm{Mbps} $ to $ 50 $ $\mathrm{Mbps}$ in the forward link (from Earth to Mars) at the maximum distance between Mars and Earth should be investigated \cite{2018_deep_Space_Relay_Terminals}. Data rate requirements of a crewed mission to Phobos and Mars are given in Table \ref{table:crewed_Mission_to_Phobos}. It can be claimed that there is no limit on achievable data rates due to the nature of these missions, which is exploration. Also, crewed missions require a different kind of attention because human lives will be put forward, which is more precious than anything else. During the mission, every aspect of astronauts' physical, mental, and psychological health should be extensively monitored and analyzed. Necessary psychiatric support should be served by experts from Earth. Because these missions could take months or even years, astronauts should be able to talk with their families and loved ones just as envisioned by directors of various sci-fi movies.
		On the other hand, technical support by engineers on Earth should always be accessible during the mission. Software updates for mission-critical functions at times of unexpected situations should be transmitted with minimum delay and maximum throughput. When all these things are considered, the data rates given by Table \ref{table:crewed_Mission_to_Phobos} are sensible but also open for improvements. For example, more than the shown minimum data rates on the Earth-to-Mars link may be needed for high-quality video streaming.
	Moreover, data rates of $ 1 $ $ \mathrm{Gbps} $ are desired by 2030, according to NASA. Data rate requirements, efficiency, mass, cost, available frequencies, space, and power will all play a key role in determining the best antenna technology for future Mars exploration missions \cite{Miranda_antenna_tech}.
	\begin{table}[!t]
		\caption{Min. and Max. Data Rate Requirements of the Two Mars Missions \cite{Wallace_Mars_2018}.}
		\label{table:crewed_Mission_to_Phobos}
		\centering
		\resizebox{\columnwidth}{!}{
			\renewcommand{\arraystretch}{1.5}
			\begin{tabular}{|l|c|c|c|c|}
				\hline 
				Mission                                                                   & \begin{tabular}[c]{@{}c@{}}Earth-to-Mars\\ (Min)\end{tabular} & \begin{tabular}[c]{@{}c@{}}Mars-to-Earth\\ (Min)\end{tabular} & \begin{tabular}[c]{@{}c@{}}Earth-to-Mars\\ (Max)\end{tabular} & \begin{tabular}[c]{@{}c@{}}Mars-to-Earth\\ (Max)\end{tabular} \\ \hline
				\begin{tabular}[c]{@{}l@{}}Crewed Mission To Phobos\end{tabular} & 1.7 Mbps                                                      & 111.7 Mbps                                                    & 24.7 Mbps                                                     & 200.8 Mbps                                                    \\ \hline
				\begin{tabular}[c]{@{}l@{}}Mars Short Stay Mission\end{tabular}  & 11.6 Mbps                                                     & 93.7 Mbps                                                     & 28.4 Mbps                                                     & 211.9 Mbps                                                    \\ \hline
			\end{tabular}
		}
	\end{table}

	\section{Mars and Effect of Martian Atmosphere on Radio Wave Propagation} \label{sec:mars_atmosphere}
	In this section, we provide statistical features of Mars with comparisons to Earth and discuss Martian atmospheric effects on radio propagation. 
	
	Mars is similar to Earth in various ways. It is the seventh biggest planet in our solar system, with a radius of  \SI{3389.5}{\kilo\meter} (\SI{2106}{\mile}), and is roughly half the diameter of Earth. It has \SI{37.9}{\percent} of Earth's surface gravity. The mass of Mars is roughly \SI{11}{\percent} of Earth's. Every $ 24.6 $ Earth hours, Mars circles on its axis, determining the duration of a $ \mathrm{sol} $. Statistical parameters of Mars are given in Table \ref{table:Mars_Statistical_Paramters}. Because of its \ang{25.2} inclination from the plane of its orbit around the Sun, Mars experiences seasons that are comparable to those on Earth. Spring and summer are experienced by whichever hemisphere is oriented to the Sun, whereas fall and winter are experienced by the hemisphere oriented away from the Sun. Both hemispheres get equal illumination at two exact times each year, known as the equinoxes \cite{greshko_mars_2021}. 
	\begin{table}[!t]
		\caption{Numerical Parameters for Mars \cite{Radio_wave_handbook}, \cite{williams_mars_fact_sheet_2021}. }
		\label{table:Mars_Statistical_Paramters}
		\centering
		\resizebox{\columnwidth}{!}{
			\renewcommand{\arraystretch}{1.5}
			\begin{tabular}{|l|l|}
				\hline
				\rowcolor[HTML]{93C0FF} 
				\textbf{Parameter}                                                                            & \textbf{Value}                                                                                         \\ \hline
				\rowcolor[HTML]{DAE8FC} 
				Equatorial Radius                                                                         &  $ 3396.2 $ \si{\km} $ (0.532 \times $Earth)                                                                                 \\ \hline
				\rowcolor[HTML]{ECF4FF} 
				Polar Radius                                                                              &  $ 3376.2 $ \si{\km} $ (0.531 \times $Earth)                                                                                 \\ \hline
				\rowcolor[HTML]{DAE8FC} 
				Length of Day                                                                            & $ 24.6597 $ $ \mathrm{hours} $ $ (1.027 \times $Earth)                                                                                \\ \hline
				\rowcolor[HTML]{ECF4FF} 
				Length of Year                                                                                & $ 687 $ Earth days                                                                                         \\ \hline 
				\rowcolor[HTML]{DAE8FC} 
				Mass $\left(\num{e24}\right)$                                                              & $ 0.64169 $ \si{\kilogram} $ (0.107 \times $Earth)                                                                                \\ \hline 
				\rowcolor[HTML]{ECF4FF} 
				Mean Density                                                                       & $ 3934 $ \si[per-mode = symbol]{\kilo\gram\per\meter\cubed} $ (0.714 \times $Earth)                                                                                   \\ \hline
				\rowcolor[HTML]{DAE8FC} 
				Tilt of Axis                                                                                  & \ang{25;;12}                                                                             \\ \hline
				\rowcolor[HTML]{ECF4FF} 
				Minimum Distance from Sun                                                                 & $ 205 \times 10^{6} $ \si{\km}                                                                             \\ \hline
				\rowcolor[HTML]{DAE8FC} 
				Maximum Distance from Sun                                                                 & $ 249 \times 10^{6} $ \si{\km}                                                                             \\ \hline
				\rowcolor[HTML]{ECF4FF} 
				Surface Gravity                                                                     & $ 3.71 $ \si[per-mode = symbol]{\meter\per\second\squared} $ (0.379 \times $Earth)                                                                                   \\ \hline
				\rowcolor[HTML]{DAE8FC} 
				Temperature                                                                                   & \SI{-82}{\degreeCelsius} to \SI{0}{\degreeCelsius} (\ang{-116}$ \mathrm{F} $ to \ang{32}$ \mathrm{F} $) \\ \hline
				\rowcolor[HTML]{ECF4FF} 
				Average Temperature                                                                           & $\sim$\SI{210}{\kelvin} ($-63^{\circ}$C)                                                                  \\ \hline
				\rowcolor[HTML]{DAE8FC} 
				\begin{tabular}[c]{@{}l@{}}Minimum Distance from Earth \end{tabular}           & $ 54.6 \times 10^{6} $ \si{\km}                                                                        \\ \hline
				\rowcolor[HTML]{ECF4FF} 
				\begin{tabular}[c]{@{}l@{}}Maximum Distance from Earth \end{tabular} & $ 401.4 \times 10^{6} $ \si{\km}                                                                       \\ \hline
				\rowcolor[HTML]{DAE8FC} 
				Natural Satellites                                                                            & Deimos, Phobos                                                                                         \\ \hline
			\end{tabular}
		}
	\end{table}

	Besides these similarities, there are also differences between Mars and Earth. Martian orbit is more elliptical than Earth orbit, causing characteristics such as climate and surface to be significantly different from Earth, as shown in Fig. \ref{fig:orbit_of_mars}. When Mars is at the closest point to Sun, the southern hemisphere faces the Sun, resulting in a shorter and hot summer than the northern hemisphere. When Mars is at the farthest point from Sun, the northern hemisphere faces the Sun, resulting in long and cold summer. Thus, the maximum temperature in the southern hemisphere is greater than in the northern hemisphere. The water level in the atmosphere of Mars is one-thousandth of Earth's, but it seems enough to form clouds around the globe \cite{Radio_wave_handbook}. 

	Martian atmosphere is made up of \SI{95.2}{\percent} of carbon dioxide (\ch{CO2}), \SI{2.7}{\percent} of nitrogen (\ch{N2}) and \SI{1.6}{\percent} of argon (\ch{Ar}). At the surface, atmospheric pressure is approximately \SI{0.6}{\percent} of Earth's surface pressure. The temperature on the planet changes from \ang{-199}$\mathrm{F} $ (\SI{-128}{\degreeCelsius}) to \ang{80}$ \mathrm{F} $ (\SI{27}{\degreeCelsius}). \ang{199}$ \mathrm{F} $ can be recorded in the polar regions at night and \ang{80}$ \mathrm{F} $ can be recorded in equator during noon. Average surface temperature is \ang{-82}$ \mathrm{F} $ (\SI{-63}{\degreeCelsius}). Comparison of surface atmospheric parameters of two planets is given in Table \ref{tab:surface_atmospheric_parameters_mars_earth}. Water can not stay liquid on the Martian surface because the weather is arid and the atmosphere is very thin, making it easy to vaporize. On the other hand, it can remain in ice form because the average temperature is very low \cite{nasa_maven_press_kit_2013}.
	\begin{table}[!t]
		\caption{Mars and Earth's Surface Atmospheric Parameters \cite{Radio_wave_handbook}.}
		\label{tab:surface_atmospheric_parameters_mars_earth}
		\centering
		\renewcommand{\arraystretch}{1.5}
		\begin{tabular}{|
				>{\columncolor[HTML]{E1E1E1}}l |l|l|}
			\hline
			\cellcolor[HTML]{FFC702}{\textbf{Parameter\textbackslash{}Planet}} & \cellcolor[HTML]{FFC702}\textbf{Mars} & \cellcolor[HTML]{FFC702}\textbf{Earth} \\ \hline
			\textbf{Surface Pressure}                                            & $\sim6.1$ \si{\milli\bar}                          & $ 1013 $ \si{\milli\bar} (avr)                          \\ \hline
			\textbf{Surface Density}                                             & $\sim0.020$ \si[per-mode = symbol]{\kilogram\per\meter\cubed}   & $\sim1.29$  \si[per-mode = symbol]{\kilogram\per\meter\cubed}    \\ \hline
			\textbf{Scale height}                                                & $\sim11.1 $ \si{\km}                         & $\sim9.5 $ \si{\km}                           \\ \hline
			\textbf{Average temp.}                                               & $\sim210 $ \si{\kelvin}                           & $\sim300 $ \si{\kelvin}                             \\ \hline
			\textbf{Diurnal temp. range}                                         & $ 184 $ \si{\kelvin}  to $ 242 $ \si{\kelvin}                         & $ 210 $ \si{\kelvin}  to $ 320 $ \si{\kelvin}                          \\ \hline
			\textbf{Mean molecular weight}                                       & $ 43.34$ \si[per-mode = symbol]{\gram\per\mole}                         & $ 28.61$ \si[per-mode = symbol]{\gram\per\mole}                            \\ \hline
		\end{tabular}
	\end{table}
	Apart from \ch{CO2} and carbon monoxide (\ch{CO}), all gaseous concentrations on Mars are lower than on Earth. Despite the fact that \ch{CO2} is the most prevalent gas on Mars, its density is just $ 25 $ times that of Earth. On both planets, the amount of water vapor is far more variable. Water vapor concentration on Mars varies with season and latitude, ranging from $ 100 $ to $ 400 $ parts per million ($ \mathrm{ppm} $) with an average of $ 300 $ $ \mathrm{ppm} $. On the other hand, water vapor concentration on Earth ranges from $ 40 $ to $ 40,000 $ $ \mathrm{ppm} $, with a mean value of $ 10,000 $ $ \mathrm{ppm} $ \cite{Radio_wave_handbook}. A comparison of atmospheric compositions between two planets is given in Table \ref{tab:atmsp_comp}.
	
	\subsection{Radio Propagation}
	If the classical radio wave theory is considered, radio propagation at Mars is determined with two parameters: ionospheric (plasma) refractive index and its tropospheric (atmosphere) refractive index. The propagation direction, intensity, and polarization of radio waves are all controlled by these parameters. The refractive index is mostly determined by electron density and the background magnetic field. Besides, atmospheric pressure, absolute temperature, and water vapor pressure control tropospheric radio refractivity. The troposphere generally has an effect on radio waves with frequencies greater than \SI{1}{\giga\hertz}. Attenuation mainly caused by dust storms, atmospheric aerosols, clouds, and morning fogs may also have significant effects on electromagnetic waves. Additionally, radio waves can be reflected and diffracted by surface structures such as crater domes, polar ice caps, and canyons \cite{Radio_wave_handbook}.
	\begin{table}[!t]
		\caption{Comparison of Mars and Earth's Atmospheric Elements Based on Their Volumes \cite{Radio_wave_handbook}, \cite{williams_mars_fact_sheet_2021}.}
		\label{tab:atmsp_comp}
		\centering
		\resizebox{\columnwidth}{!}{
			\renewcommand{\arraystretch}{1.5}
			\begin{tabular}{|
					>{\columncolor[HTML]{FFFFFF}}l cc|}
				\hline 
				\multicolumn{1}{|l|}{\cellcolor[HTML]{FFFFFF}\textbf{Atmospreic Composition}} & \multicolumn{1}{l|}{\cellcolor[HTML]{FFFFFF}\textbf{Mars}} & \multicolumn{1}{l|}{\cellcolor[HTML]{FFFFFF}\textbf{Earth}} \\ \hline
				
				\multicolumn{3}{|l|}{\cellcolor[HTML]{FFFFFF}\textbf{Major:}}                                                                                                                                                        \\ \hline
				\multicolumn{1}{|l|}{\cellcolor[HTML]{FFFFFF}argon (\ch{Ar})}                                  & \multicolumn{1}{c|}{\SI{1.94}{\percent }}                                 & \SI{0.93}{\percent}                                                      \\ \hline
				\multicolumn{1}{|l|}{\cellcolor[HTML]{FFFFFF}carbon dioxide $\left(\ch{CO2}\right)$}                        & \multicolumn{1}{c|}{\SI{95.1}{\percent}}                               & \SI{0.03}{\percent}                                                      \\ \hline
				\multicolumn{1}{|l|}{\cellcolor[HTML]{FFFFFF}carbon monoxide (\ch{CO})}                        & \multicolumn{1}{c|}{\SI{0.06}{\percent}}                                & -                                                           \\ \hline
				\multicolumn{1}{|l|}{\cellcolor[HTML]{FFFFFF}nitrogen $\left(\ch{N2}\right)$}                               & \multicolumn{1}{c|}{\SI{2.59}{\percent}}                                 & \SI{78.09}{\percent}                                                     \\ \hline
				
				\multicolumn{1}{|l|}{\cellcolor[HTML]{FFFFFF}oxygen $\left(\ch{O2}\right)$}                                 & \multicolumn{1}{c|}{\SI{0.16}{\percent}}                                & \SI{20.95}{\percent}                                                     \\ \hline
				
				\multicolumn{3}{|l|}{\cellcolor[HTML]{FFFFFF}\textbf{Minor (units in parts per million {[}ppm{]}):}}                                                                                                                 \\  \hline
				\multicolumn{1}{|l|}{\cellcolor[HTML]{FFFFFF}carbon Monoxide (\ch{CO})}                        & \multicolumn{1}{c|}{-}                                     & 0.2                                                         \\ \hline
				\multicolumn{1}{|l|}{\cellcolor[HTML]{FFFFFF}helium (\ch{He})}                                 & \multicolumn{1}{c|}{-}                                     & 5.2                                                         \\ \hline
				\multicolumn{1}{|l|}{\cellcolor[HTML]{FFFFFF}hydrogen $\left(\ch{H2}\right)$}                               & \multicolumn{1}{c|}{-}                                     & 1.0                                                         \\ \hline
				\multicolumn{1}{|l|}{\cellcolor[HTML]{FFFFFF}hydrogen-deuterium-oxygen (\ch{HDO})}             & \multicolumn{1}{c|}{0.85}                                  & -                                                           \\ \hline
				\multicolumn{1}{|l|}{\cellcolor[HTML]{FFFFFF}krypton (\ch{Kr})}                                & \multicolumn{1}{c|}{0.3}                                   & 1.1                                                         \\ \hline
				\multicolumn{1}{|l|}{\cellcolor[HTML]{FFFFFF}methane $\left(\ch{CH4}\right)$}                               & \multicolumn{1}{c|}{-}                                     & 1.5                                                         \\ \hline
				\multicolumn{1}{|l|}{\cellcolor[HTML]{FFFFFF}neon (\ch{Ne})}                                   & \multicolumn{1}{c|}{2.5}                                   & 20                                                          \\ \hline
				
				\multicolumn{1}{|l|}{\cellcolor[HTML]{FFFFFF}nitrogen Oxide (\ch{NO})}                         & \multicolumn{1}{c|}{100}                                   & -                                                           \\ \hline
				\multicolumn{1}{|l|}{\cellcolor[HTML]{FFFFFF}nitrous oxide $\left(\ch{N2O}\right)$}                         & \multicolumn{1}{c|}{-}                                     & 0.6                                                         \\ \hline
				\multicolumn{1}{|l|}{\cellcolor[HTML]{FFFFFF}ozone $\left(\ch{O3}\right)$}                                  & \multicolumn{1}{c|}{0.04-0.2}                              & \textless 0.05                                              \\ \hline
				\multicolumn{1}{|l|}{\cellcolor[HTML]{FFFFFF}water vapor $\left(\ch{H2O}\right)$}                           & \multicolumn{1}{c|}{$\sim$100-400}                         & $\sim$40-40,000                                             \\ \hline
				\multicolumn{1}{|l|}{\cellcolor[HTML]{FFFFFF}xenon (\ch{Xe})}                                  & \multicolumn{1}{c|}{0.08}                                  & 0.09                                                        \\ \hline
				
			\end{tabular}
		}
	\end{table}
	\subsubsection{Ionosphere}
\begin{figure}[!t]
		\centering
		\includegraphics[scale = 0.55]{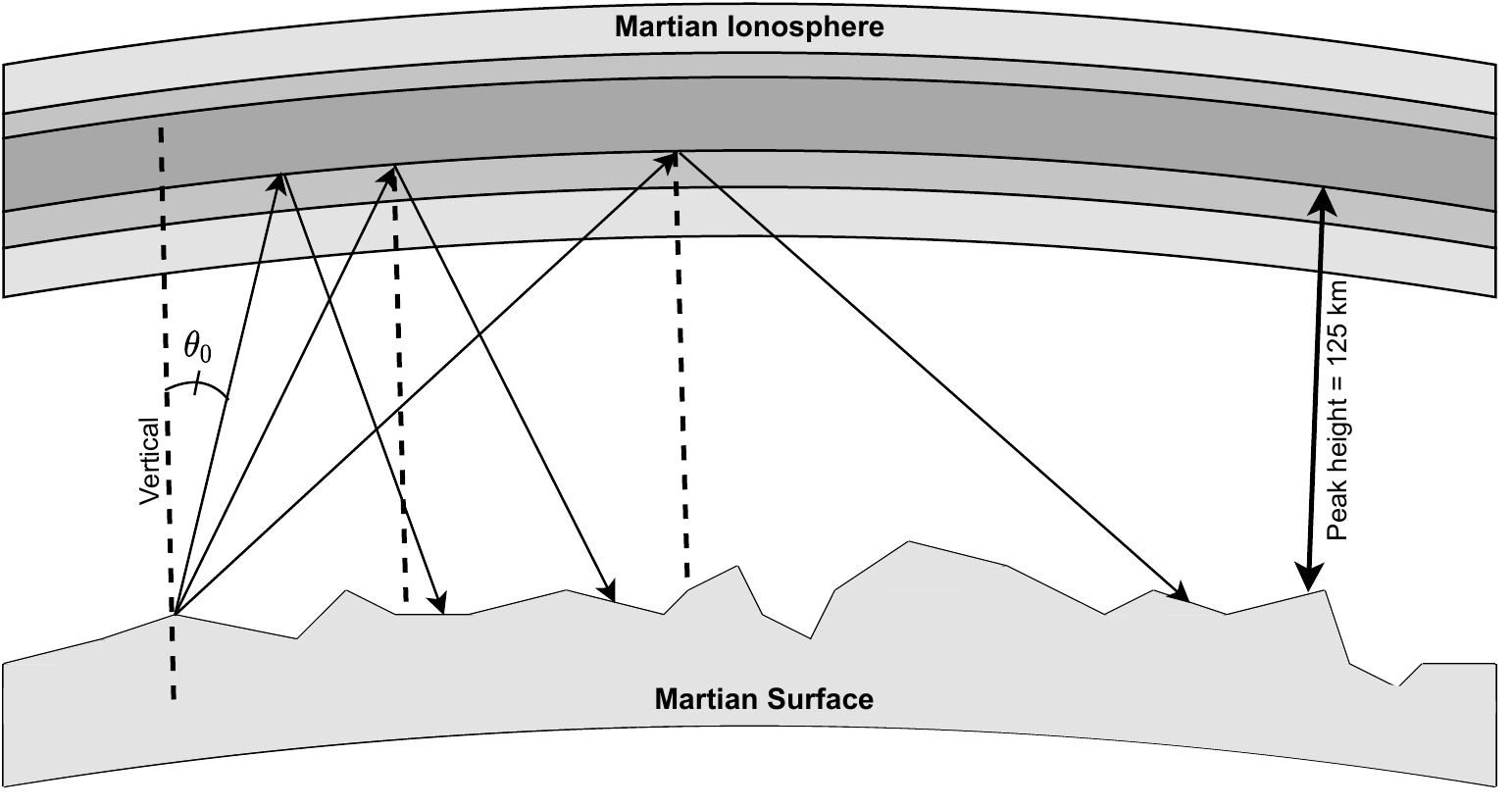}
		\caption{Visual representation of ionosphere communication for different launch angles \cite{Radio_wave_handbook}.}
		\label{fig:hop_distances}
	\end{figure}

	Martian ionosphere also has substantial effects on radio wave propagation. The photo-ionization of Mars's upper atmosphere creates the planet's dayside ionosphere. Solar wind pressure determines the height of the ionosphere. The peak height and peak density of the ionosphere are generally static. Martian ionosphere has lower plasma density than Earth because it is farther from Sun than Earth. Martian ionosphere has only one layer that forms between one hundred kilometers and several hundred kilometers above the surface, unlike Earth's ionosphere, which has four layers.
	A powerful planetary magnetic field protects Earth's ionosphere from the solar wind. On the contrary, not having a strong magnetic field on Mars makes the Martian ionosphere vulnerable to the solar wind.
	In the 80s, there were some arguments about the Martian magnetic field. Some thought that the weak magnetic field measured by Mars missions is the internal magnetic field of Mars. Some others said that it could be an external magnetic field formed by solar winds flowing across the solar system called an interplanetary magnetic field (\acrshort{imf}). In the late 90s, measurements by the MGS verified that Mars had no internal magnetic dipole field. The existing magnetic field was $ 800 $ times lower in terms of strength than Earth’s magnetic field, which probably proved the other scientists right about the solar wind smashing into the Martian ionosphere, causing complex magnetic fields to form. 
	However, Mars's ground-to-ground low-frequency communication will undoubtedly benefit from the Martian ionosphere. The Martian ionosphere has the potential to keep communication links online between surface operators in situations where there is no LOS communication available. Maximum usable frequencies and one-hop distances are shown in Table \ref{tab:usable_critical_Freq_hop_dist} relative to launch angle $ \left(\theta_{0}\right) $. The maximum usable frequencies considerably increase as $ \theta_{0} $ increases. The use of the Martian ionosphere for communication throughout the globe can be seen in Fig. \ref{fig:hop_distances}. Also, dust storms may increase the ionosphere's peak height, increasing hop distance. 
	Waves with frequencies lower than \SI{4.5}{\mega\hertz} cannot penetrate the ionosphere. However, radio waves with frequencies greater than \SI{450}{\mega\hertz} do not have any problem going through the ionosphere.
	\subsubsection{Troposphere}
	The troposphere is the lower layer of the atmosphere where signal attenuation occurs the most for high-frequency signals. Aerosols, fog, gases, water vapor, ice, cloud, and dust are the main elements that form the propagation medium in the Martian troposphere. The Martian atmosphere, however, is thin in comparison to that of Earth. As a result, a substantially less effect by the troposphere on radio wave transmission is anticipated. Multipath caused by reflections and solar scintillation are the two primary effects of the Martian troposphere. Fading, dispersion, and other effects are ignored in a clear Martian atmosphere. 
	\begin{table}[!t]
		\caption{Max. Usable Frequencies and One-hop Distances for Multiple Launch Angles \cite{Radio_wave_handbook}.}
		\label{tab:usable_critical_Freq_hop_dist}
		\centering
		\resizebox{\columnwidth}{!}{
			\renewcommand{\arraystretch}{1.3}
			\begin{tabular}{l|llllll|}
				\cline{2-7}
				\multirow{2}{*}{}                                                                                    & \multicolumn{6}{c|}{\textbf{Launch Angle $\left(\mathbf{\theta_{0}}\right)$}}                                                                                                                                                \\ \cline{2-7} 
				& \multicolumn{1}{l|}{ $\mathbf{0^{\circ}}$ } & \multicolumn{1}{l|}{ $\mathbf{15^{\circ}}$ } & \multicolumn{1}{l|}{ $\mathbf{30^{\circ}}$ } & \multicolumn{1}{l|}{ $\mathbf{45^{\circ}}$ } & \multicolumn{1}{l|}{ $\mathbf{60^{\circ}}$ } &  $\mathbf{75^{\circ}}$  \\ \hline
				\multicolumn{1}{|l|}{\textbf{\begin{tabular}[c]{@{}l@{}}Max. Usable \vspace{-1mm}\\ Frequency (MHz)\end{tabular}}} & \multicolumn{1}{l|}{4.0}        & \multicolumn{1}{l|}{4.14}        & \multicolumn{1}{l|}{4.62}        & \multicolumn{1}{l|}{5.66}        & \multicolumn{1}{l|}{8.0}         & 15.5        \\ \hline
				\multicolumn{1}{|l|}{\textbf{One-hop Distance (km)}}                                                 & \multicolumn{1}{l|}{0}          & \multicolumn{1}{l|}{67.0}        & \multicolumn{1}{l|}{144.3}       & \multicolumn{1}{l|}{250.0}       & \multicolumn{1}{l|}{433.0}       & 933.0       \\ \hline
			\end{tabular}
		}
	\end{table}
	\subsubsection{Cloud and fog}
	The amount of water in the Martian atmosphere is very low compared to Earth’s, but the available water is enough to form clouds because of low pressure and temperature. According to the observations, clouds are usually formed in equatorial and low-latitude regions. Many young gullies, potentially caused by flowing water, have been discovered in recent MGS detailed photos. MGS discovered water in several clouds using its thermal emission spectrometer. Because of the extreme cold, Martian clouds are most likely made up of ice crystals. The attenuation caused by those ice clouds is approximately half that of water clouds. Some Martian clouds appear before sunrise and disappear quickly, while others appear only in the middle of the day. There are more clouds in northern regions than in southern regions. Fog on Mars is quite common in low-lying places like valleys, canyons, and craters. Cloud and fog attenuation is mostly determined by their water content.  Martian clouds are predicted to have a lower water-liquid composition due to their low optical depth. The optical depth of Martian fog and aerosols has likewise been discovered to be low. \cite{Radio_wave_handbook}. Comparison of visual optical depths of cloud and fogs between the two planets are given in Table \ref{table:visual_optical_depths_cloud}.

	\subsubsection{Dust}
	Martian dust particles are thought to be the primary contributor to Martian aerosols. At dusk, the Martian sky seems pink and black. Martian aerosol dust has an optical depth of roughly $ 0.5 $, as shown in Table \ref{table:visual_optical_depths_cloud}. It has a lesser attenuation impact on radio wave transmission than the Martian clouds. On the other hand, dust plays an important role in the transport and cycling of energy in the atmosphere due to its absorption capability of solar radiation.
	There are also dust devils on Mars. Multiple signs of a dust devil were recorded by Pathfinder. They are formed due to the warming up of air regionally above a flat desert floor resulting in a vertical updraft of air \cite{Radio_wave_handbook}.
	
	\subsubsection{Gases}
	At centimeter and millimeter wavelengths, radio waves are absorbed and scattered by molecules as they move through the atmosphere. Water vapor and oxygen are the main reasons for the gaseous absorption on Earth. \ch{H2O} has $ 29 $ absorption lines (up to \SI{1097}{\giga\hertz}) while \ch{O2} has $ 44 $ absorption lines (up through \SI{834}{\giga\hertz}). Other gases can play a part at frequencies larger than \SI{70}{\giga\hertz} in the absence of water vapor, but their spectral lines are typically too small to affect propagation \cite{Radio_wave_handbook}.
	The major gases in the Martian atmosphere are \ch{CO2} and \ch{N2}, both of which have low microwave absorption because these gases do not generate electric or magnetic dipoles under ordinary circumstances. \ch{O2} and \ch{H2O} are responsible for the majority of the attenuation. In comparison to Earth, the attenuation values related to oxygen on Mars are lowered by a factor of $1.4\times10^4$. The water vapor attenuation on Mars is $ 3068 $ times lower than on Earth. The overall zenith attenuation of the Martian atmosphere is expected to be less than \SI{0.01}{\decibel} at \SI{32}{\giga\hertz} (Ka-band). On the other hand, attenuation of Earth's atmosphere is roughly \SI{0.3}{\decibel}. The atmospheric attenuation goes to \SI{0.1}{\decibel} for higher frequencies such as \SI{100}{\giga\hertz}. Those attenuations are usually insignificant for radio communication at Mars \cite{Radio_wave_handbook}. 
	\begin{table}[!t]
		\caption{Cloud and Fog's Visual Optical Depths \cite{Radio_wave_handbook}.}
		\label{table:visual_optical_depths_cloud}
		\centering
		\resizebox{\columnwidth}{!}{%
			\renewcommand{\arraystretch}{1.3}
			\begin{tabular}{|l|cc|cl|}
				\hline
				\multicolumn{1}{|c|}{\multirow{2}{*}{\textbf{\begin{tabular}[c]{@{}c@{}}Atmospheric\\ Condition\end{tabular}}}} & \multicolumn{2}{c|}{\textbf{Earth}} & \multicolumn{2}{c|}{\textbf{Mars}} \\ \cline{2-5} 
				\multicolumn{1}{|c|}{} & \multicolumn{1}{c|}{\textbf{Optical Depth}} & \textbf{Distribution} & \multicolumn{1}{c|}{\textbf{Optical Depth}} & \multicolumn{1}{c|}{\textbf{Distribution}} \\ \hline
				Clouds $\left(\ch{H2O}\right)$ & \multicolumn{1}{c|}{$\sim 5 $} & \begin{tabular}[c]{@{}c@{}}\SI{50}{\percent} \vspace{-1mm}\\ coverage\end{tabular} & \multicolumn{1}{c|}{$\sim 1.0 $} & \begin{tabular}[c]{@{}l@{}}Winter polar; \vspace{-1mm}\\ behind high places\end{tabular} \\ \hline
				Clouds $\left(\ch{CO2}\right)$ & \multicolumn{1}{c|}{None} & None & \multicolumn{1}{c|}{\begin{tabular}[c]{@{}c@{}} $\sim 0.001$ \vspace{-1mm}\\ $\sim 1.0 $\end{tabular}} & \begin{tabular}[c]{@{}l@{}}Many places \vspace{-1mm}\\ Winter polar\end{tabular} \\ \hline
				Fog & \multicolumn{1}{c|}{$\sim 3 $} & Many places & \multicolumn{1}{c|}{\begin{tabular}[c]{@{}c@{}}$\sim 0.2 $ \vspace{-1mm}\\$\sim 1.0 $\end{tabular}} & \begin{tabular}[c]{@{}l@{}}Morning\\ \quad Valleys \& \vspace{-1mm}\\ \quad crater bottoms\end{tabular} \\ \hline
				Aerosol Dust & \multicolumn{1}{c|}{Variable} & Variable & \multicolumn{1}{c|}{0.5} & Everywhere \\ \hline
				Dust Storms & \multicolumn{1}{c|}{Variable} & Variable & \multicolumn{1}{c|}{10.0} & \begin{tabular}[c]{@{}l@{}}Southern Hemisphere \vspace{-1mm}\\ or global\end{tabular} \\ \hline
			\end{tabular}%
		}
	\end{table}
	
	\subsubsection{Dust storms}
	Martian winds may produce massive dust storms on a consistent basis, even though Mars has a thin atmosphere. It is particularly true in the southern hemisphere during the late spring or early summer seasons. When Mars is at the closest distance to Sun, the southern hemisphere is rapidly heated, resulting in a significant difference in temperature from the northern hemisphere. This difference causes strong dust storms to flow from the southern hemisphere to the northern hemisphere. This phenomenon has a remarkable influence on the atmosphere and terrain characteristics of Mars. There are different types of dust storms. A global dust storm may cover almost all of the planet. Local storms with a geographical range of several hundred kilometers may occur frequently. Dust storms with an opacity of $ 4 $–$ 6 $ can reach heights of \SI{50}{\kilo\meter}. However, due to the small size of the dust particles, it is anticipated that Martian dust storms will have quite a low effect on radio wave transmission. In the worst-case scenario, Martian dust storms might cause a \SI{3}{\decibel} attenuation in Ka-band. Attenuation will be significantly higher with optical communication. The average size of Martian dust is $ 1 $-$ 10 $ \si{\micro\meter}, which is at least four times smaller than Earth dust, as shown in Table \ref{tab:comp_of_dust_btw_earth_mars}. It can be easily claimed that dust on the Martian surface is substantially less disruptive than on Earth. It is suggested to be careful when planning a spacecraft mission to the southern hemisphere during late spring or early summer. Additionally, the southern polar-hood area is where the strongest dust storms usually occur, which makes communication at the Ka-band experience severe attenuation. Therefore, sufficient margins are required \cite{Radio_wave_handbook}. 

	Total tropospheric losses of cloud, fog, gaseous attenuation, and tropospheric scattering are predicted to be around \SI{0.4}{\decibel} at Ka-band. Thus, for a vertically propagating wave, the attenuation from the troposphere and a dust storm is approximately $ 1.4 $ to \SI{2}{\decibel} under typical conditions. In the worst-case scenario, total attenuation will be around \SI{3.4}{\decibel}. Table \ref{tab:radio_wave_att_around_mars_var_freq} shows the corresponding attenuation values for different frequency bands \cite{Radio_wave_handbook}.
	\begin{table}[!t]
		\caption{Comparison of Earth and Mars's Dust Storm Parameters \cite{Radio_wave_handbook}.}
		\label{tab:comp_of_dust_btw_earth_mars}
		\centering
		\renewcommand{\arraystretch}{1.3}
		\begin{tabular}{|l|c|c|}
			\hline
			\textbf{Parameter \textbackslash Planet} & \textbf{Earth} & \textbf{Mars} \\ \hline
			Path Length (\si{\kilo\meter}) & 10 & 10 \\ \hline
			\begin{tabular}[c]{@{}l@{}}Particle Number Density \\ $(m^{-3})$\end{tabular} & $10^{8}$ & $3 \times 10^{7}$ \\ \hline
			\begin{tabular}[c]{@{}l@{}}Dust Material Mass Density \\ (\si[per-mode = symbol]{\gram\per\meter\cubed})\end{tabular} & $2.6 \times 10^{6}$ & $3.0 \times 10^{6}$ \\ \hline
			Mean Size (\si{\micro\meter}) & 30 - 40 & 1 - 10 \\ \hline
			Max. Size (\si{\micro\meter}) & 80 - 300 & 20 \\ \hline
			Visibility (\si{\meter}) & 5.1 - 3.8 & 184 \\ \hline
			Attenuation at 32 GHz (dB) & 65 & 3 \\ \hline
			Mass Loading (\si[per-mode = symbol]{\gram\per\meter\cubed}) & 40-60 & 0.4 \\ \hline
		\end{tabular}%
		
	\end{table}

	\subsubsection{Terrain}
	Mars's surface geomorphologic structures are highly complex. When the satellite is at low elevation angles, these structures interfere with radio links between surface equipment, aircraft, and satellites. When topography or rocks disrupt direct radio wave beams, a surface operator may also connect to a satellite via multipath rays. Compared to direct signals, the strengths of diffracted wave signals are lowered by at least a factor of five. The intensity of the reflected signal is directly affected by the reflection coefficients of the surface elements. When the phase difference between multipath and direct signals is approximately \ang{180}, the combined signal strength can be reduced dramatically \cite{Radio_wave_handbook}.
	
	\subsubsection{Blackout}
	Another challenge needed to be considered when planning a surface operator mission to Mars is the communication blackout during the atmospheric entry phase. Because the flight speed of a spacecraft is much faster than the local speed of sound, a shock layer forms in front of the capsule body as it reaches the Martian atmosphere. Communications are interrupted during the entrance phase due to the high plasma density surrounding the capsule. This event is known as \textit{blackout}. Because this process includes numerous chemical processes, there is no straightforward answer for the plasma density distribution surrounding the capsule. The rate at which electrons are produced as a result of the collision is determined by the spacecraft's speed, capsule design, air density, and composition. During the EDL phase of the Viking mission, the two Viking landers encountered one-minute communication blackouts on the UHF band. Some solutions for blackout include mounting the antenna on the sheltered side where plasma density is dissolved and using relay orbiters to communicate, adding some chemicals to eliminate the plasma, and increasing the carrier frequency from X-band into Ka-band \cite{Radio_wave_handbook}. 
	\begin{table}[!t]
		\caption{Attenuation of Radio Waves on Various Frequency Bands \cite{Radio_wave_handbook}.}
		\label{tab:radio_wave_att_around_mars_var_freq}
		\centering
		\resizebox{\columnwidth}{!}{
			\renewcommand{\arraystretch}{1.25}
			\begin{tabular}{l|l|l|l|l|}
				\cline{2-5}
				& \multicolumn{1}{c|}{\textbf{\begin{tabular}[c]{@{}c@{}}VHF\\ (100-500 MHz)\end{tabular}}} & \multicolumn{1}{c|}{\textbf{\begin{tabular}[c]{@{}c@{}}S-Band\\ (2-4 GHz)\end{tabular}}} & \multicolumn{1}{c|}{\textbf{\begin{tabular}[c]{@{}c@{}}X-Band\\ (10-12 GHz)\end{tabular}}} & \multicolumn{1}{c|}{\textbf{\begin{tabular}[c]{@{}c@{}}Ka-Band\\ (30-38 GHz)\end{tabular}}} \\ \hline
				\multicolumn{1}{|l|}{\textbf{Ionosphere (dB)}}                                                       & 0.5                                                                                       & 0.15                                                                                     & 0.1                                                                                        & 0.05                                                                                        \\ \hline
				\multicolumn{1}{|l|}{\textbf{Troposphere (dB)}}                                                      & 0                                                                                         & 0                                                                                        & 0                                                                                          & negligible                                                                                  \\ \hline
				\multicolumn{1}{|l|}{\textbf{Gaseous (dB)}}                                                          & 0                                                                                         & 0                                                                                        & 0                                                                                          & 0                                                                                           \\ \hline
				\multicolumn{1}{|l|}{\textbf{Cloud (dB)}}                                                            & 0                                                                                         & 0                                                                                        & 0.05                                                                                       & 0.1                                                                                         \\ \hline
				\multicolumn{1}{|l|}{\textbf{Rain (dB)}}                                                             & 0                                                                                         & 0                                                                                        & 0                                                                                          & 0                                                                                           \\ \hline
				\multicolumn{1}{|l|}{\textbf{Fog (dB)}}                                                              & 0                                                                                         & 0                                                                                        & 0                                                                                          & 0.1                                                                                         \\ \hline
				\multicolumn{1}{|l|}{\textbf{Aerosol (dB)}}                                                          & 0                                                                                         & 0                                                                                        & 0                                                                                          & 0.1                                                                                         \\ \hline
				\multicolumn{1}{|l|}{\textbf{Dust (dB)}}                                                             & 0.1                                                                                       & 0.3                                                                                      & 1.0                                                                                        & 3.0                                                                                         \\ \hline
				\multicolumn{1}{|l|}{\textbf{\begin{tabular}[c]{@{}l@{}}Total Vertical \vspace{-1mm}\\ Losses (dB)\end{tabular}}} & 0.5                                                                                       & 0.45                                                                                     & 1.15                                                                                       & 3.35                                                                                        \\ \hline
			\end{tabular}
		}
	\end{table}

	\section{Channel Modeling} \label{sec:channel_modeling}
	This section provides an overview of channel models anticipated for deep space radio links and studies on 3D channel modeling in the Mars environment from literature. 
	
	The propagation medium between Mars and Earth that microwave signals travel is essentially free space because there are no obstacles to them apart from rocks, meteors, and planets. On the other hand, the distance between Earth and Mars is approximately $ 225\times10^{7} $ \si{\km}, which is a big challenge for the latency of signals even if they travel at the speed of light. So, the link budget is mainly made up of path loss and atmospheric losses. Even atmospheric losses can be neglectable because, at the S-band and X-band frequencies like \SI{2.3}{\giga\hertz} and \SI{8.4}{\giga\hertz}, which are being used to communicate with Mars, Earth's atmosphere has almost no attenuation effect on signals \cite{yuen1983deep_space_telecom}.
	
	Generalized received power consisting of possible losses along with path loss and atmospheric loss is given by the following expression:
	\begin{equation}
		P_{R} = P_{T} L_{A} L_{S} L_{T} L_{TP} L_{R} L_{RP} L_{P} G_{T} G_{R} 
	\end{equation}
	where $ P_{R} $ is the received signal power, $ P_{T} $ is the total transmitted power, $ L_{A} $ is the atmospheric attenuation, $ L_{S} $ is the space loss, $ L_{T} $ is the circuit loss between transmit antenna terminals and radio case due to cabling, $ L_{TP} $ is the pointing loss of the transmit antenna, $ L_{R} $ is the circuit loss between receive antenna and receiver due to cabling, $ L_{RP} $ is the pointing loss of the receive antenna, $ L_{P} $ is the polarization loss between transmit and receive antennas due to mismatch in polarization patterns, $ G_{T} $ is the transmit antenna gain, and $ G_{R} $ is the receive antenna gain \cite{Yuen1983_deep_space}. 
	The transmit antenna gain $ G_{T} $ is given in terms of the effective antenna aperture $ A_{T} $ as 
	\begin{equation} \label{eq:GT}
		G_{T} = \frac{ 4 \pi A_{T}}{\lambda^{2}}
	\end{equation}
	where $ \lambda $ is the wavelength. Here, $ A_{T} $ is given as
	\begin{equation}
		A_{T} = \mu A_{t}
	\end{equation}
	where $ A_{t} $ is the actual antenna aperture and $ \mu $ is the antenna efficiency. The receive antenna gain is also defined similar to \eqref{eq:GT}. Detailed information about pointing and polarization losses are given in \cite{Rahmat-Samii1983}.

	In \cite{Pan_review_of_channel_mod_ds}, Earth to Mars channel link is divided into three parts; the near-Earth link, the interstellar link, and the near planet link. Characteristics of those links are different in terms of noise and interference. So, there must be different channel models for each link. In what follows, we investigate these near-Earth, interstellar, and near-planet links.
	\subsection{Near-Earth Link}
	Earth's atmosphere is the main component of the near-Earth link. Water vapor and oxygen molecules composing Earth's atmosphere have an attenuation effect on electromagnetic waves. On the other hand, the impact of those molecules can be ignored when the signal's frequency is in the  $ 0.3 $-$ 10 $ \si{\giga\hertz} range. Also, fluctuations can be observed in the amplitude and phase of the received signal because of atmospheric scintillation resulting from tropospheric atmospheric turbulence. Several statistical models were developed to characterize this scintillation, such as the STHV2 model, ITU-R model, Ortgies-N model, Karasawa model, and van de Kamp model.
	
	There is also rain and fog attenuation in the near-Earth link. The scattering absorption of rainwater, whose attenuation is highly dependent on local precipitation, is the primary source of rain attenuation. The Corazza model, composed of Rician and Lognormal distributions, can characterize cloudy weather that causes multipath and shadow effects due to vast volumes of clouds. Similarly, the dominant factor of signal attenuation in the case of foggy weather is the density of the fog. As a result, the Nakagami model can characterize it with a parameter representing the density of the fog.
	
	\subsection{Interstellar Link}
	The interstellar link is the most dominant one when communicating with Mars and other planets. Because electromagnetic waves can not be entirely concentrated in a specific direction, the energy collected at a certain point decreases exponentially as the total distance that electromagnetic waves travel to reach that point increases. The shortest distance between the two planets is around $55\times10^6 $ \si{\km} when Earth passes between Mars and the Sun. The maximum distance between the two planets is close to $400\times10^6 $ \si{\km} when they are on opposite sides of the Sun. Assuming that isotropic antennas are used, the loss caused by such long distances is called free-space path loss, and it is expressed with the famous Friis formula
	\begin{equation}
		L_{FS}  = \left( \frac{ 4 \pi d }{ \lambda } \right)^{2}
	\end{equation}
	where $ d $ is the distance between transmitter and receiver in meters and $ \lambda $ is the wavelength in meters. Free space losses calculated for various frequencies are given in Table \ref{tab:free_space_losses}. Free space path loss can also be given in terms of \si{\decibel} as 
	\begin{equation}
		\left(L_{FS}\right)_{\si{\decibel}} = 20 \log f + 20 \log d - 27.55
	\end{equation}
	where $ f $ is the frequency in \si{\mega\hertz}. 
	
\begin{figure}[!t]
		\centering
		\includegraphics[scale = 0.7]{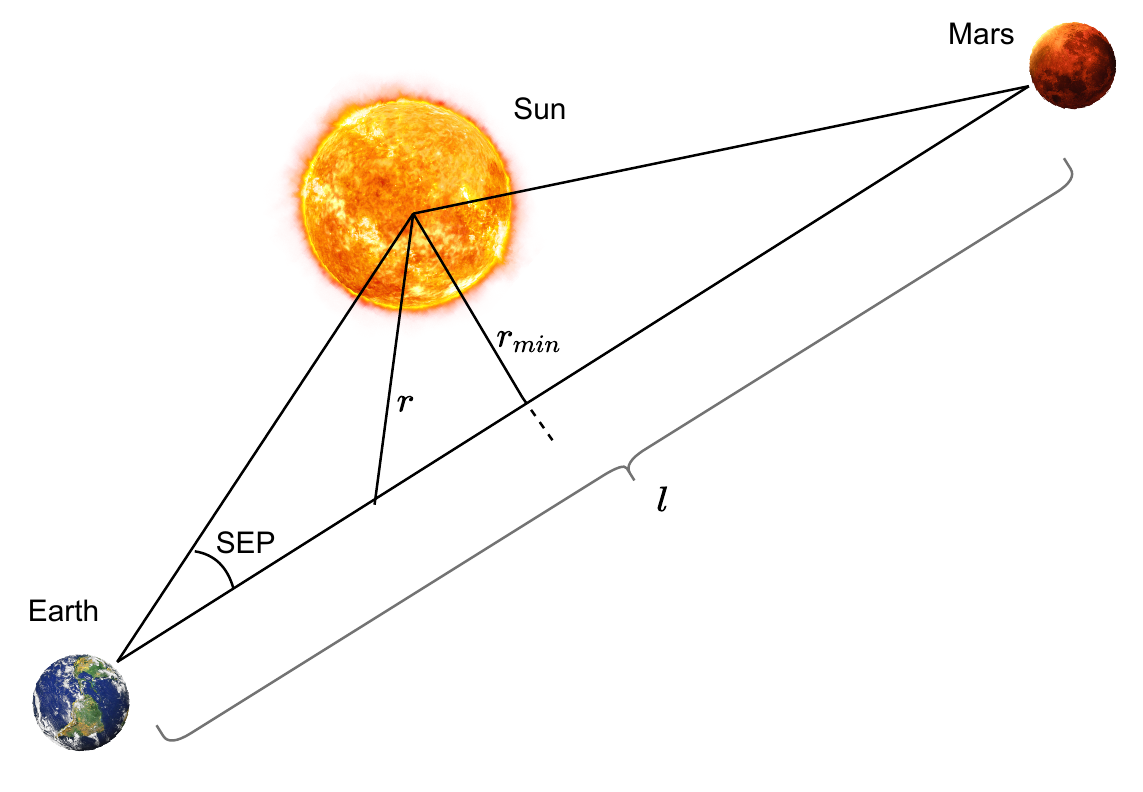}
		\caption{Earth and Mars's relative positions to the Sun during their superior conjunction \cite{morabito2003solar}.}
		\label{fig:superior_conj_geometry}
	\end{figure}

	\begin{table}[!t]
		\caption{Free Space Losses Between Mars and Earth at Several Frequencies \cite{Radio_wave_handbook}.}
		\label{tab:free_space_losses}
		\centering
		\renewcommand{\arraystretch}{1.3} 
		\begin{tabular}{l|c|c|}
			\cline{2-3}
			& \textbf{Min. Distance}     & \textbf{Max. Distance}      \\ \hline
			\multicolumn{1}{|l|}{\textbf{Distance (km)}}      & $ 55 \times 10^{6} $ & $ 400 \times 10^{6} $ \\ \hline
			\multicolumn{1}{|l|}{\textbf{VHF-Band (300 MHz)}} & $\sim$237 dB               & $\sim$254 dB                \\ \hline
			\multicolumn{1}{|l|}{\textbf{S-Band (3 GHz)}}     & $\sim$257 dB               & $\sim$274 dB                \\ \hline
			\multicolumn{1}{|l|}{\textbf{X-Band (10 GHz)}}    & $\sim$267 dB               & $\sim$284 dB                \\ \hline
			\multicolumn{1}{|l|}{\textbf{Ka-Band (32 GHz)}}   & $\sim$277 dB               & $\sim$294 dB                \\ \hline
		\end{tabular}%
	\end{table}
	During the travel of signals in interstellar space, solar scintillation also has an undeniable effect. Charged particles spreading from the Sun via solar winds scatter the electromagnetic waves if the link between spacecraft and Earth ground station is close to the Sun. Scattering waves may cause to multipath effect. In 2003, Morabito developed the Rician channel model after discovering a link between the scintillation index and the Rician factor. It has constantly improved since then.
	Fluctuations in the received signal power can decrease telemetry performance. The amount of fluctuations caused by solar scintillation in superior conjunction varies with several factors. The minimum distance from the Sun to ray path of the signal in the Eart-Mars link, expressed as $ r_{min} $ in Fig. \ref{fig:superior_conj_geometry}, is one of the important factors. When $ r_{min} $ is low enough Sun–Earth–probe (\acrshort{sep}) angle can be directly used to determine the ray distance ($ l $ in Fig. \ref{fig:superior_conj_geometry}) that signal travel between two planets according to trigonometry. As the SEP angle decreases, fluctuations in the densities of electrons increase. In other words, the communication link is substantially disrupted by solar scintillation when the planets are in superior conjunction. However, if the SEP angle is in a range where the fluctuations saturate, the amount of increase in the scintillation intensity fades away \cite{morabito2003solar}. 	
	The solar scintillation intensity, which denotes the plasma concentration in the solar wind, is measured using the index $ m $. If $ m $ is less than $ 0.3 $, the scintillation is weak. If $ m $ is between $ 0.3 $ and $ 1 $, it is in the transition area. If $ m $ equals to $ 1 $, the scintillation is strong. The relationship between $ m $ and the Rician factor can be expressed as
	\begin{equation}
		K = \frac{\sqrt{1-m^{2}}}{1-\sqrt{1-m^{2}}} .
	\end{equation}
	NASA has modeled the relation between $ m $ and the SEP angle in the X-band in \cite{morabito2003solar}.

	The antenna pointing and polarization error loss are the two main factors of antenna losses. Antenna pointing error loss occurs when the receiver antenna axis cannot be pointed in the optimal direction that the energy of the transmitted signal is maximum. Spacecraft and ground stations have different reasons to face this error. Control accuracy is the main parameter of the pointing error loss for spacecraft, and rotation accuracy and weather conditions are the main parameters of the loss for ground stations. 
	The antenna's real gain and pointing error have the following relation \cite{Pan_review_of_channel_mod_ds}:
	\begin{equation}
		G\left(\theta\right) = G_{m} e^{-2.77 \left( \theta / \theta_{0.5} \right)^{2}},
	\end{equation}
	where $ \theta $ is the antenna pointing error, $ \theta_{0.5} $ is the beamwidth, and $ G_{m} $ is the maximum gain of antenna. Hence, the following expression can be given for the pointing loss \cite{Pan_review_of_channel_mod_ds}:
	\begin{equation}
		\left[L_{pe}\right] = \left[G_{m}\right] - \left[ G\left(\theta\right) \right] = 12 \left( \theta / \theta_{0.5}\right)^{2}.
 	\end{equation}
	Additionally, changes in satellite altitude create antenna polarization loss, preventing polarization matching by the transceiver antenna. There are also complex cosmic rays coming from other solar systems, which is considered as Gaussian distributed random variable with a constant power spectral density \cite{Pan_review_of_channel_mod_ds}.
	
	\subsection{Near-Planet Link}
	Channel analysis is similar to Earth in the Near-Planet link because it is determined by atmospheric conditions as in Earth. In the Mars case, the attenuation of electromagnetic waves is much smaller than on Earth because the atmosphere of Mars is thinner than Earth. Cloud formation, snowfall, temperatures, the gaseous structure of the atmosphere, terrains, and the occurrence of dust storms are the major subjects of the near planet link channel model. The Rayleigh and Mie scattering principles are used according to the particle size and wavelength comparison. If the particle size is smaller than the wavelength, Rayleigh is chosen, and Mie otherwise. Throughout dust storms, it is discovered that Mars is far more visible than Earth \cite{Shekh_effect_of_sandstorm}.
	
	The lognormal distribution is used to model the distribution of dust particles on Mars. The Mie scattering theory is used to determine the connection between visibility and attenuation per unit distance. When dust particles are analyzed at the micro-level, it can be seen that their size is widely spread in a range of diameters from tens to hundreds of microns. According to some research studies, lognormal distribution can determine the distribution of sizes of dust particles. When the decay rates of water vapor and oxygen molecules in Earth's and Mars's atmospheres are compared in terms of attenuation, it is observed that the decay rate on Earth's atmosphere is roughly \SI{e4}{\percent} and \SI{e5}{\percent}, that of Mars, respectively \cite{Diao_compr_of_inf_2021}.
	\subsection{Mars Environment Simulations: An Overview of the Literature}
	There are a number of works on 3D Martian channel modeling in the literature from the past couple of years. This subsection will provide a brief overview of these works to understand Martian channel characteristics better. 
	
	In \cite{Pan_review_of_channel_mod_ds}, a computer simulation considering ground station-orbiter and orbiter-rover link is made on STK. The ground station is considered to have \SI{35}{\meter} antenna aperture, \SI{8.5}{\giga\hertz} received frequency, \SI{2}{\mega\hertz} bandwidth, and \SI{50}{\percent} antenna efficiency. The orbiter, on the other side, is considered to have $ 44.3 $ $ \mathrm{dBW} $ transmit power, \SI{8.5}{\giga\hertz} frequency for ground station, \SI{415}{\mega\hertz} frequency for the rover, \SI{2.5}{\meter} antenna aperture, and \SI{50}{\percent} antenna efficiency. Finally, the rover is considered to have \SI{415}{\mega\hertz} receive frequency and \SI{2}{\mega\hertz} bandwidth. The computer simulation period is set from March to April 2021. The received power from the orbiter to the ground station and from the ground station to the orbiter for five day period is provided in this study. The received power of the ground station has been reported to be reducing and reasonably steady, indicating that Earth and Mars are steadily drifting apart during the computer simulation period. The rationale for this steady behavior is claimed to be that the distance between Earth and Mars is so large that changes in the relative distance have minimal influence. While the influence of the relative distance variation on the orbiter is said to be very significant, so is the fluctuation of received power. Furthermore, there were certain discrete points created by the orbiter's fast drift at its closest point to Mars, and the period was quite short, making communication extremely difficult.

	In \cite{Chukkala_modeling_the_radio_freq_env}, a channel modeling computer simulation is made using HertzMapper software. Gusev crater is selected as a simulation environment, one of four possible landing sites for rovers selected by the Mars exploration community. For the environment parameters, permittivity is selected between $ 2.5 $ and $ 9 $, conductivity as \SI[per-mode=symbol]{e-8}{\siemens\per\meter}, and pressure as \SI{6.1}{\milli\bar}. For the communication parameters, \SI{1}{\meter} height omnidirectional antenna with \SI{1}{\watt} radiated power at \SI{2.4}{\giga\hertz} is selected. For the receiver, a minimum requirement of $ -93 $ $ \mathrm{dBm} $ received power for a $ 1 $ $ \mathrm{Mbps} $ link and a minimum of $ -84 $ $ \mathrm{dBm} $ for a $ 11 $ $ \mathrm{Mbps} $ link is assumed. According to the results, even with low radiated power and low antenna heights, areas that are several thousand meters could be covered at \SI{2.4}{\giga\hertz}. It is claimed that with such a range, a wireless microcell on Mars might be built, with the lander functioning as a base/relay station and rovers/sensors utilizing low-cost, IEEE 802.11b wireless networking technology. It is also claimed that the data rates might be enough to send pictures and even low-resolution videos. 
	
	In \cite{Chukkala_simulation_and_analysis}, the authors used the same parameters as in \cite{Chukkala_modeling_the_radio_freq_env} to calculate PDPs for different distances such as \SI{100}{\meter}, \SI{500}{\meter} and \SI{1000}{\meter}. Results showed that, the RMS delay spread values calculated for the \SI{100}{\meter} points had less than \SI{0.72}{\micro\second} in \SI{92}{\percent} of the cases. For the \SI{100}{\meter} scenario, the highest delay spread value determined was \SI{0.7565}{\micro\second}. The RMS delay spread values calculated for the \SI{500}{\meter} points had less than \SI{0.72}{\micro\second} in \SI{70}{\percent} of the cases. For the \SI{500}{\meter} scenario, the highest delay spread value determined was \SI{3.1257}{\micro\second}. Finally, the RMS delay spread values calculated for the \SI{1000}{\meter} points had less than \SI{0.72}{\micro\second} in \SI{72}{\percent} of the cases. For the \SI{1000}{\meter} scenario, the highest delay spread value determined was \SI{3.0825}{\micro\second}. The results also revealed that landscape might cause large RMS delay spread values for an IEEE 802.11b 11 Mbps link, which can exceed \SI{3}{\micro\second} or more than \SI{10}{\percent} of the symbol duration in some situations.
	
	In \cite{Pucci_perf_evol_of_802_15_4}, the authors considered IEEE 802.15.4 (ZigBee) standard for Mars because of its low cost and low power consumption across wireless sensor network (\acrshort{wsn}) devices. Simulink and OMNET++ software are used for computer simulations. The Ricean channel model having both LOS and non-line-of-sight (\acrshort{nlos}) components with a Rice factor $ K=10 $ is considered for low-medium rock density areas, while the Rayleigh channel model is considered for high rock density areas. Additionally, three different channel models are used for dust storms: "faint dust storm channel," which had low particle density, faint wind, and mean particle radius of \SI{1}{\micro\meter}, "strong dust storm channel," which had medium particle density, strong wind, and mean particle radius of \SI{10}{\micro\meter}, and "heavy dust storm channel," which had high particle density, heavy wind, and mean particle radius of \SI{20}{\micro\meter}. The bit error rate (\acrshort{ber}) of offset quadrature phase shift keying (\acrshort{oqpsk}) modulation featured in the IEEE 802.15.4 standard is measured. Acceptable BER values for "standard case," "faint sandstorm," and "strong sandstorm" are observed relative to other cases. The BER trends in the scenarios of high rock density ("worst case") and "heavy sandstorm" are said to be quite similar and could be assumed to represent the worst cases among the given channel models. Also, computer simulations on OMNET++ considering an IEEE 802.15.4 network transmission system consisting of a mobile rover and $ 40 $ sensors that are randomly placed in a \SI{100}{\meter\squared} area are made. Simulation results revealed that an IEEE802.15.4-based WSN might be deployed in the Mars exploration. It is claimed that such a WSN could perform nicely, even when transmitting inside terrains with a high density of rocks.	
	
	In \cite{Daga_terrain_based_sim_of_802_11_a}, the use of IEEE 802.11a/b wireless local area network (\acrshort{w-lan}) standards for proximity wireless networks on the Martian surface are investigated. Gusev crater and Meridiani Planum (Hematite) are chosen to simulate the RF environment on the Martian surface. Then, path loss models obtained from these computer simulations are used in a PHY simulation. Results are reported to indicate that Martian terrain can cause multipath effects, which influence the performance of 802.11a and b. However, over distances up to a few hundred meters, orthogonal frequency division multiplexing (\acrshort{ofdm})-based IEEE 802.11a standard could have quite strong PHY performance in terms of BER and packet error rate (\acrshort{per}) with a very low radiated power on the order of \si{\milli\watt} and antenna heights of $ 1-2 $ \si{\meter}. $ 2.4 $ and $ 5.25 $ \si{\giga\hertz} carrier frequencies are used for IEEE 802.11b and 802.11a, respectively. With the help of those parameters, PDP, received power, and RMS delay spread at various distances to the transmitter are computed. 
	In the study, it is noted that despite increasing distances, there are some circumstances when the RMS delay spread is lower, and the received power is higher owing to advantageous topographical conditions. 
	
	In \cite{Sacchi_From_LTE-A_to_LTE-M}, a Long-Term Evolution (\acrshort{lte})-based Martian wireless network architecture called LTE-M is proposed. In the study, a space mission's communication objectives are considered to be two-fold as follows: first, the various uncrewed (and, in the near future, human) exploration entities should share information among themselves; second, the gathered and analyzed data should be relayed to Earth. The study claimed that LTE-M would provide stable and adaptable communication with high bandwidth accessibility for rover and lander communications, as well as effective human-to-human data transmission when crewed missions are planned. Simulating the RF Martian environment with significant propagation limitations is used to assess the adaptability of terrestrial LTE uplink and downlink transmission. The obtained data is utilized to provide some guidelines for future LTE-M implementations. In the design, LTE user equipment (\acrshort{ue}s) with lower radiated power, lower energy consumption, smaller size, and weight are given to rovers and, in the future, to human astronauts. The LTE-M cellular network is said to manage in-situ bi-directional data transfers between rovers, landers, and people. As with terrestrial LTE, the designated bandwidth for LTE-M is considered to be in the ultra-high frequency domain. The aerial satellite is stated to be in charge of the long-distance link to Earth as well as a relatively short-distance link to surface operators spread on the Martian surface.
	According to the study, satellites may operate in the X-band or contemplate using higher frequency bands such as Ka-band or mmWave bands to take advantage of bigger spectral bands. The path loss model of \cite{Chukkala_modeling_the_radio_freq_env} and the multipath channel model of \cite{Chukkala_simulation_and_analysis} are used in this work. Gusev crater and Meridiani Planum are considered Martian computer simulation environments. The delay spread observed in the Gusev crater is calculated as \SI{0.2}{\micro\second}, whereas that of Meridiani Planum is \SI{0.625}{\micro\second}, using a frequency of \SI{2.4}{\giga\hertz} and a reference distance of \SI{50}{\meter}. SIMULINK computer simulations are used to investigate the possibility of LTE transmission on Mars. Specifications given in Table \ref{tab:LTE_M_link_sim_par} are used for uplink and downlink. Channel coding is noted to be not taken into account in these simulations.
	It is reported that without FEC coding, 4-QAM and 16-QAM uplinks may operate properly in the Gusev Crater, reaching full spectral efficiencies of $ 2 $ $ \mathrm{b}/\mathrm{s}/\mathrm{Hz} $ and $ 4 $ $ \mathrm{b}/\mathrm{s}/\mathrm{Hz} $, respectively. 64-QAM uplink, on the other hand, needs at least $ 5/6 $-rate turbo coding to achieve the desired BER, according to the study. It is reported that Meridiani Planum is a more challenging environment for transmission and required a stronger turbo code for both modulations. In the case of downlink, all tested modulations are claimed to work without coding in Gusev crater, but when examining the downlink in Meridiani Planum, only 4-QAM could function without coding.
	
	In \cite{Bonafini_eval_of_large_scale_propag}, a realistic Martian channel model is developed by analyzing large-scale propagation phenomena on the Martian surface. The Digital Elevation Model (\acrshort{dem}) of the Gale crater is used in the computer simulations. The LOS power is estimated over the 3D environment created from DEM using a ray-tracing technique that models the movement of an EM signal produced by an isotropic antenna. For the \SI{2.4}{\giga\hertz} operating frequency under consideration, a complex dielectric permittivity value is determined, which affects the propagation of electromagnetic waves. Cole-Cole equations and The Johnson Space Center (\acrshort{jsc}) Mars-1 Martian regolith simulant are used to calculate permittivity. The angle of incidence, polarization, and permittivity is used to get the Fresnel coefficients. Then, path loss samples are calculated for different distances between the transmitter and receiver. For each distance, the samples are averaged, and the path loss exponent, which indicates the roughness of the terrain, is calculated by locating the path loss curve that best fits the averaged samples. Simulations are made on two subareas of the Gale crater; one is considerably flat, and the other is rocky. As a result, the flat area's path loss exponent is found to be $ 2.12 $, whereas the rocky one's parameter is $ 2.37 $. Finally, the standard deviation of the shadow fading is obtained as $ 11.41 $ \si{\decibel} for the flat area and $ 13.26 $ \si{\decibel} for the rocky area. 
	\begin{table}[!t]
		\caption{computer simulation Parameters of the LTE-M link adopted from \cite{Sacchi_From_LTE-A_to_LTE-M}}
		\label{tab:LTE_M_link_sim_par}
		\centering
		\renewcommand{\arraystretch}{1.5}
		\begin{tabular}{|l|c|}
			\hline
			\multicolumn{1}{|c|}{\textbf{Parameter}} & \textbf{Value} \\ \hline
			Carrier frequency & \SI{2.4}{\giga\hertz} \\ \hline
			Number of subcarriers & $2048$ \\ \hline
			Subcarrier spacing & \SI{15}{\kilo\hertz} \\ \hline
			Modulation constellation & 4-QAM, 16-QAM, 64-QAM \\ \hline
			Baud-rate & $ 30.72 $ $ \mathrm{Mbaud/sec} $ \\ \hline
			Number of users & $4$ \\ \hline
			Cyclic prefix length & $512$ \\ \hline
		\end{tabular}%
	\end{table}

	In \cite{Bonafini_Design_of_a_3D_ray-tracing}, the same ray-tracing technique developed in \cite{Bonafini_eval_of_large_scale_propag} is implemented, considering the DEM of Gale crater. Various outputs covering large and small-scale effects such as outage probability, shadowing parameters, path losses, and PDPs are presented. Operating frequencies of $ 2.5 $ and \SI{39.0}{\giga\hertz}, corresponding to the S and EHF bands, are used. Calculations are made on distances of $ 100 $ and \SI{200}{\meter} between transmitter and receiver. The path loss exponent of the flat area is found to be $ 2 $, which corresponds to free space path loss. However, the path loss of the rocky area is calculated around $ 2.6 $, which corresponds to terrestrial urban area.
	Additionally, compared to the flat area, the shadowing parameter of the rocky one was significantly higher, reaching approximately \SI{12}{\decibel}. Although mean delays were close to each other for the flat and rocky areas, the standard deviation for the rocky area was greater than the flat area considering the \SI{200}{\meter} distance. Also, path attenuation in the rocky area was more significant than in the flat area in most cases. 
	
	\begin{table*}[!t]
		\caption{An overview of the studies on Martian channel modeling}
		\label{tab:literature_summary}
		\centering
		\resizebox{\textwidth}{!}{%
			\renewcommand{\arraystretch}{1.6}
			\begin{tabularx}{\textwidth}{|c|c|X|}
				\hline
				\rowcolor[HTML]{DADCFB} 
				\multicolumn{1}{|l|}{\cellcolor[HTML]{DADCFB}\textbf{Study}} &
				\multicolumn{1}{l|}{\cellcolor[HTML]{DADCFB}\textbf{Computer Simulation Software}} &
				\textbf{Summary} \\ \hline
				\rowcolor[HTML]{EDEDFD} 
				\text{\cite{Pan_review_of_channel_mod_ds}} &
				STK &
				A computer simulation is made, considering the ground station-orbiter and orbiter-rover link. Various results are provided, such as losses, Doppler shift, and scintillation index. \\ \hline
				\rowcolor[HTML]{DADCFB} 
				\text{\cite{Chukkala_modeling_the_radio_freq_env}} &
				HertzMapper &
				A channel modeling computer simulation is made on Gusev crater. Results have shown that even with low radiated power and low antenna heights, areas that are several thousand meters could be covered at \SI{2.4}{\giga\hertz}. \\ \hline
				\rowcolor[HTML]{EDEDFD} 
				\text{\cite{Chukkala_simulation_and_analysis}} &
				HertzMapper &
				The authors used the same parameters as in \cite{Chukkala_modeling_the_radio_freq_env} to calculate PDPs for different distances. The results revealed that landscape might cause large RMS delay spread values for an IEEE 802.11b link. \\ \hline
				\rowcolor[HTML]{DADCFB} 
				\text{\cite{Pucci_perf_evol_of_802_15_4}} &
				SIMULINK \& OMNET++ &
				The authors considered the ZigBee standard for Mars because of its low cost and power consumption. Simulation results revealed that an IEEE802.15.4-based WSN might perform well, even when transmitting inside terrains with a high density of rocks. \\ \hline
				\rowcolor[HTML]{EDEDFD} 
				\text{\cite{Daga_terrain_based_sim_of_802_11_a}} &
				MATLAB &
				The use of IEEE 802.11a/b W-LAN standards for proximity wireless networks on the Gusev crater and Meridiani Planum are investigated. Results are reported to indicate that Martian terrain can cause multipath effects, which influence the performance of 802.11a and b. \\ \hline
				\rowcolor[HTML]{DADCFB} 
				\text{\cite{Sacchi_From_LTE-A_to_LTE-M}} &
				SIMULINK &
				An LTE-based Martian wireless network architecture called LTE-M is proposed and simulated on the Gusev crater and Meridiani Planum. The study claimed that LTE-M would provide stable and adaptable communication for rover and lander communications. \\ \hline
				\rowcolor[HTML]{EDEDFD} 
				\text{\cite{Bonafini_eval_of_large_scale_propag}} &
				MATLAB &
				A realistic Martian channel model for Gale crater was developed by analyzing large-scale propagation phenomena on the Martian surface. Path loss exponent and shadow fading parameters are presented for each scenario. \\ \hline
				\rowcolor[HTML]{DADCFB} 
				\text{\cite{Bonafini_Design_of_a_3D_ray-tracing}} &
				MATLAB &
				The same ray-tracing technique developed in \cite{Bonafini_eval_of_large_scale_propag} is implemented, considering the DEM of Gale crater. Outage probability, shadowing parameters, path losses, and PDPs are presented. \\ \hline
				\rowcolor[HTML]{EDEDFD} 
				\text{\cite{Bonafini_3D_Ray-tracing_analysis}} &
				MATLAB &
				The results of the 3D ray-tracing analysis on Gale crater in \cite{Bonafini_Design_of_a_3D_ray-tracing} are expanded. LOS, first and second reflections, and total received powers are plotted versus distance for each scenario. \\ \hline
				
			\end{tabularx}%
		}
	\end{table*}
	In \cite{Bonafini_3D_Ray-tracing_analysis}, the results of the 3D ray-tracing analysis on flat and rocky Gale crater areas in \cite{Bonafini_Design_of_a_3D_ray-tracing} are expanded. LOS, first and second reflections, and total received powers are plotted versus distance for both flat and rocky areas. According to the results, there are more received power samples below a certain threshold in the rocky area. Then, the outage probability is calculated as the ratio of the number of samples received below the threshold and the total number of samples. In the flat area, the average outage probability was less than \SI{7}{\percent} for both operating frequencies, \SI{2.5}{\giga\hertz} and \SI{39.0}{\giga\hertz}, while for the second sub-areas, the outage probability could reach \SI{40}{\percent}.
	
	A summary of the channel modeling studies reviewed in this subsection is provided in Table \ref{tab:literature_summary}.
	
	\subsection{New Computer Simulations for Mars Channel Modeling}
\begin{figure*}[!t]
		\centering
		\subfloat[3D model of Hogwallow Flats, Sol 461]{\includegraphics[trim={6cm 0 4cm 0},clip,width=3.5in]{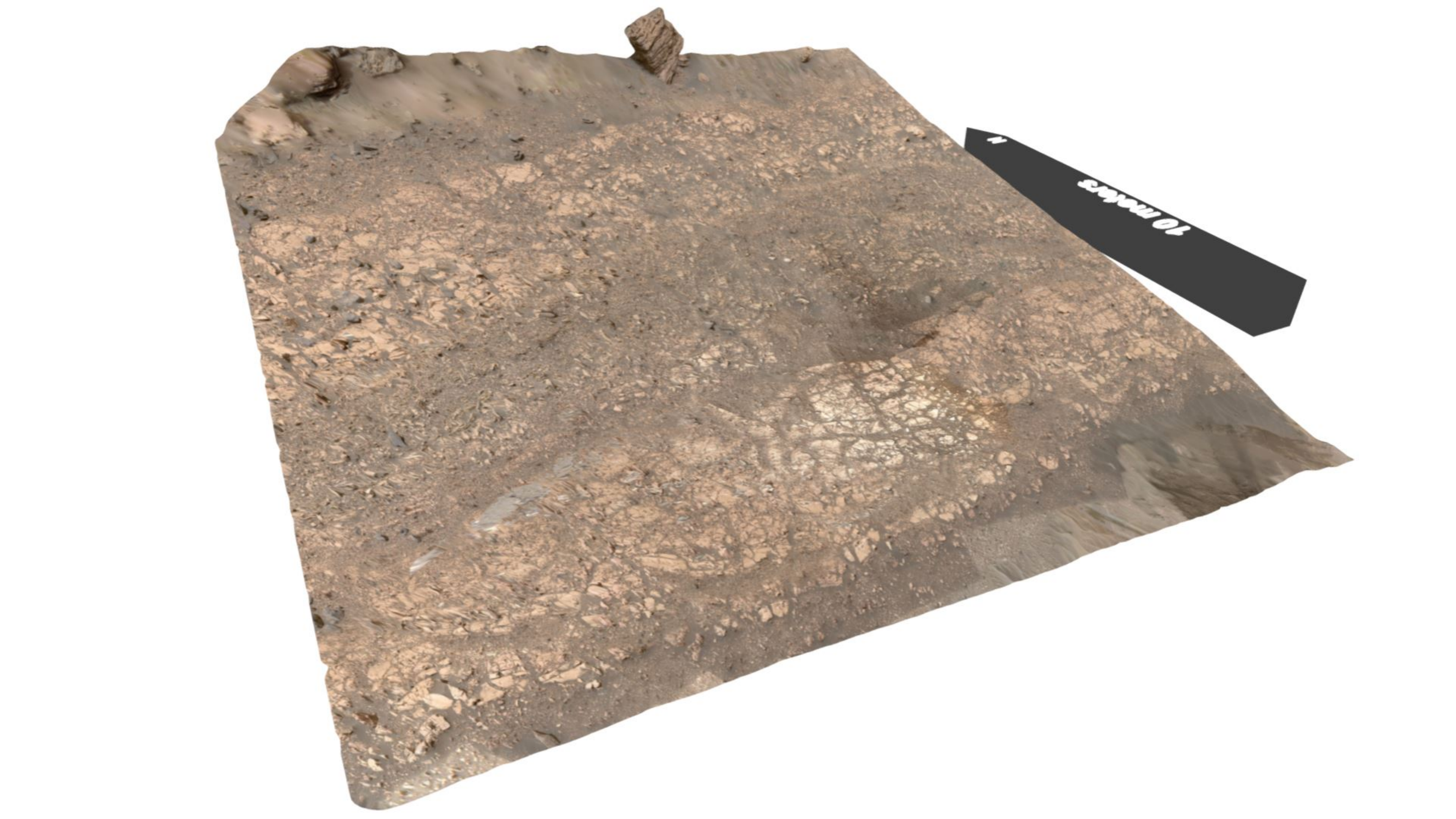}%
			\label{fig:model_of_hogwallow_flats}}
		\hfil
		\subfloat[3D model of Rocky Top, Sol 466]{\includegraphics[trim={2.5cm 0 2.5cm 0},clip,width=3.5in]{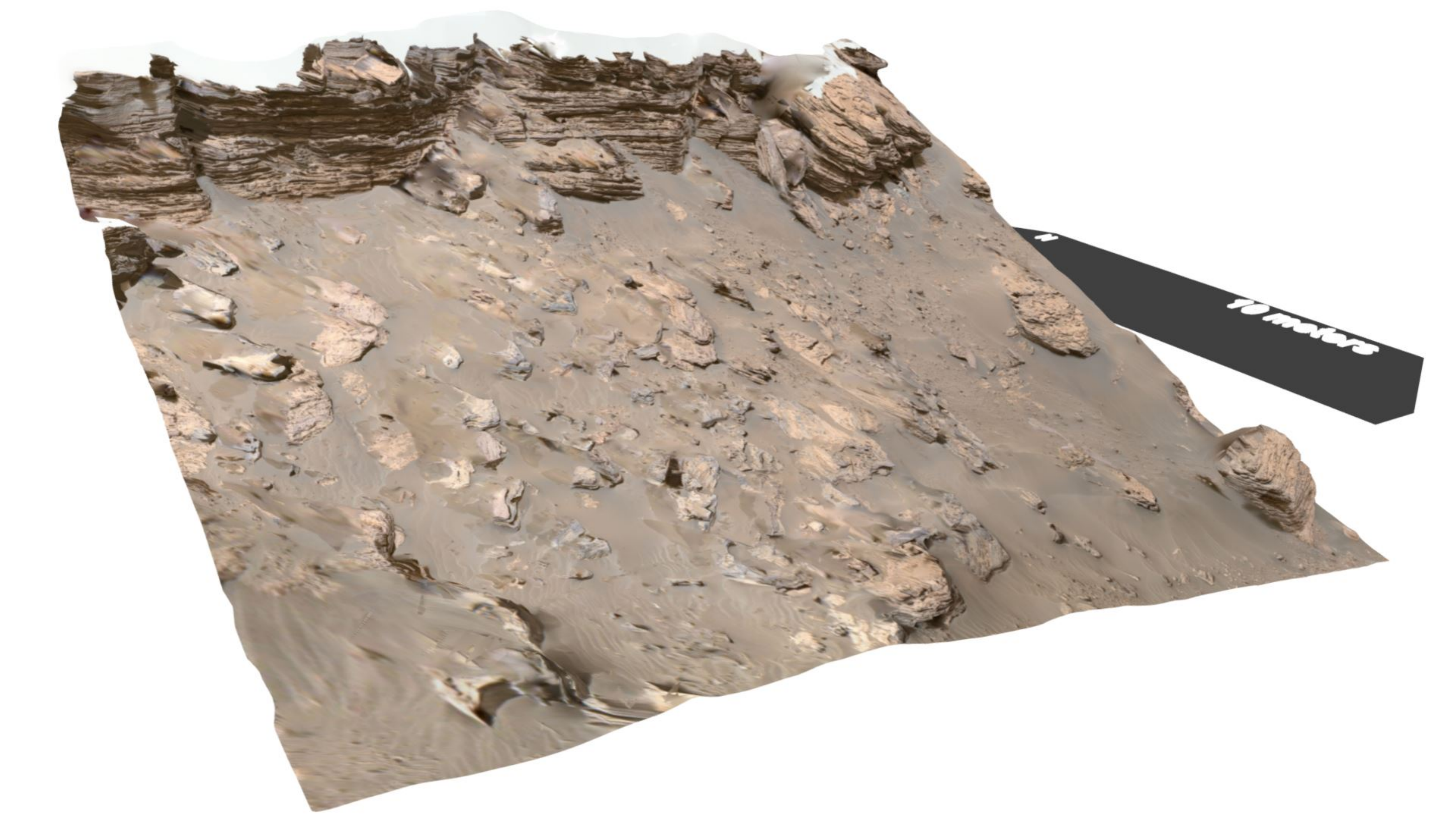}%
			\label{fig:model_of_rocky_top}}
		\caption{3D models created from high-resolution photos captured by the Perseverance rover \cite{Hogwallow_flats}. (a) Hogwallow Flats: combination of images taken on Sols 449-461\cite{Hogwallow_flats}. (b) Rocky Top: combination of images taken on Sols 459-470 \cite{Rocky_top}.}
		\label{fig:models_of_flats_and_rocky_togh}
	\end{figure*}

	Against this background, we have performed extensive computer simulations on the Mars environment using Remcom Wireless Insite software \cite{remcom_wireless_insite}. Wireless Insite is an enhanced ray tracing-based channel modeling software capable of calculating various channel-related parameters of the target scenario built using the tools given by the program. In challenging urban, indoor, rural, and mixed path situations, this RF propagation program offers practical and precise calculations of communication channel parameters. Insite has a linear approach to building a communication system. First, the simulation environment should be built within the software or imported as a supported file format. Different materials can be assigned to the environment or subparts of it. Then, the waveforms and antennas that will be used in the simulations should be created. As the most critical part, transmitters, and receivers should be placed using the project view as a visual reference. Different sets of transmitters and receivers might be used, such as points, polygon, route, trajectory, and XY grid. Lastly, a study area should be created where the user determines various parameters such as the propagation model, the number of reflections and diffractions, atmosphere parameters, and output requests. 
	We chose two small 3D areas created from images captured by the Perseverance Rover's Mastcam-Zs, Navcams, and Hazcams as a starting point. One of the areas is flat, roughly \SI{16}{\meter} $ \times $ \SI{16}{\meter} in $ x $-$ y $ axis and the other one considerably inclined and rocky, as shown in Fig. \ref{fig:models_of_flats_and_rocky_togh}. The scalebar in these models is \SI{10}{\meter} long and \SI{1}{\meter} wide, and it heads north. To increase the contrast of the models, the colors are brightened and gamma-corrected by the creator\footnote{This work is licensed under the Creative Commons Attribution 4.0 International License. To view a copy of this license, visit http://creativecommons.org/licenses/by/4.0/ or send a letter to Creative Commons, PO Box 1866, Mountain View, CA 94042, USA.}.

\begin{figure*}[!t]
		\centering
		\subfloat[]{\includegraphics[trim={0cm 2.8cm 0cm 1.8cm},clip,width=3.5in]{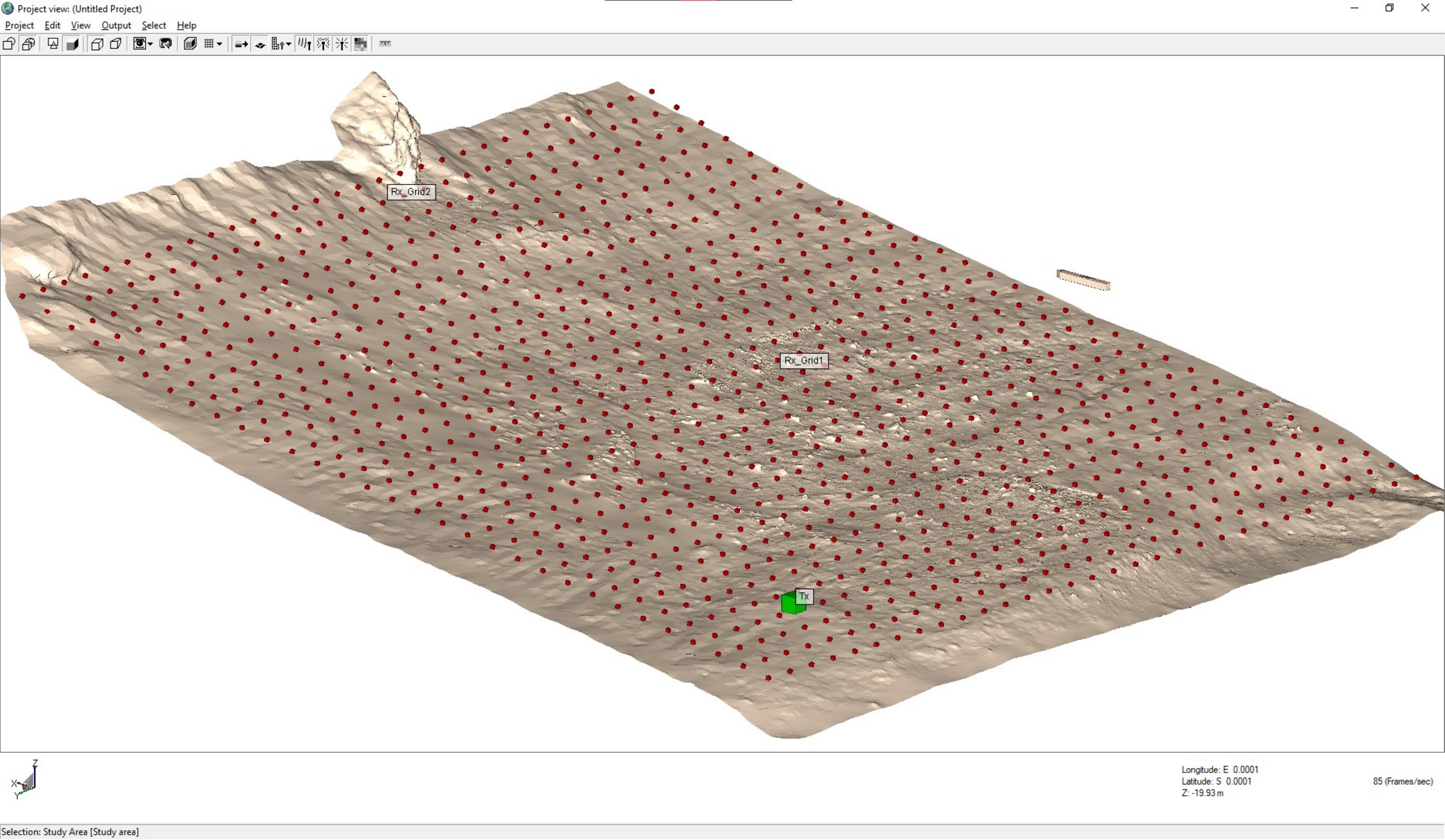}%
			\label{fig:placement_Tx_Rx_hogwallow}}
		\hfil
		\subfloat[]{\includegraphics[trim={0cm 2.8cm 0cm 1.8cm},clip,width=3.5in]{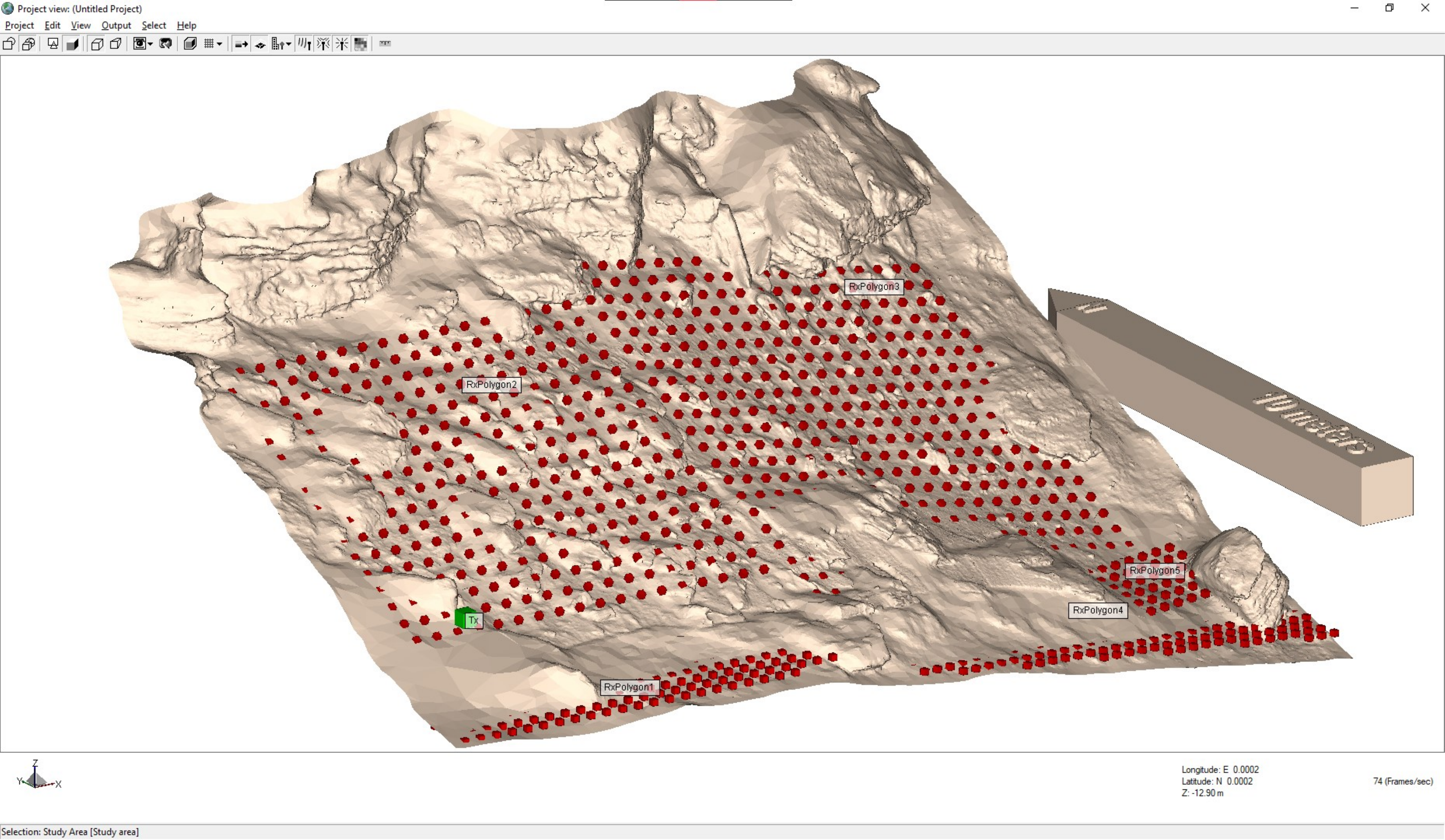}%
			\label{fig:placement_Tx_Rx_rocky}}
		\caption{Placement of transmitters and receivers on 3D models. Green and red cubes represent transmitters and receivers, respectively. (a) Placement of transmitter and receivers on Hogwallow Flats. (b) Placement of transmitters and receivers on Rocky Top.}
		\label{fig:placement_Tx_Rx}
	\end{figure*}

	There are four parameters that need to be set to create a new material for the Martian surface; thickness (\si{\meter}), roughness (\si{\meter}), conductivity (\si[per-mode=symbol]{\siemens\per\meter}), and permittivity. Thickness is set to \SI{1}{\meter} just for display purposes since Insite does not use this parameter in calculations. The mean global total vertical roughness for Mars was calculated in \cite{Garvin_global_vertical_roughness}. Conductivity and permittivity are set to \SI[per-mode = symbol]{e-8}{\siemens\per\meter} and \SI[per-mode = symbol]{4}{\farad\per\meter}, respectively as in \cite{Daga_terrain_based_sim_of_802_11_a}. We created a new material called "Dry Mars" using these three parameters. When creating a new waveform, the Insite menu requires three essential parameters: carrier frequency, effective bandwidth, and phase \cite{Insite_ref_manual}. These values are set to \SI{2.4}{\GHz}, \SI{40}{\MHz}, and \ang{0}, respectively. Isotropic antennas with 0 dBi of maximum gain are assigned to transmitters and receivers. X3D is utilized as the propagation model. The number of reflections and diffractions are set to $ 3 $ and $ 1 $, respectively, which may be enough for Martian outdoor propagation. Atmosphere parameters of temperature (\si{\celsius}), pressure (\si{\milli\bar}), and humidity $\left(\si{\percent}\right)$ are set to \num{-63}, \num{6.1}, and \num{20}. These parameters are commonly used in all computer simulations presented in this work. The placement of transmitters and receivers on the two 3D models is shown in Fig. \ref{fig:placement_Tx_Rx}. 
	The height of the transmitters on both models is \SI{2}{\meter}, which is approximately equal to the height of the Perseverance rover. Two grids of receivers were enough to cover the Hogwallow area because of its flat nature. On the other hand, even five grids of receivers were not enough to cover the Rocky area. Because the software does not automatically place receivers relative to given terrain, some of the receivers remain under the terrain. The received powers of those points are automatically assigned to \SI{-250}{\decibel} because there is not even a single path from the transmitter to them. So, we eliminated those values in the calculations. 

	Two outputs are essential to our work. One is the received power given in \si{\decibel} with its corresponding distance. It enabled us to calculate the path loss exponent using the minimum mean square error (MMSE) method. The other one is complex impulse response, which gives received powers of ray paths in $ \mathrm{dBm} $ with their corresponding mean time of arrivals in second. This data helped us to deliver PDPs with corresponding RMS delay spread (\si{\second}) values. 
	
\begin{figure*}[!t]
		\centering
		\subfloat[]{\includegraphics[scale = 0.6]{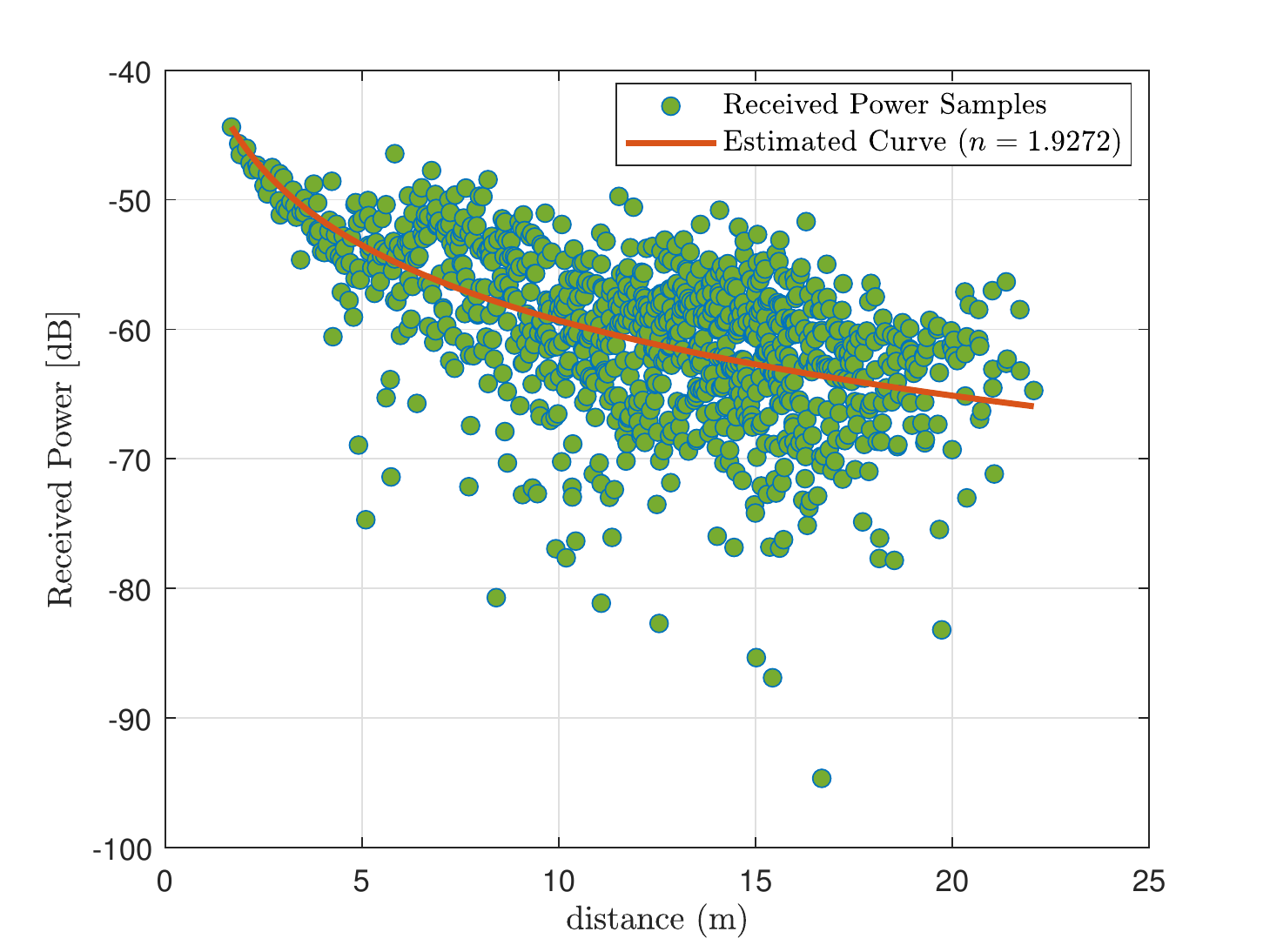}%
			\label{fig:rcv_pwr_hogwallow}}
		\hfil
		\subfloat[]{\includegraphics[scale = 0.6]{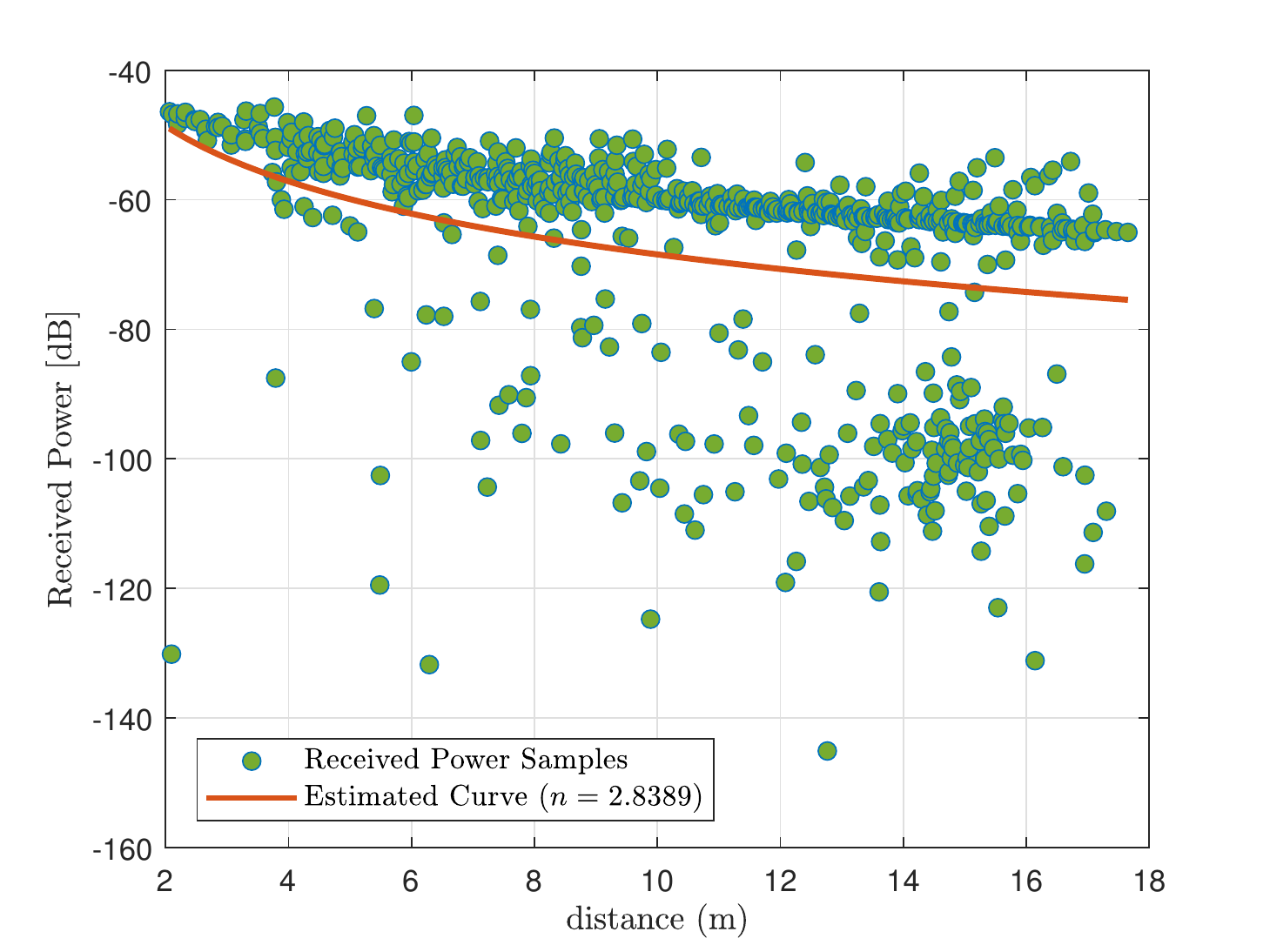}%
			\label{fig:rcv_pwr_rocky}}
		\caption{Received power samples versus distance. (a) Hogwallow Flats. (b) Rocky Top.}
		\label{fig:rcv_power_hogwallow_rocky}
	\end{figure*}

\begin{figure*}[!t]
		\centering
		\subfloat[]{
			\includegraphics[scale = 0.59]{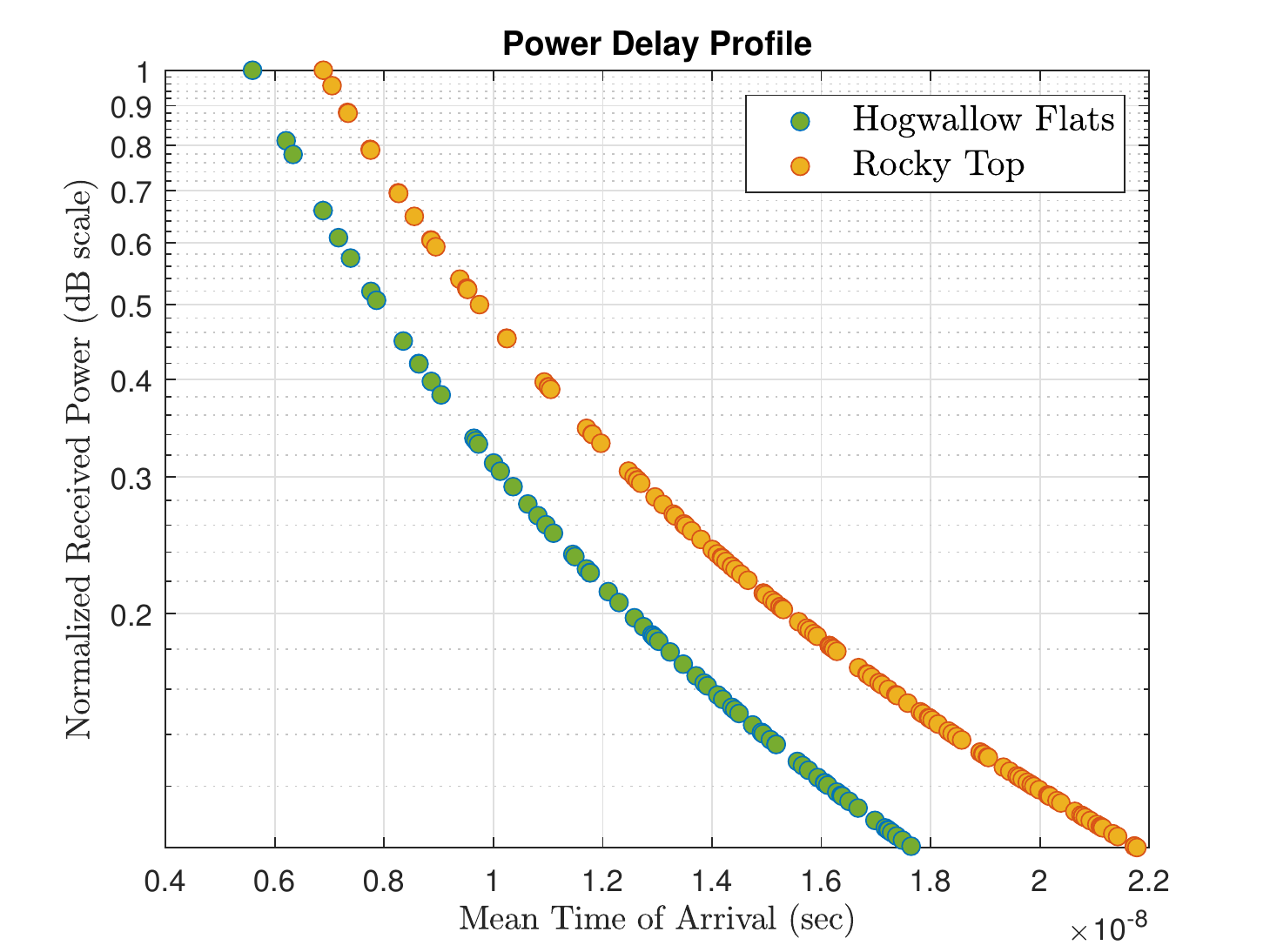}
			\label{fig:pdp_hogwallow_rocky_top}}
		\hfil
		\subfloat[]{
			\includegraphics[scale = 0.59]{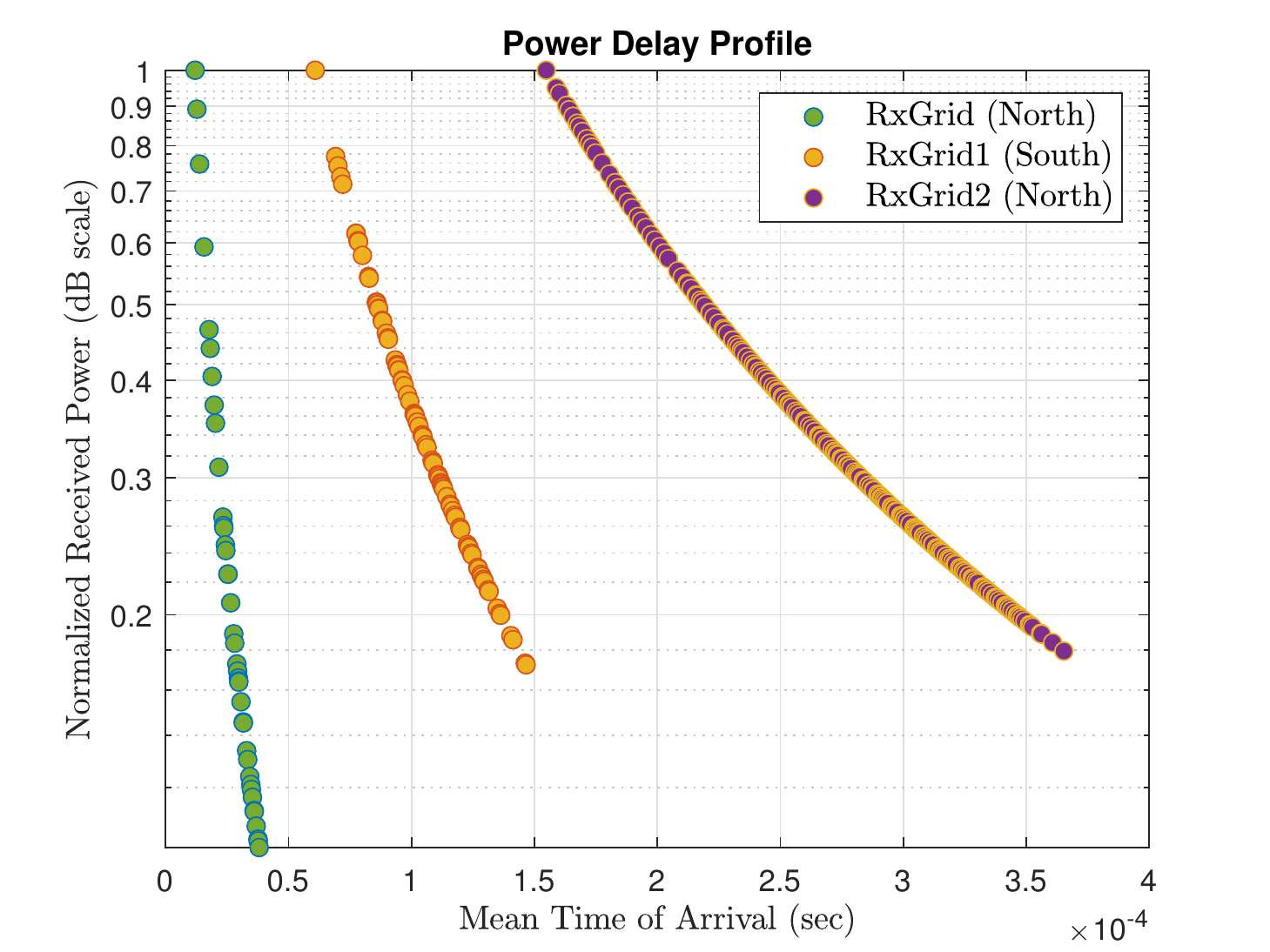}
			\label{fig:PDP_Gale_All}}
		\caption{PDPs of different Mars scenarios. (a) PDP of Hogwallow Flats and Rocky Top. (b) PDPs of various Gale crater scenarios.}
		\label{fig:all_PDPs}
	\end{figure*}

	The path loss exponent is estimated as $ 1.9272 $ for Hogwallow Flats, as shown in Fig. \subref*{fig:rcv_pwr_hogwallow}. This value is slightly lower than the path loss exponent value of the free space, which is $ 2 $. It makes sense because the area is small and flat; therefore, there is strong LOS. Also, the reflected ray paths may have a constructive effect on the received signal. The shadowing parameter is calculated as $ \SI{5.3141}{\decibel} $ for this area, which can be considered to be low as well because there are not many obstacles between the transmitter and the receivers.
	On the other side, the path loss exponent of the rocky area is higher with respect to the flat area as expected, reaching $ 2.8389 $. This value is in the range of terrestrial urban areas, as shown in Table 3.2 in \cite{Rappaport_wireless_comm}. The shadowing parameter is $ \SI{16.2670}{\decibel} $, higher than the flat area because of NLOS components and blockage from rocks. So, path loss exponent and shadowing parameters are comparable to the rocky area of \cite{Bonafini_Design_of_a_3D_ray-tracing}.
	PDPs of the two areas are graphically represented in Fig. \subref*{fig:pdp_hogwallow_rocky_top}. When plotting PDPs, we selected a noise threshold that neglects the samples that are ten times lower than the strongest multipath. The mean time-of-arrivals of multipath components are lower in Hogwallow Flats compared to Rocky Top, as predicted. 
	PDP can be used to calculate the mean excess delay $ \left( \bar{\tau} \right) $, RMS delay spread $ \left( \sigma_{\tau} \right) $, and excess delay spread of a multipath channel. The mean excess delay and RMS delay spread are often used to quantify the time dispersive features of wideband multipath channels. The mean excess delay is defined as the first moment of the PDP \cite{Rappaport_wireless_comm}:
	\begin{equation} \label{eq:mean_excess_delay}
		\bar{\tau} = \frac{ \displaystyle\sum_{k}^{} P \left( \tau_{k} \right) \tau_{k} }{\displaystyle\sum_{k}^{} P \left( \tau_{k} \right)} .
	\end{equation}
	The RMS delay spread is defined as the square root of the second central moment of the PDP \cite{Rappaport_wireless_comm}:
	\begin{equation} \label{eq:rms_delay_spread}
		\sigma_{\tau} = \sqrt{ \bar{ \tau^{2} } - \left( \bar{\tau} \right)^{2} }
	\end{equation}
	where
	\begin{equation} \label{eq:tau_square_bar}
		\bar{\tau^{2}} = \frac{ \displaystyle\sum_{k} P \left( \tau_{k} \right) \tau^{2}_{k} }{ \displaystyle\sum_{k} P \left( \tau_{k} \right) }.
	\end{equation}
	Here, $ \tau_{k} $ denotes the time delay and $ P_{k} $ denotes the received power for the $k$th multipath. Using \eqref{eq:mean_excess_delay}, \eqref{eq:rms_delay_spread}, and \eqref{eq:tau_square_bar}, we calculated the RMS delay spread of flat and rocky areas as $ \SI{3.4859}{\nano\second} $ and $ \SI{4.3235}{\nano\second} $, respectively. Typically, RMS delay spread is order on the order of \si{\nano\second} in indoor environments and on the order of \si{\micro\second} in outdoor environments. Our results are similar to indoor channels because the considered areas are relatively small. 	
	
\begin{figure*}[!t]
		\centering
		\subfloat[]{\includegraphics[trim={0cm 2.8cm 0cm 1.8cm},clip,width=3.5in]{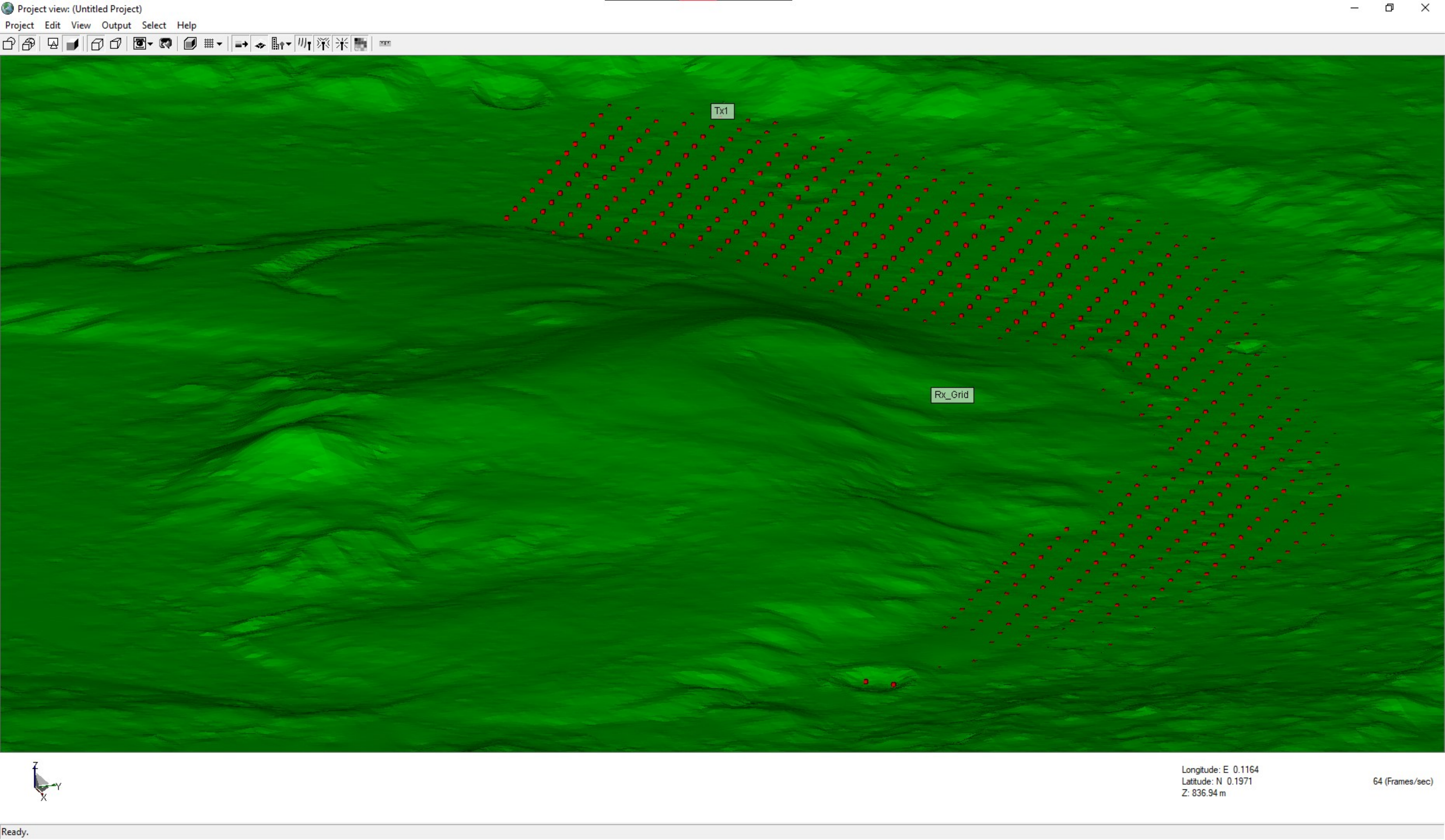}%
			\label{fig:Gale_curiosity}}
		\hfil
		\subfloat[]{\includegraphics[trim={0cm 2.8cm 0cm 1.8cm},clip,width=3.5in]{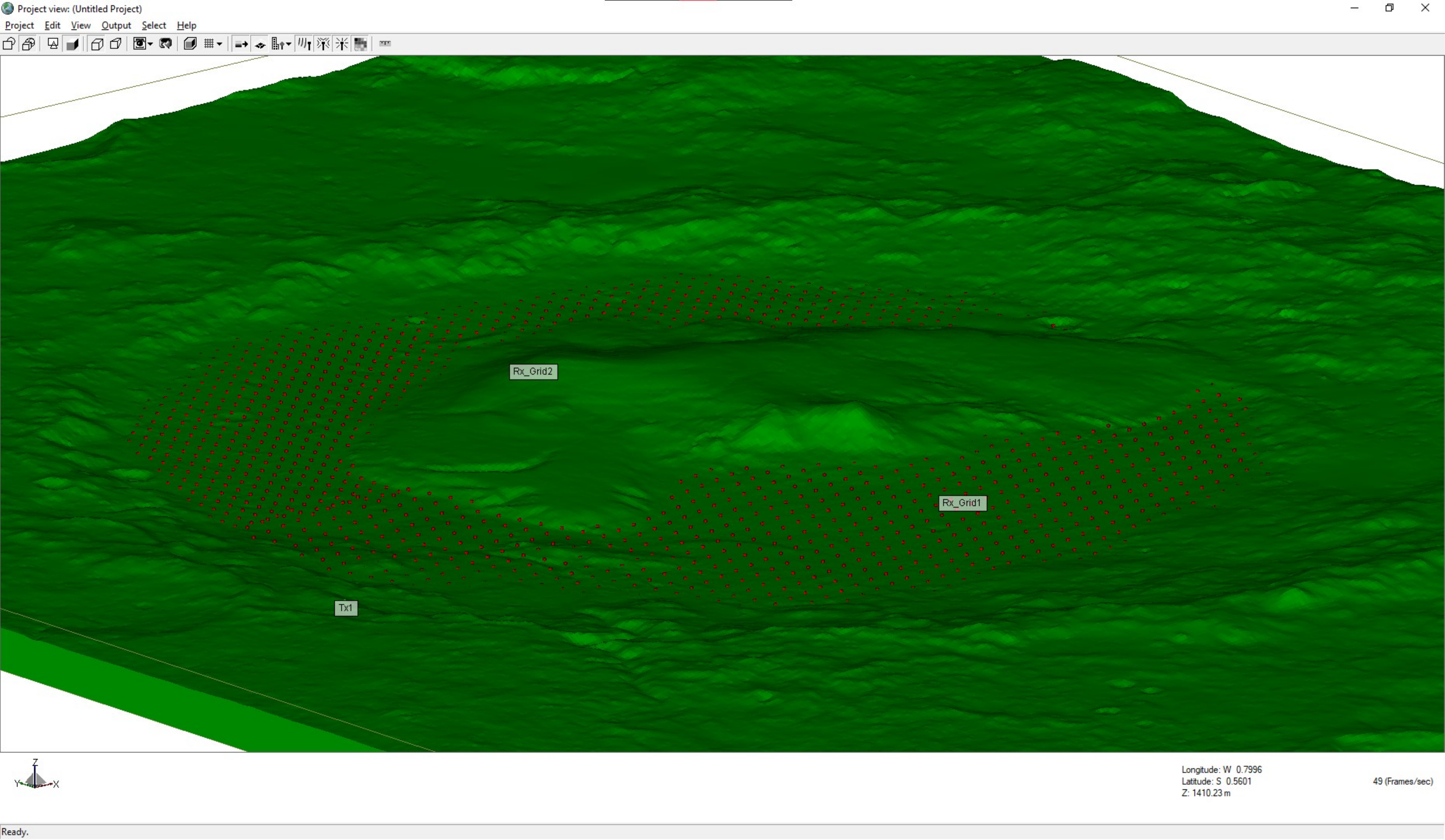}%
			\label{fig:Gale_large_grid}}
		\hfil
		\subfloat[]{\includegraphics[trim={0cm 2.8cm 0cm 1.8cm},clip,width=3.5in]{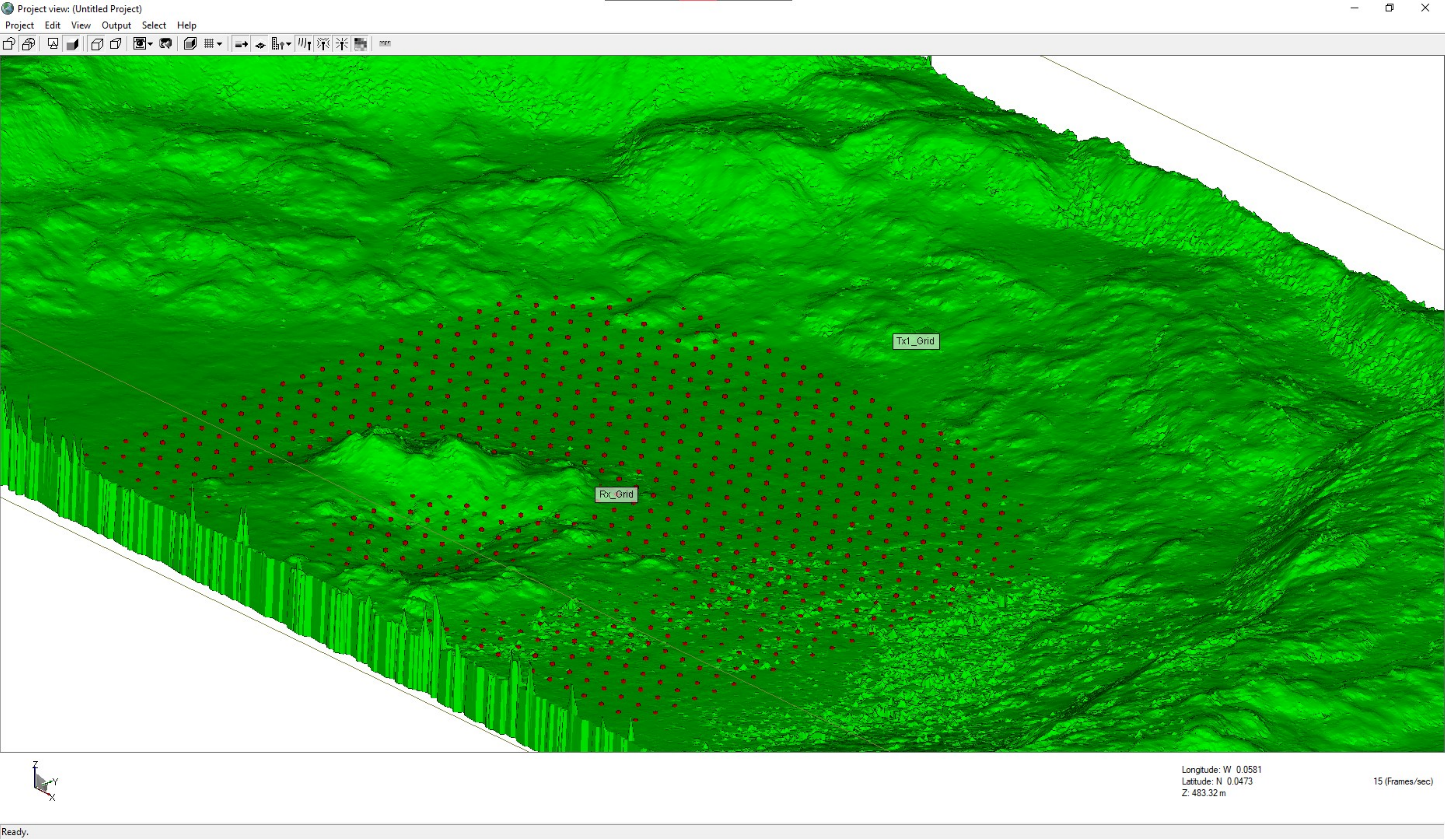}%
			\label{fig:Beagle}}
		\hfil
		\subfloat[]{\includegraphics[trim={0cm 2.8cm 0cm 1.8cm},clip,width=3.5in]{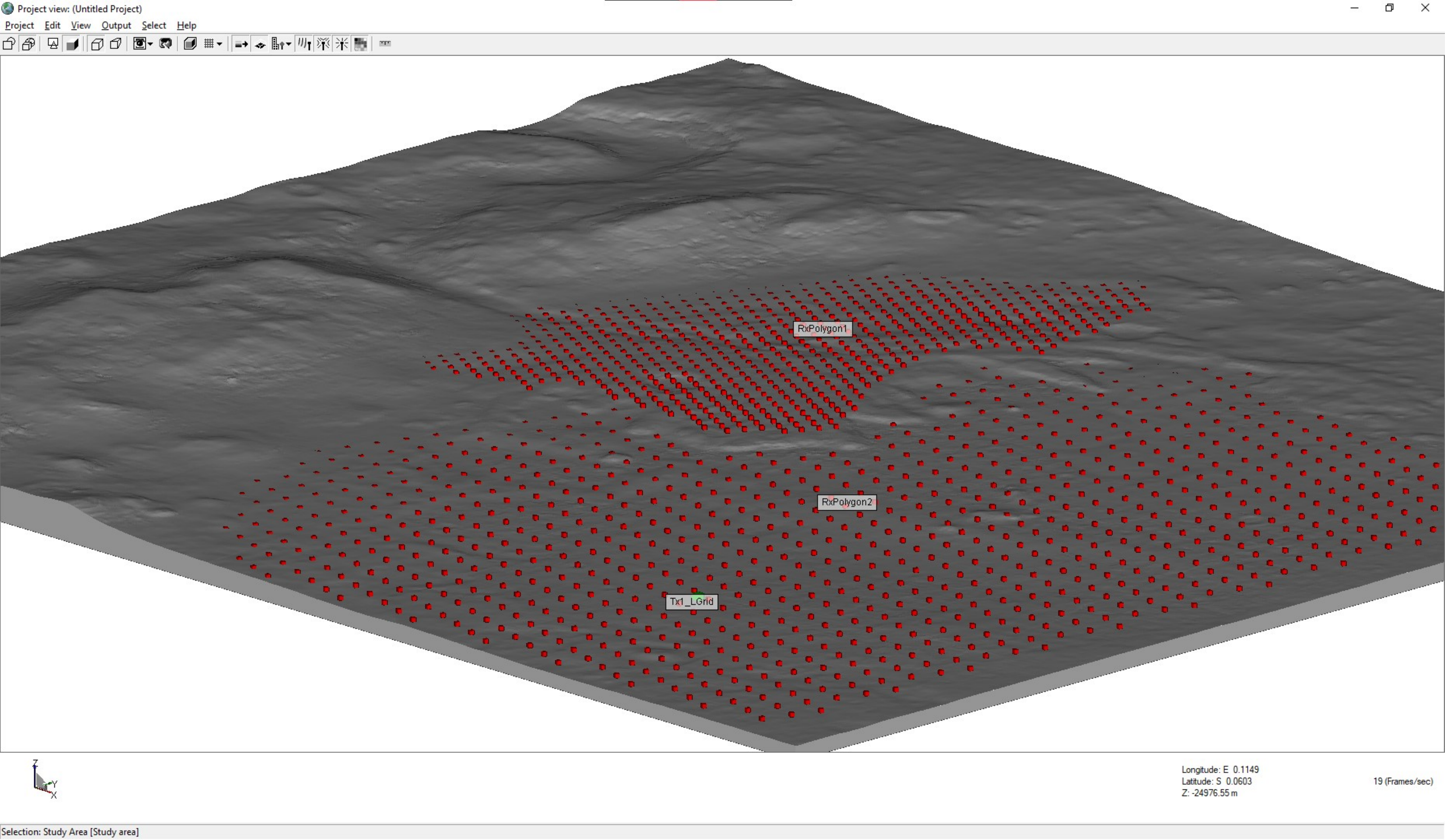}%
			\label{fig:Jezero}}		
		\caption{Placements of transmitters and set of receivers for different communication scenarios in Mars craters; (a) Gale crater - Rx\_Grid. (b) Gale crater - Rx\_Grid1 \& Rx\_Grid2. (c) Beagle crater. (d) Jezero crater.}
		\label{fig:Impcat_craters}
	\end{figure*}

	IIn the next step, we have expanded our work by analyzing the impact craters on Mars, considering the future large-scale deployment of crewed missions. For this, we have chosen the Gale crater, the landing site of the Curiosity rover. One of our reasons for choosing the Gale crater is that it has been studied for over a decade, which means a good deal of scientific data is available about this crater. Another reason is that some of the studies we covered before also used this crater in their computer simulations. Gale crater is roughly 155 \si{\km} ($ 96 $ \si{\mile}) in diameter and \SI{5.5}{\kilo\meter} (\SI{18,000}{\feet}) in height relative to the minimum terrain level. We used a 3D model of the Gale crater provided by NASA \cite{Nasa_Gale_3D_stl}. Then, we rescaled it on the $ x $, $ y $, and $ z $-axis at the same amount in a way that its depth fits into the actual measures. However, this was not enough because the models were vertically exaggerated to increase visuality. So, we rescaled it one more time on the $ x $ and $ y $-axis at the same amount in a way that its diameter fits into the actual measures, as shown in Fig. \ref{fig:Impcat_craters}. 

	The Curiosity rover landed on Bradbury Landing site, north of the Gale crater, and made its way to Kimberley, Pahrump Hills, and Ogunquit Beach. After that, it climbed Ver Rubin Ridge and headed to Mount Sharp. In the first scenario, we wanted to simulate the north side of the Gale crater and the discovery route of the Curiosity rover at the same time. We covered this large area with a flat polygon called "Rx\_Grid," as shown in Fig. \subref*{fig:Gale_curiosity}. There are $ 972 $ receiver points in this area. We placed the transmitter in the middle of the Salagnos crater and Peace Vallis.

\begin{figure*}[!t]
		\centering
		\subfloat[]{\includegraphics[scale = 0.6]{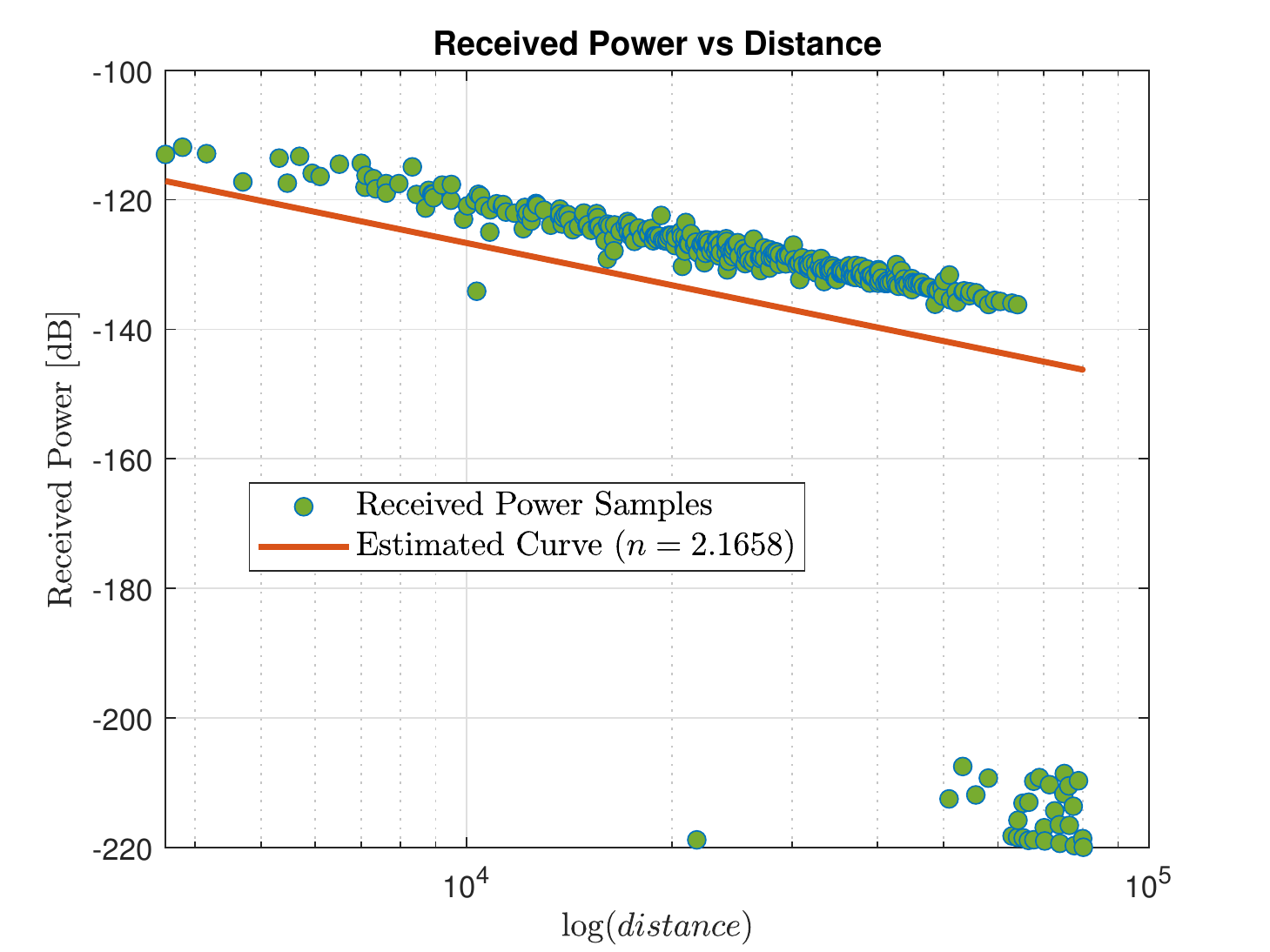}%
			\label{fig:Rcv_pwr_Rx_Grid}}
		\hfil
		\subfloat[]{\includegraphics[scale = 0.6]{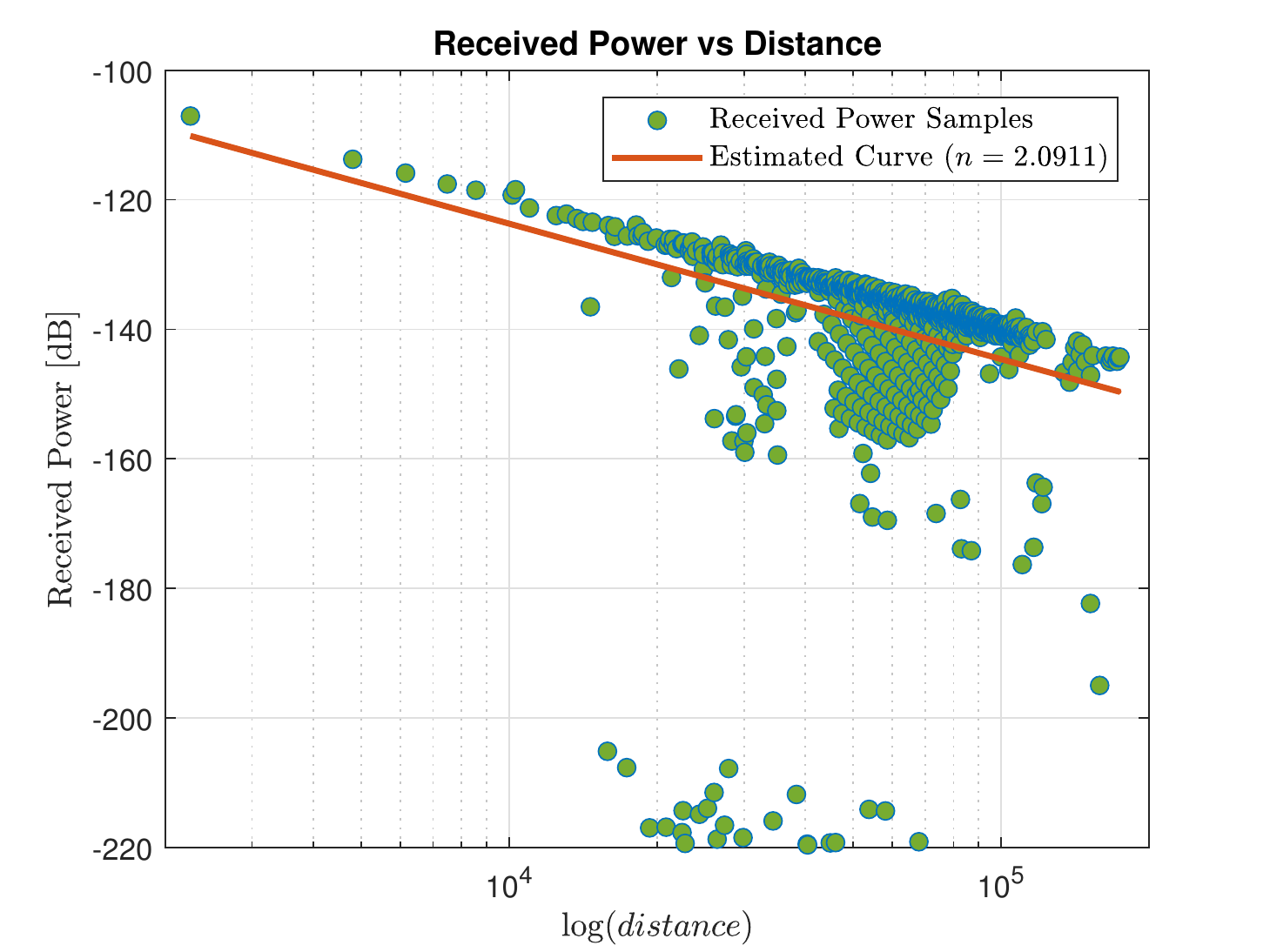}%
			\label{fig:Rcv_pwr_Gale_Rx_Grid}}
		\hfil
		\subfloat[]{\includegraphics[scale = 0.6]{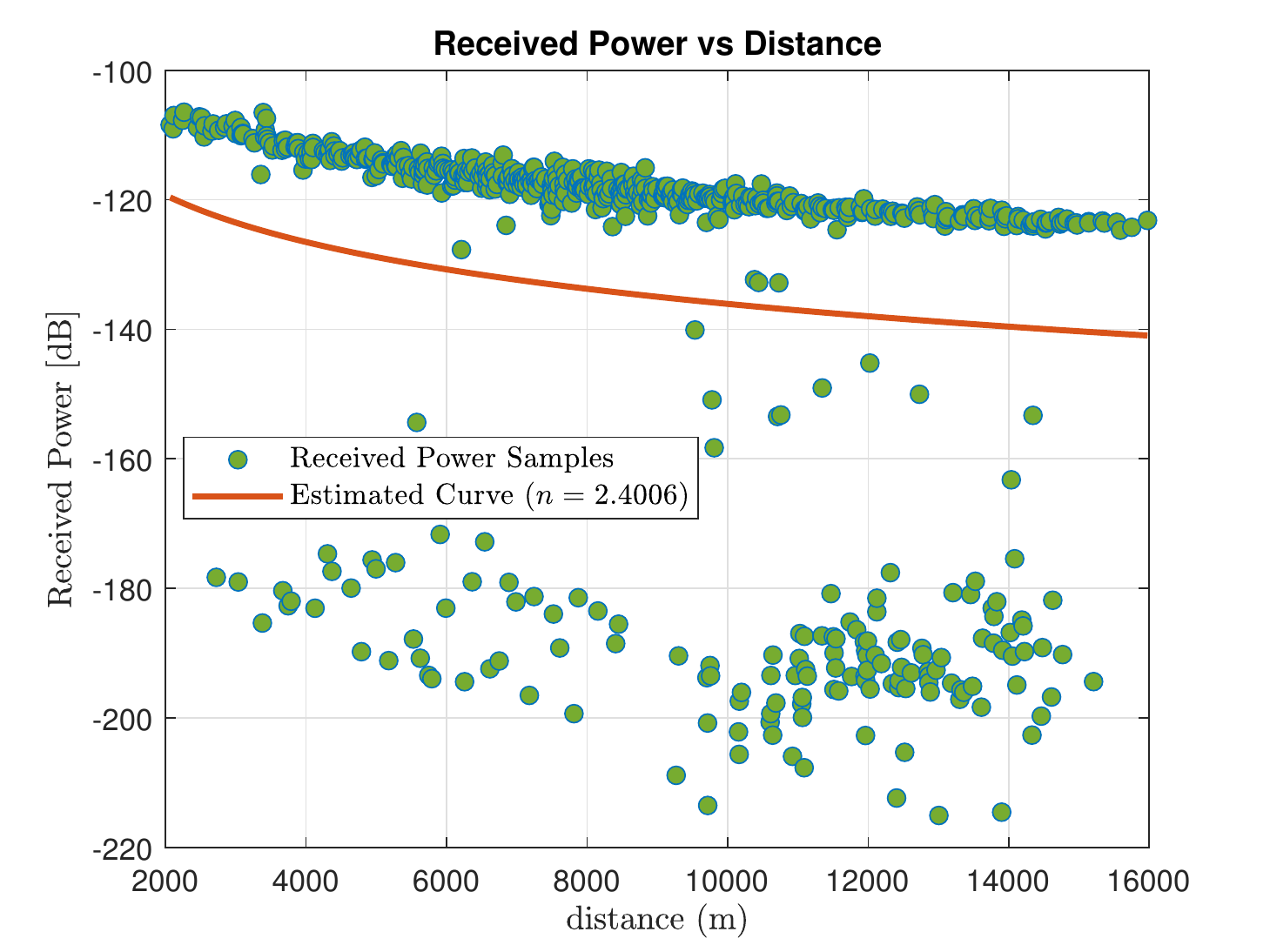}%
			\label{fig:Rcv_pwr_Beagle_Rx_Grid}}
		\hfil
		\subfloat[]{\includegraphics[scale = 0.6]{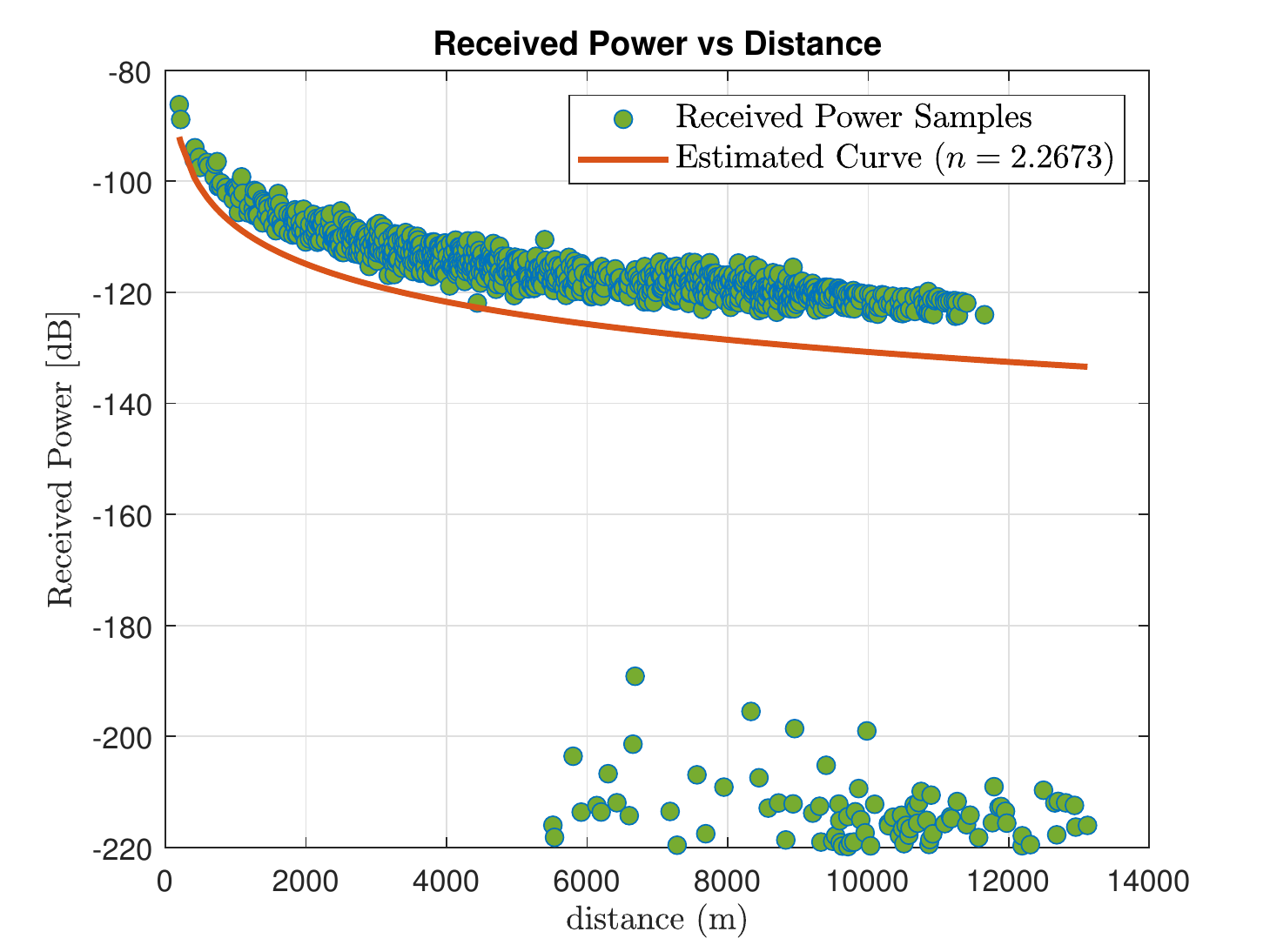}%
			\label{fig:Rcv_pwr_Jezero_Rx_Grid}}
		\caption{Received power samples versus distance of different communication scenarios. (a) Gale crater - Rx\_Grid. (b) Gale crater - Rx\_Grid1 \& Rx\_Grid2. (c) Beagle crater. (d) Jezero crater.}
		\label{fig:Rcv_pwr_all}
	\end{figure*}

	Received power samples and estimated path loss exponent, which is $ 2.1658 $, are shown in Fig. \subref*{fig:Rcv_pwr_Rx_Grid}. This value is slightly higher than the free-space path loss exponent because most of the points are actually in the LOS. The shadowing parameter is \SI{21.7214}{\decibel}, which is considerably high because the farthest points are in the blockage. Additionally, received power values are noticeably low in this simulation because the area is extremely huge such that the farthest point to the transmitter is roughly \SI{100}{\kilo\meter} far away. So, this area can be considered an extremely large macro cell.

	In the second scenario, we simulated an even bigger area covering almost all the bases of the Gale crater. We used two flat polygons; "Rx\_Grid1" to cover the south and "Rx\_Grid2" to cover the north, as shown in Fig. \subref*{fig:Gale_large_grid}. We placed the transmitter on the hills near Farah Vallis to increase the coverage, even though it may not be appropriate to place a base station (\acrshort{bs}) at the edge of a crater because of the dangerous impacts that dust storms could have on the equipment. Actually, those hills are the highest point around the Gale crater, which is the reason why the path loss exponent is again close to the free-space path loss exponent, as shown in Fig. \subref*{fig:Rcv_pwr_Gale_Rx_Grid}. Shadowing parameter is \SI{14.3744}{\decibel} in this case. When calculating these parameters, we combined the results of two grids and reordered them in terms of distance to analyze the area as a whole. 
	In general, the path loss exponent and shadowing parameter increase as the number of receivers with no direct link to the transmitter increases. 
	
	PDPs of three different receiver grids on the Gale crater are given in Fig. \subref*{fig:PDP_Gale_All}. Delay in Rx\_Grid is low as expected because the transmitter is close to the receivers compared to other grids. Rx\_Grid1, the south side of the Gale crater, is worse than Rx\_Grid because of its higher distance. On the other side, Rx\_Grid1 is better than Rx\_Grid2 because of its LOS dominance. Rx\_Grid2 is the worst because Aeolis Mons (Mount Sharp) interrupts LOS paths and creates blind zones at the edge. RMS delay spreads of Rx\_Grid, Rx\_Grid1, Rx\_Grid2 are calculated as \SI{7.7848}{\micro\second}, \SI{19.354}{\micro\second}, and \SI{56.158}{\micro\second}, respectively.
		
	Additionally, we have made computer simulations on other craters, such as the Beagle crater located east of the landing site for the ESA’s Beagle 2 lander. Beagle crater is roughly 42 \si{\km} in diameter \cite{Hawai_Beagle_3D_model}. The Beagle crater scenario can be seen in Fig. \subref*{fig:Beagle}. There are plenty of conceptual works on how humans should colonize Mars and how they should build their settlements in the most optimal and beneficial way. Those works meet on a common point: to use Mars craters as base points. One of those works is given in \cite{star_city}, which has won the Mars Colony Design contest held by Mars Society. The Star City comprises villages placed inside the side wall of a crater and public institutions such as schools, hospitals, universities, and arts in the center of the crater. Communication infrastructures built to provide a connection between villages and public institutions should also be crucial in the design. Taking inspiration from this concept, we placed the transmitter on the side wall of the crater where the villages would conceptually be placed. The receiver grid is placed so that it covers the center of the crater. Although the Beagle crater is considerably large and irregular for such a settlement, this simulation might give us an insight into how such a strategy would perform. The optimal path loss exponent is $ 2.4006 $, and shadowing parameter is \SI{28.7769}{\decibel} for this scenario, as shown in Fig. \subref*{fig:Rcv_pwr_Beagle_Rx_Grid}. RMS delay spread, on the other hand, is calculated as \SI{4.3938}{\micro\second}. 

	We also consider the Jezero crater, the landing site of the Perseverance rover. According to scientists, the region was formerly underwater and was the site of an old river delta. A flat plain named Isidis Planitia found slightly north of the Martian equator, is where the Jezero crater is located. The Jezero crater is \SI{45}{\kilo\meter} (\SI{28}{\mile}) wide. The distance between Curiosity's landing site in Gale crater to the landing site of the Perseverance is approximately \SI{3,700}{\kilo\meter} (\SI{2,300}{\mile}) \cite{nasa_Jezero}.
	
	After landing and completing the flight test of Ingenuity, Perseverance started to discover the south of the landing site, namely "First Abrasion Site," "First Drill Site," and "1st and 2nd Rock Coring Site". Then, it headed north and turned west from Port Angeles. Now, it is near Belva and may set a route to Neretva Vallis, thought to be the place where water spilled into a crater in the future.  
	
	We used the 3D model of the Jezero crater from UT Austin Planetary Surface Processes Group Dataverse, as shown in Fig. \subref*{fig:Jezero} \cite{Jezero_3D_model_Goudge}. Considering the discovery route of the Perseverance, we used two different grids with different heights, namely "RxPolygon1" and "RxPolygon2". RxPolygon1 covers the possible future-discovery route, and RxPolygon2 covers the landing site, flight test area, and other experiment locations of the rover. The transmitter is placed near the landing site. The optimal path loss exponent is obtained as $ 2.2673 $, and shadowing parameter is calculated as \SI{25.5992}{\decibel}, as shown in Fig. \subref*{fig:Rcv_pwr_Jezero_Rx_Grid}. RMS delay spread is \SI{0.1991}{\micro\second}.
	All main results revealed in our computer simulations are given in Table \ref{tab:3D_analysis_results_table}. 

	\section{Emerging Communication Technologies for Mars} \label{sec:emerging_tech}
	This section provides an overview of emerging communication technologies for deep space and Mars communication. Emerging space-based technologies are specifically investigated in other surveys such as \cite{Kua_2021} and \cite{Nessel_2020}. Apart from these studies, we give our own insights and perspectives in this section. There is a wide range of study areas on potential technologies and applications that can improve Mars communication. Development of novel wireless local area networks like wireless fidelity (\acrshort{wi-fi}) as well as metropolitan area network (\acrshort{man}) standards for Mars by considering the communication needs of future Mars settlements is crucial to achieving reliable communication. Also, the development of novel cellular communication standards, design of novel Mars waveforms, multiple access schemes, and modulation formats for the need of Martians should be analyzed to increase the throughput and reliability. Furthermore, investigation of intra-Mars internet, interplanetary internet, and delay tolerant networking (\acrshort{dtn}) technologies will help to increase the connectivity between Mars and Earth.
	
	In addition to these, integration of emerging technologies such as RISs, cognitive radio, SDR, environment-aware communication, non-orthogonal multiple access (\acrshort{noma}), and artificial intelligence/deep learning for future Mars communication networks, exploration of millimeter-wave and \si{\tera\hertz} bands, and investigation of hybrid optical and RF link designs for Earth-Mars communications will shape the course of extraterrestrial evolution of humans. 
	
	In Table \ref{tab:major_dev}, we provide major developments and contributions on the related topics discussed in this section. In what follows, we focus on a number of promising technologies.
	
	\subsection{Optical Communication}
	In its ongoing evolution, deep-space communications are either reaching or at least getting close to a turning point. Even though RF and microwave communications still have some room for development, most of the capability that was anticipated decades ago has already been reached. Most people will concur that going from the RF zone of the electromagnetic spectrum to the optical zone will result in the next major advancement \cite{Cesarone2011DeepSpaceOC}. There are several reasons why this significant change will occur.	
	Compared to radio frequency (\acrshort{rf}) systems, optical communication systems can offer substantially greater data rates, smaller, lighter, and less power-consuming systems, less vulnerability to interference, and more secure links. These benefits, along with the larger bandwidth of laser communications, can accelerate both autonomous and crewed exploration of Mars. Despite advances in optical communication in terms of performance, system availability, and cost, clouds will always be a problem for optical links to Earth. The link will occasionally be lost when a cloud passes between the spacecraft and the ground station.
	On the other hand, optical communication has a higher potential on Mars because of Mars's thinner atmosphere, except for the occurrence of dust storms. Laser beams can be attenuated by Martian aerosols. However, the effect is minor when compared to that of Earth aerosols. The quality of the Mars-Earth optical link is significantly impacted by orbital fluctuations \cite{Khatri_2004_link_anayl_optical}.
	\begin{table}[!t]
		\caption{RMS delay spread, path loss exponent, and shadowing parameters of various communication scenarios obtained by computer simulations.}
		\label{tab:3D_analysis_results_table}
		\centering
		\resizebox{\columnwidth}{!}{%
			\renewcommand{\arraystretch}{1.5}
			\begin{tabular}{ll|c|c|c|}
				\hline
				\multicolumn{2}{|c|}{\textbf{Environment}} & \textbf{\begin{tabular}[c]{@{}c@{}}RMS Delay \vspace{-1mm}\\ Spread\end{tabular}} & \textbf{\begin{tabular}[c]{@{}c@{}}Path Loss \vspace{-1mm}\\ Exponent ($n$)\end{tabular}} & \textbf{\begin{tabular}[c]{@{}c@{}}Shadowing \vspace{-1mm}\\ Parameter ($\sigma$)\end{tabular}} \\ \hline
				\multicolumn{2}{|l|}{\textbf{Hogwallow Flats}} & $ 3.4859 \times 10^{-9} $ \si{\second}& 1.9272 & 5.3141 \si{\decibel} \\ \hline
				\multicolumn{2}{|l|}{\textbf{Rocky Top}} & $ 4.3235 \times 10^{-9} $ \si{\second}& 2.8389 & 16.2670 \si{\decibel} \\ \hline
				\multicolumn{1}{|l|}{\multirow{3}{*}{\textbf{Gale}}} & Rx\_Grid & $ 7.7848 \times 10^{-6} \si{\second}$ & 2.1658 & 21.7214 \si{\decibel} \\ \cline{2-5} 
				\multicolumn{1}{|l|}{} & Rx\_Grid1 & $ 19.354 \times 10^{-6} $ \si{\second}& \multirow{2}{*}{2.0911} & \multirow{2}{*}{14.3744 \si{\decibel}} \\ \cline{2-3}
				\multicolumn{1}{|l|}{} & Rx\_Grid2 & $ 56.158 \times 10^{-6} $ \si{\second}&  &  \\ \hline
				\multicolumn{2}{|l|}{\textbf{Beagle}} & $ 4.3938 \times 10^{-6} $ \si{\second}& 2.4006 & 28.7769 \si{\decibel} \\ \hline
				\multicolumn{2}{|l|}{\textbf{Jezero}} & $ 0.1991 \times 10^{-6} $ \si{\second}& 2.2673 & 25.5992 \si{\decibel} \\ \hline
			\end{tabular}%
		}
	\end{table}
	
	NASA's Laser Communications Relay Demonstration (\acrshort{lcrd}) aimed to show the advantages of space-to-ground optical communications. LCRD is claimed to transmit and receive data from its orbit to Earth at a rate of $ 1.2 $ $ \mathrm{Gbps} $. Specifically, it is aimed that it would be possible to download a movie within a minute at this data rate. After launch and verification that the spacecraft is functioning properly in space, LCRD is reported to start communicating with ground stations in California and Hawaii via infrared lasers from its position in geosynchronous orbit, which is approximately \SI{35,786}{\kilo\meter} (\SI{22,000}{\mile}) above Earth \cite{margetta_2021_LCRD}. LCRD is planned to devote the next two years to collecting data, analyzing how weather and other alterations in Earth's atmosphere can affect laser communications, and monitoring link performance to improve its functional abilities. Some tests are designed to simulate relay strategies between the Moon and Earth in order to better understand how laser communications could be employed in NASA's Artemis missions in the future \cite{murphy_LCRD}. Detailed parameters of an optical link are given in \cite{2018_deep_Space_Relay_Terminals}. It is reported that, as the first users begin to appear during the post-experiment stage, LCRD will also present how the integration of both RF and optical trunklines enables the maximum capabilities of optical communications user links. \cite{David_2016_considerations_for}. The first user of the LCRD is announced to be NASA's Integrated LCRD Low-Earth Orbit User Modem and Amplifier Terminal (\acrshort{illuma-t}), which is expected to be released in 2023 and complete an end-to-end (\acrshort{e2e}) optical communications relay system. High-resolution data is envisioned to be transmitted from the International Space Station (\acrshort{iss}) to the LCRD using $ 1.2 $ $ \mathrm{Gbps} $ optical links. Then, Hawaii and California optical ground stations are scheduled to receive this data via optical beams \cite{ESC_2020_Annual_report}. Mars might be mapped at a rate more than a hundred times faster with the help of laser communications than it is currently achievable \cite{Hemmati_07}.
	\begin{table*}[!t]
		\caption{Major Developments and Contributions on emerging deep space and mars communication technologies}
		\label{tab:major_dev}
		\centering
		\resizebox{\textwidth}{!}{%
			\renewcommand{\arraystretch}{1.5}
			\begin{tabularx}{\textwidth}{|l|X|}
				\hline
				\rowcolor[HTML]{FFAEAB} 
				\multicolumn{1}{|c|}{\cellcolor[HTML]{FFAEAB}\textbf{Topic}} &
				\multicolumn{1}{c|}{\cellcolor[HTML]{FFAEAB}\textbf{Related Studies}} \\ \hline
				\rowcolor[HTML]{FFE0E0} 
				Optical Communication &
				\text{\cite{2018_deep_Space_Relay_Terminals}, \cite{Cesarone2011DeepSpaceOC},\cite{Khatri_2004_link_anayl_optical}, \cite{margetta_2021_LCRD}, \cite{murphy_LCRD}, \cite{David_2016_considerations_for}, \cite{Hemmati_07}, \cite{Kaushal_2017}, \cite{Li_2022}, \cite{hemmati2006deep}, \cite{Farzana_2004}, \cite{Hemmati_2011}, \cite{Karafolas_2009}, \cite{Paul_2009}, \cite{Chen_2022}, \cite{Daniel_2012_RF_Opt}, \cite{app12020619} }, \text{\cite{Chan_2006}, \cite{Khalighi_2014}, \cite{DREISCHER20091772}, \cite{Brandl_2009}} \\ \hline
				\rowcolor[HTML]{FFAEAB} 
				\si{\tera\hertz} Communication &
				\text{\cite{Zhuo_2021_Comparison_of_the_influene}, \cite{Ozgur_2021_Terahertz_in_space}, \cite{Tekbiyik_2020_A_Holistic_Investigation_of_THz}, \cite{Han_2022}, \cite{Sumen_2022}, \cite{Dong_2011}, \cite{Nie_2021}, \cite{Alqaraghuli_2021}, \cite{Choudhury_2016}, \cite{Nagatsuma_2018}, \cite{Mehdi_2018}, \cite{Diao_2021}, \cite{Wedage_2022}, \cite{Sarieddeen_2020}} \\ \hline
				\rowcolor[HTML]{FFE0E0} 
				RIS &
				\text{\cite{Khan_2022_When_RIS_Meets_GEO}, \cite{Tian_2022_Enabling_NLoS_LEO_Satellite}, \cite{Kisseleff_2021_RIS_in_chall_env}, \cite{Tekbiyik_2020_RIS_empw_THz}, \cite{Zhang_RIS_2021}, \cite{Zheng_RIS_2022}, \cite{Xu_RIS_2021}} \\ \hline
				\rowcolor[HTML]{FFAEAB} 
				Multiple Spacecraft \& Uplink per Antenna &
				\text{\cite{David_2018_MUPA}, \cite{Tkacenko_2019}, \cite{Towfic_2017}} \\ \hline
				\rowcolor[HTML]{FFE0E0} 
				Mars Planetary Network &
				\text{\cite{2018_deep_Space_Relay_Terminals}, \cite{Wallace_Mars_2018}, \cite{howell_2015}, \cite{2018_Space_and_earth_terminal_sizing}, \cite{Bappy_2019}, \cite{Alhilal_2019}, \cite{Wan_2019}, \cite{Wang_IPI_2009}, \cite{Zhao_Netw_2018}, \cite{Wan__Orb_2020}} \\ \hline
				\rowcolor[HTML]{FFAEAB} 
				Cognitive Radio &
				\text{\cite{2020_cognitive_radios}, \cite{hamkins2006autonomous}, \cite{Pawelczak_2011}, \cite{Hackett_2017}, \cite{David_Les_2014}, \cite{Ferreira_2017}, \cite{Joseph_2016}, \cite{Xue_2020}, \cite{Jia_2016}, \cite{Ferreira_2018}, \cite{Simons_2021}, \cite{Raghunandan_2010}, \cite{Greta_2022}, \cite{Duncan2011SoftwareDG}, \cite{Iancu_2009}}, \text{\cite{AKEELA2018106}, \cite{Scardelletti2007SoftwareDR}} \\ \hline
				\rowcolor[HTML]{FFE0E0} 
				Acquisition, Tracking, and Navigation &
				\text{\cite{2018_Two-way_ranging}, \cite{Deutsch_2021}, \cite{Kinnian_2003}, \cite{Reynolds_2002}, \cite{Vilnrotter_2019}, \cite{Shin_2014}, \cite{Divsalar_2019}, \cite{Divsalar_Adaptive_2020}} \\ \hline
				\rowcolor[HTML]{FFAEAB} 
				Delay Tolerant Networking &
				\text{\cite{Mukherjee_2013_Interplanetary_Internet}, \cite{monaghan_2020_DTN}, \cite{Cola_2011}, \cite{Madoery_2022}, \cite{Gu_2020}, \cite{Zhao_Perf_2016}, \cite{Dhara_2019}, \cite{Caini_2011}, \cite{Alessi_2018}, \cite{Alessi_2022}, \cite{Bertolazzi2019MarsTE}, \cite{Wu_2015}, \cite{Caini_2020}, \cite{Fraire_2019}, \cite{Wang_2017}, \cite{Burleigh_2003} }, \text{\cite{Hu_2014}, \cite{Sun_2013}, \cite{Yang_Queueing_2018}, \cite{Araniti_2015}, \cite{SAMARAS2010863}, \cite{Marco_2019}, \cite{Shi_2017}, \cite{Wang_LTP_2013}} \\ \hline
				\rowcolor[HTML]{FFE0E0} 
				Multiple Access Schemes &
				\text{\cite{Babuscia_2017_CDMA}, \cite{Zhiguo_2017_application_of_NOMA}, \cite{Liu_2017_NOMA_for_5G}, \cite{Quirk_2001}, \cite{Babuscia_2016}, \cite{Divsalar_Opt_2018}, \cite{Xiaojuan_2018}, \cite{Jiao_2020}, \cite{Ruixing_2021}, \cite{Zhixiang_2020}, \cite{Xinwei_2022}, \cite{Yan_Perf_2018}, \cite{Le_Tran_2021}, \cite{Divsalar_Wave_2020}} \\ \hline
				\rowcolor[HTML]{FFAEAB} 
				Environment-Aware Communications &
				\text{\cite{Zeng_toward_env_aware_2021}, \cite{Haoyun_2022}, \cite{Wu_Env_2022}, \cite{Ding_2021} } \\ \hline
				\rowcolor[HTML]{FFE0E0} 
				Artificial Intelligence &
				\text{\cite{Basar_deep_learning_2022}, \cite{Ferreira_2019}, \cite{app12105106}, \cite{Oche_2021}, \cite{Hackett_Imp_2018}, \cite{Gomez_2022}, \cite{Miguel_2021}, \cite{Zhou_Machine_2021}, \cite{Zhang_Multi_2022}, \cite{Kothari_2020}, \cite{TIPALDI20221}} \\ \hline
			\end{tabularx}%
		}
	\end{table*}
	
	Deep space optical communications has a growing interest from the literature. Couple of comprehensive surveys on use of optical links for space communication applications and interplanetary networks architectures along with their challenges are presented in \cite{Kaushal_2017}, and \cite{Li_2022}. Also, NASA published a very detailed open book on deep space optical communications \cite{hemmati2006deep}. 
	In \cite{Farzana_2004}, a high capacity Mars to Earth link is studied on a system-level.
	In \cite{Hemmati_2011}, the recent deep-space optical data transmission and networking technologies, along with ongoing experiments, potential future developments, and applications are investigated.
	One of the earliest demonstrations of optical communication between inter-satellite links is done by ESA in 2001, discussed in \cite{Karafolas_2009}. A link budget analysis of Moon-Earth and Mars-Earth optical links are investigated in \cite{Paul_2009}. In \cite{Chen_2022}, optical communication is examined regarding the near-Earth, deep space, and lunar links. In this work, simulation parameters such as average received photons per pulse, intensity, and data throughput are provided. 
	Further, optical links can be combined with RF links to further improve link quality. 
	Early experiments of hybrid RF/optical interplanetary links are carried out by NASA in \cite{Daniel_2012_RF_Opt}. In \cite{app12020619}, performance of a hybrid dual-hop RF/FSO deep space communication system is studied under the effect of solar scintillation. Results in the article are claimed to shown that the hybrid RF/FSO system enhances the BER performance in a deep space environment by a factor of 10 to 30. In \cite{Chan_2006}, an analysis made on decreasing the cost of optical links via using photon-counting receivers, coherent systems, and multiple access techniques. A survey on free space optical communications concerning their various aspects such as channel models, transciever architectures, modulation, channel coding, diversity techniques, etc. is provided in \cite{Khalighi_2014}. An integrated RF-optical telemetry, tracking and control transponder system is discussed and hardware test results are provided in \cite{DREISCHER20091772}. A link budget analysis on possible telecommunications requirements of an interplanetary network is presented in \cite{Brandl_2009}. Routing deisgn analysis for optical links in future deep space optical networks is made in \cite{Wang_2022}.

	\subsection{\si{\tera\hertz} Communication}
	\si{\tera\hertz} waves offer a variety of applications in space communication due to their wideband, high-penetration capability in the Martian atmosphere, and powerful directional antennas. \si{\tera\hertz} links in the vicinity of Mars are primarily influenced by the atmosphere's absorption attenuation and the scattering of severe dust storms. In \cite{Zhuo_2021_Comparison_of_the_influene}, the Martian atmosphere's $ 0.1 $–$ 1 $ \si{\tera\hertz} wave absorption is examined. The \si{\tera\hertz} wave combines the benefits of optical and microwave communication. \si{\tera\hertz} communication has stronger confidentiality and anti-interference resilience than microwave communication. It has a higher data transmission throughput and efficiency, making it more advantageous for the real-time transfer of high-definition video and images. Because the \si{\tera\hertz} wavelength is less than the microwave wavelength, a smaller antenna can still perform the same task. Compared to optical communication, the \si{\tera\hertz} wave has a better ability to penetrate dust and smoke, allowing it to function normally in harsh environments with heavy winds. \si{\tera\hertz} channels have various characteristics that set them apart from lower frequency bands, such as noise, propagation, and molecular absorption. Thus, it is critical to investigate the propagation parameters of the Mars channel on \si{\tera\hertz} wave transmission for the upcoming Martian exploration era. Further information about attenuation of water vapor, carbon monoxide, oxygen, carbon dioxide, and oxygen gas molecules on \si{\tera\hertz} waves is given in \cite{Zhuo_2021_Comparison_of_the_influene}.
	Multiple challenges exist to overcome before activating \si{\tera\hertz} space links. Because electromagnetic waves expand as they travel, the spreading loss dramatically rises with frequency. It restricts communication distance on Earth to a few meters because of the immature \si{\tera\hertz} source technology. Furthermore, the use of the high \si{\tera\hertz} band is constrained by severe molecule absorption, which causes high atmospheric attenuation in Earth-to-space communications. Since water molecules, the primary reason for atmospheric attenuation, are limited on Mars, atmospheric attenuation is anticipated to be lower on Mars than on Earth. These open the possibility of using the high \si{\tera\hertz} band. Since \si{\tera\hertz} lines are more resistant to weather conditions and pointing losses, they can be utilized as an alternative in the event of an outage caused by beam pointing errors or severe atmospheric attenuation. Furthermore, \si{\tera\hertz} and optical communications can be integrated. A Mars rover, for example, can communicate with a relay orbiter via a \si{\tera\hertz} link, and the relay orbiter relays messages via optical links. In the future, CubeSats may be installed with \si{\tera\hertz} transceivers so they can transmit data to Earth using optical links and communicate with the surface operators via low-latency \si{\tera\hertz} communications. The Planetary Spectrum Generator (\acrshort{psg}), a precise radiative transfer tool, is used in \cite{Ozgur_2021_Terahertz_in_space} to simulate the penetration scale of Mars's atmosphere in clean and dusty conditions.
	
	As another potential application of \si{\tera\hertz} systems, aerial access points (\acrshort{ap}s) such as high-altitude platform stations (\acrshort{haps}s), networked flying platforms (\acrshort{nfp}s), and uncrewed aerial vehicles (\acrshort{uav}s) have gained a lot of interest recently. There has been exceptional upward directional growth in the wireless network. Moreover, it appears that transmission between satellites and Earth will be greater in the following decade. Even though the Starlink and Keppler projects have begun to develop a comprehensive satellite network, research in this field is still in its early stages. For inter-satellite and satellite-to-X links, free-space optics (\acrshort{fso}) and mmWave systems have been suggested. CubeSats, on the other side, are built for Low Earth orbit (\acrshort{leo}) operations and must travel at high speeds to maintain their orbits. This velocity needs fast beam steering; thus, \si{\tera\hertz} allows fast-tracking of a moving receiver using electronic steering.
	UAV communication presents another difficulty. Because of flight aerodynamics, UAVs vibrate a little bit even when they hang in the air. The extremely narrow antenna beam of \si{\tera\hertz} communication is seriously affected by these vibrations. \si{\tera\hertz} antenna misalignment is known to cause significant loss. After the antenna alignment breaks down, the beam search could take a very long period. Channel modeling and real-time measurements of \si{\tera\hertz} waves are given in \cite{Tekbiyik_2020_A_Holistic_Investigation_of_THz}. Although the \si{\tera\hertz} communication offers a significant improvement in data rate and throughput, these challenges should be extensively studied before deploying any \si{\tera\hertz} communication system into deep space. There is an increasing interest on use of \si{\tera\hertz} band signals for satellite and space applications. In \cite{Han_2022}, a comprehensive overview of \si{\tera\hertz} wireless links is provided. Integration of \si{\tera\hertz} waves with satellite technologies is studied in a number of studies such as \cite{Sumen_2022}, \cite{Dong_2011}, \cite{Nie_2021}, and \cite{Alqaraghuli_2021}. Specifically, \si{\tera\hertz} communication is studied for space applications in \cite{Choudhury_2016}, \cite{Nagatsuma_2018}, \cite{Mehdi_2018}. The absorption of $ 0.1 $-$ 1 $ \si{\tera\hertz} waves in the Martian and Earth's atmosphere are investigated and compared in \cite{Diao_2021}. Also, the path loss performance of THz links in different atmospheric gas concentrations and ecological conditions on Mars and Earth is compared in \cite{Wedage_2022}. A comprehensive and visionary perspective on the future of \si{\tera\hertz} communications is presented in \cite{Sarieddeen_2020}.
	
	\subsection{Reconfigurable Intelligent Surfaces}
\begin{figure*}[t]
		\centering
		\includegraphics[scale = 0.6]{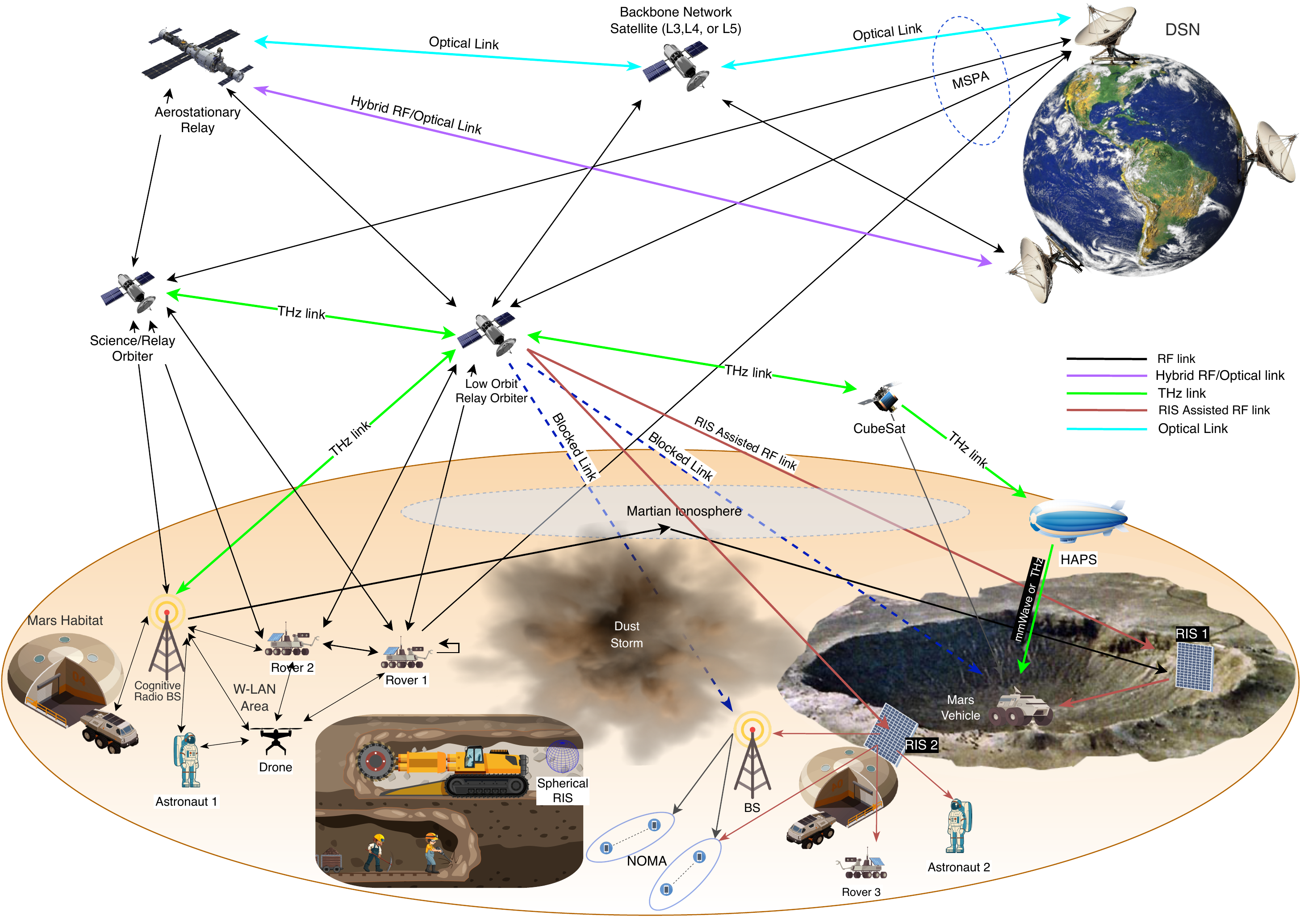}
		\caption[Caption for Tech]{Emerging technologies for Mars communication\footnotemark.}
		\label{fig:emerging_tech}
	\end{figure*}

	RIS-empowered wireless communication has received growing interest over the years and is considered one of the emerging technologies for 6th generation (\acrshort{6g}) networks. RIS may also have a potential for future deep space habitat (\acrshort{dsh}) missions. RIS-based communication is a low-cost and low-energy option for achieving high spectral efficiency. The main idea of RIS is to dynamically adjust signal propagation utilizing passive reflecting devices. In other words, the aim of the RIS is to reflect signals into relevant points with configurable phase shifts. These surfaces can be placed on building walls, ceilings, and other surfaces because they are thin and can be dynamically adjusted via a software controller. The key benefits of RIS include increased wireless coverage, improved data security, cost-effectiveness, low hardware complexity, and inexpensive installation costs. RISs cannot create communication links on their own. So, they are most beneficial when they are used to assist existing infrastructures. One of the use cases of RIS-assisted communications is geostationary orbit (\acrshort{geo}) satellite communications (\acrshort{satcom}). There is an increasing demand for high-capacity GEO SatCom to supply broadband services in inaccessible portions of terrestrial networks. In \cite{Khan_2022_When_RIS_Meets_GEO}, a GEO SatCom that uses multiple carriers to send data to the ground terminal is taken into consideration. To increase the signal gain, a direct link and a RIS-assisted link, in which a RIS is put on the top of the building and supports the data transmission from the satellite to the ground terminal, are studied. Similar to this work, a RIS can be used in Mars communication when the direct link between transmitter and receiver is blocked due to extreme conditions such as a dust storm or an irregular terrain. Also, it can be used to increase the throughput of an orbiter-rover link where the rover operates in a crater, as shown in Fig. \ref{fig:emerging_tech}. Another work on RIS-assisted communication consisting of two LEO SatCom to achieve full coverage in an urban area is given in \cite{Tian_2022_Enabling_NLoS_LEO_Satellite}. Additionally, \cite{Kisseleff_2021_RIS_in_chall_env} studied various use cases of RIS-assisted schemes in challenging environments, which include underwater elements; submarines and ships, Internet-of-underground things; smart agriculture applications, and mines and tunnels. The idea of placing RISs in the walls to enhance the directivity of signals in a tunnel can be integrated into the idea of Star City, tunneling on Martian craters to build a liveable habitat.
	
	In \cite{Tekbiyik_2020_RIS_empw_THz}, the concept of RIS and \si{\tera\hertz} waves are integrated into LEO satellite networks. This work provided a solution to both frequency constraint and low complexity system design by combining the RIS and \si{\tera\hertz} bands in LEO inter-satellite links (\acrshort{isl}s). It is demonstrated that by utilizing the \si{\tera\hertz} wave's ultra-wide bandwidth and the RIS's ability to act like a virtual multiple-input multiple-output (\acrshort{mimo}) system, achievable rates could be increased. The computer simulation results and derived mathematical expressions in the study have showen that a greater number of RISs can lower the necessary transmit power while still achieving the same error probability. \cite{Zhang_RIS_2021}, \cite{Zheng_RIS_2022}, and \cite{Xu_RIS_2021} stand out as example studies on integration of RISs with satellite netwroks.
	
	\subsection{Multiple Spacecraft \& Uplink per Antenna}
	The multiple spacecraft per antenna (\acrshort{mspa}) method is being utilized by the DSN, which uses a single antenna to track multiple downlinks from spacecraft within its beam, including when several spacecraft are orbiting Mars at \SI{8.4}{\giga\hertz} (X-band). \cite{David_2018_MUPA} suggested that this technique can be extended to the uplink when a single station sends a signal to several spacecraft to make better use of terrestrial resources. It is claimed that it could be applied to multiple smallsat constellations under consideration for upcoming missions, as well as spacecraft destined for Venus and Mars, most of which are within the half-power beamwidth of a single $ 34 $-\si{\meter} diameter antenna. In one concept, the instruction datasets of each spacecraft are designed to be time-multiplexed onto a single uplink frequency. Each spacecraft is envisioned to focus on the uplink signal and only take instructions specifically addressed to it using unique identification codes. Also, the downlink signal that each spacecraft would send to Earth is said to be coherent with the uplink signal but could have an exclusively dedicated frequency channel and unique identification data. Detailed analysis of this new scheme called multiple uplinks per antenna (\acrshort{mupa}) is given in \cite{David_2018_MUPA}, which can be employed for efficient use of resources in interplanetary links. Additionally, NASA demonstrated a new beam-sharing technique called Opportunistic Multiple Spacecraft Per Antenna (OMSPA) for small satellites and CubeSats \cite{Tkacenko_2019}, \cite{Towfic_2017}.
	\footnotetext{This figure has been designed using images from Freepik.com}
	
	\subsection{Mars Planetary Network}
	Space exploration efforts made up until now and will be made in the future are envisioned to open the way for an interplanetary space network potentially spanning the whole solar system. Mars is the first step in configuring such a vast and complex topology. First, orbit definitions should be studied. A geosynchronous orbit (\acrshort{gso}) is a high Earth orbit that enables satellites to synchronize with the rotation of the planet. This position, which is \SI{35,786}{\kilo\meter} above the equator of the planet Earth, is advantageous for keeping an eye on the weather, surveillance, and communications \cite{howell_2015}. An areostationary orbit is a circular areo­synchronous orbit (\acrshort{aso}) that is \SI{17,032}{\kilo\meter} above the surface of Mars. Each point on the areostationary orbit spins around Mars with the same velocity and direction as the Martian surface. As a result, the areostationary orbit can be thought of as the Martian equivalent of the GSO.
	
	In \cite{Wallace_Mars_2018}, a hypothetical Mars planetary network is described as an E2E system that includes an Earth network with the DSN and the deep space facilities of various agencies, the Mars surface network with a W-LAN at the exploration zone, and the Mars relay network composed of one or more relay orbiters on connecting science orbiters. The hypothetical Mars relay network is reported to consist of three orbiters: two areostationary orbiters, one of which is always accessible from the Mars exploration area, and the DSH in a $ 48 $-$ \mathrm{hours} $ elliptical orbit. The three spacecraft are supposed to be visible to each other most of the time and could communicate via X and Ka-band. The Mars landing location is assumed to have a Wi-Fi hotspot that gathers and transfers data between the Mars ground equipment and the orbiters. In order to allow inter-vehicle and astronaut communications in the exploration zone, a Mars surface network is said to function as a W-LAN that connects all devices.
	
	According to \cite{Wallace_Mars_2018}, the Mars planetary network would include the following features: 
	\begin{enumerate}
		\item The integration of Ka-band and optical communications for generating Mars-Earth high-rate links by utilizing the optical/RF hybrid antennas in DSN;
		\item The combined use of the MSPA technology for return link and the MUPA technique for forward link in order to decrease the amount of \SI{34}{\meter} antennas required for the period;
		\item Three arrayed \SI{34}{\meter} antennas integrated with MSPA/MUPA techniques;
		\item Launching two Mars relay orbiters that are both areostationary and areosynchronous, one of which might serve as a fictitious DSH;
		\item Launching a science orbiter with equal distances to Mars and Earth might also act as a relay during the Mars superior solar conjunction.
	\end{enumerate}
	
	The last item is explained in more detail. When the data transfer is critical within a mission, the Sun occasionally blocks the direct link between Earth and Mars, implying the requirement for a relay. Multiple strategies depending on laser or radio-frequency relays positioned in deep space between Earth and Mars are investigated in \cite{2018_deep_Space_Relay_Terminals}, involving periodic orbits in the Sun-Mars and Sun-Earth rotating cycles, as well as sun-centered and eccentric orbits. The Sun-Mars system's L4 and L5 Lagrange points, in other words, long-period orbits, allow feasible communications topology consistently for quite long periods of time. In such an orbit, a deep space relay terminal (\acrshort{dsrt}) is anticipated to be able to support Mars superior conjunctions providing optical data rates of $ 28 $ to $ 44 $ $ \mathrm{Mbps} $ for return links and $ 30 $-$ 36 $ $ \mathrm{Mbps} $ for forward links with the help of two \SI{75}{\centi\meter} Ka-band dish antennas, two \SI{50}{\centi\meter} optical telescopes. Use of Lagrange points to enhance relay satellite constellations for the exploration of the solar system is also studied in \cite{Bappy_2019}. Solutions that can meet NASA's specific working requirements for data rates between $ 10 $ $ \mathrm{Mbps} $ and $ 250 $ $ \mathrm{Mbps} $ in the return direction and between $ 3 $ $ \mathrm{Mbps} $ and $ 50 $ $\mathrm{Mbps}$ in the forward direction are required. These data rate ranges are valid for Mars in 2030–2040 when its distance from Earth is at its maximum. To meet these requirements, a single hypothetical DSRT considered in \cite{2018_Space_and_earth_terminal_sizing} is imagined to be connected with additional relays to establish a multi-hop network in deep space, allowing for a higher transmission rate. This multi-hop network is divided into three parts: a space terminal Mars, a DSRT, and Earth terminals. In the study, series of assumptions and suggestions are made. It is suggested that the space terminal near Mars is an areostationary relay in Martian orbit. The areostationary relay is assumed to include a \SI{50}{\centi\meter} optical telescope. The MRO is used as a model for other radio properties. The DSRT is envisioned to be compact in order to make the launch easier and keep costs low. Lastly, Earth terminals are assumed to compose of \SI{34}{\meter} DSN antennas, utilized for the Ka-band forward and return links to the DSRT. For the DSRT's return link capacity, an \SI{8}{\meter} hybrid RF-optical telescope is assumed. Further analysis on Areostationary relays and space terminals can be found in \cite{2018_Space_and_earth_terminal_sizing}. These assumptions are just the beginning of what interplanetary internet networks would look like in the future. The number of relay satellites equipped with more sophisticated hardware and software is expected to increase substantially as new relay technologies and standards are developed. 
	There are also a number of studies on possible network architecures, protocols, topology designs, and routing algorithms for interplanetary internet that can be employed in the future such as \cite{Alhilal_2019}, \cite{Wan_2019}, \cite{Wang_IPI_2009}, \cite{Zhao_Netw_2018}, and \cite{Wan__Orb_2020}.
	
	\subsection{Cognitive Radio}
	Cognitive radios have been developing for over two decades and have the potential to be a redefining technology for next generation wireless communications because of their reconfigurable nature. Key advancements in the history of cognitive radios and future insights are reviewed in \cite{Pawelczak_2011}. There are also a number of works visioning cognitive radios for space applications such as \cite{Hackett_2017}, \cite{David_Les_2014}, and \cite{Ferreira_2017}.
	The radio parameters on the orbiter, the BS, and the lander will require ongoing adjustments as the space environment and positions of planets relative to one another change during a real-time crewed exploration mission. Astronauts might wait for $ 8 $ to \SI{40}{\minute} to receive directives from the command center on how to change the parameters. A better option would be for the transceivers to utilize neural networks to adjust their parameters in real-time. Neural networks can sustain and maximize a radio's capacity to remain online. An example of such a device is cognitive radio. Its neural network detects environmental alterations, modifies its parameters accordingly, and, most importantly, learns from the process. It means that a cognitive radio can experiment with new setups in different conditions, making it more robust in undiscovered environments than regular radio. NASA have tested this kind of technology which is called the adaptive coding and modulation (\acrshort{acm}) in \cite{Joseph_2016}. As a result, cognitive radios are suitable for space communications, where human interaction is not possible, and sustaining connectivity is crucial. The fundamental feature of a neural network is its ability to optimize the relations between the inputs and the outputs in time. This procedure is referred to as training. The radio may first try increasing its transmit power if a signal cannot go through in a noisy area. The radio then checks to see if the received signal is better and, if so, increases the transmit power to see if this further enhances quality. However, if the signal does not become better, the radio might attempt an alternative strategy, like changing frequencies. In each scenario, the radio gains some understanding of how to get a signal through its present environment. In space applications, orbiters and surface operators can substantially benefit from this adaptability. There are already some studies focusing on the implementation of cognitive radios on satellites, such as \cite{Xue_2020}, \cite{Jia_2016}, and \cite{Ferreira_2018}, and on surface assets such as \cite{Simons_2021}, and \cite{Raghunandan_2010}. When a cognitive radio is trained, its signal modulation, data rate, transmission power, or any other settings are continuously changed to help it become more effective. Every cognitive radio needs to be trained initially before going online. This training acts as a model for future radio improvements.
	Cognitive radios use a wireless device known as SDR to control their basic parameters. In SDR, features that are implemented with hardware in a traditional radio are achieved with software, such as amplifying, detecting, and signal filtering. Some experiments are made on SDRs, as in \cite{2020_cognitive_radios}. Detailed information about autonomous SDR receivers and implementation of a cognitive radio modem on SDR are given in \cite{hamkins2006autonomous} and \cite{Greta_2022}, respectively. NASA has already manufactured a SDR based GPS receiver for ISS \cite{Duncan2011SoftwareDG}. Future design trends and challenges of SDRs are discussed in \cite{Iancu_2009}, \cite{AKEELA2018106}, and \cite{Scardelletti2007SoftwareDR}.
	
	\subsection{Acquisition, Tracking, and Navigation}
	Location determination is crucial to enable diverse human and autonomous activities on the surface of Mars and in orbit. In addition to communication between orbiters and ground stations on Earth, a reliable communication network is needed for data transfer among Martian surface assets and orbiters. Detailed information about NASA's space navigation techniques and their automation capabilities is provided in \cite{Deutsch_2021}. Two-way ranging, which is performed by a single station on uplink and downlink, is not new; it is analysed as in \cite{Kinnian_2003} and \cite{Reynolds_2002}, manufactured, and used by NASA on various missions for almost fifty years. Constant development is required in this field due to the increasing number of spacecraft expected to be deployed in the future. For example, a new version of two-way ranging which removes the downlink pseudonoise ranging signal in conventional two-way ranging is studied in \cite{Vilnrotter_2019}. Also, three-way ranging, performed by multiple stations, is under development and studied in \cite{Shin_2014}. In \cite{2018_Two-way_ranging}, a hypothetical Mars regional navigation satellite system (\acrshort{mrnss}) is put forth that would offer missions operating close to a Mars landing site continuous or almost continuous in-situ navigation and timing services. Earth's DSN must employ OD for the Mars orbiters in order for their locations to be known to a fair level of precision. This allows the user devices to determine their locations based on orbiter-vehicle distance calculations.
	The study suggests that infrastructures on Mars' surface and in orbit would need to be extensively built up before humans could explore the planet. The studies already begun for Moon, an acquisition and tracking system between a lander/rover on the surface of the lunar south pole and an Earth station is analyzed in \cite{Divsalar_2019}. Location awareness and telecommunication availability are required to enable diverse human and robotic operations on the Martian surface, and in orbit. Additionally, adaptive acquisition techniques should be investigated in depth for dynamic conditions as in \cite{Divsalar_Adaptive_2020}.
	
	\subsection{Delay Tolerant Networking}
	DTN is a set of standard protocols that enable E2E data transmission over nodes in the network by utilizing headers added to data packets. DTN supports data transfer in the presence of disconnection, delay, and data rate mismatches. A bundle protocol (\acrshort{bp}) that employs the well-known strategy of store and forward is a key component of DTN communications. Under this protocol, a node promises to keep BP data units in memory until the following node acknowledges their successful reception. A study of BP over a relay-based deep-space communications system that includes lossy data links, asymmetric channel rates, and extended link failures accompanied by very long propagation delays is presented in \cite{Zhao_Perf_2016}. The DTN protocol can function separately or in collaboration with the terrestrial internet protocol (\acrshort{ip}) suite. Through the use of automated store-and-forward mechanisms, DTN offers guaranteed data delivery. Every data packet that is received is directly sent if possible. Still, if forwarding is not possible immediately but is likely to be possible in the future, it is saved for later transmission. Therefore, while using DTN, just the next node is required to be accessible. In addition, the DTN protocol includes security, quality-of-service, routing, and network management features that are comparable to those offered by the terrestrial IP. However, the DTN concept mainly addresses latencies in networks that can be of three main types: propagation delays through the medium; queuing delays among relay nodes, sources, and destinations; and clocking delays related to sending an atomic unit of data onto the medium \cite{Mukherjee_2013_Interplanetary_Internet}. \cite{Alessi_2022} provides a detailed examination of DTN performance in Mars-Earth communications, taking into account a realistic end-to-end scenario with various assets and data flows. The results in the study demonstrated that both bundle priority and sporadic connection in ensuring a fast and organized bundle delivery is crucial. Intensity and amount of connections grow hop-by-hop for Mars to Earth data streams; as a result, the biggest problem is on the links exiting from the lander where congestion can rapidly occur. Even though the outcomes are often impressive, network management must provide sufficient safety margins to avoid network saturation. A three-phased strategy is being utilized by the Space Communications and Navigation (\acrshort{scan}) program to integrate operational communications support for DTN across the SCaN Network. Its initial operating capability (\acrshort{ioc}), which will serve the upcoming DTN users plankton, aerosol, cloud, ocean ecosystem (\acrshort{pace}) and Korea pathfinder lunar orbiter (\acrshort{kplo}) are planned for 2021. Then more spacecraft will join the network, supporting a multi-mission environment in 2022-2026. After 2026, multi-network will start including habitats and orbiters on Moon and Mars \cite{monaghan_2020_DTN}. DTN is also envisioned for Earth satellite networks. A comprehensive overview of DTN and its potential use in future satelilite networks is presented in \cite{Caini_2011}. In \cite{Cola_2011}, the reliability options from the physical layer up to the application layer for use in deep-space missions that are included in the CCSDS Protocol Stack are reviewed. DTN sutructures are investigated specifically for space applications in \cite{Burleigh_2003}, \cite{Alessi_2018}, \cite{Wu_2015}, \cite{Hu_2014}, \cite{Wang_2017}, \cite{Sun_2013}, and \cite{Yang_Queueing_2018}. Additionally, new routing algorithms and protocols, both novel and DTN-integrated are continuously being studied in for deep space networks. For example, an information centric networking (\acrshort{icn})-based publish-subscribe networking architecture is studied for crewed deep space missions in \cite{Gu_2020}. An overview of the contact graph routing (\acrshort{cgr}) algorithm developed by NASA-JPL, which selects a single path between a source and a destination node to ensure that a bundle is delivered as quickly as possible, is provided in \cite{Araniti_2015}, and its in-depth analysis is given in \cite{Dhara_2019}. A DTN routing algorithm known as Moderate Source Routing (\acrshort{msr}), which is a version of CGR, is proposed in \cite{Caini_2020}. A deep space communication protocol called Delay-Tolerant Transport Protocol (\acrshort{dttp}) which provides fast and reliable communication is presented in \cite{SAMARAS2010863}. By utilizing a peer-to-peer directory synchronization application between DTN nodes, called DTNbox, \cite{Marco_2019} proposed a potential solution for direct file transmission in interplanetary links. For more effective data transmission in space networks, a new local data-link layer protocol to the space DTN that combines Reed-Solomon (\acrshort{rs}) coding to Licklider transmission protocol (\acrshort{ltp}) is proposed in \cite{Shi_2017}. In order to achieve the best performance over extremely asymmetric channel rates between Mars and Earth, the impact of data bundle aggregation in space communications, which are formed by asymmetric and low channel rates is investigated in \cite{Wang_LTP_2013}. An interesting study on performance of DTN when communicating with TRAPPIST-1 system is presented in \cite{Fraire_2019}. In the study, it has been demonstrated that how DTN protocols may easily be included into the interstellar relay system's architecture in order to prevent the retransmission of data across great distances and, as a result, achieve reduced end-to-end latency while minimizing the relay buffer requirements.
	
	\subsection{Multiple Access Schemes}
	Multiple access schemes such as code-division multiple access (\acrshort{cdma}) and NOMA can also be utilized in Mars communication. 
	\subsubsection{CDMA}
	In \cite{Quirk_2001}, NASA compared the effectiveness of three different CDMA transmission techniques in terms of energy-per-bit because it is preferred to reduce the energy used by the surface-to-orbit link transceiver in order to increase the battery life of a Martian surface operator.
	In \cite{Babuscia_2017_CDMA}, a CDMA-based communication network consist of a an areostationary relay satellite placed at \SI{17,000}{\kilo\meter} altitude, a science orbiter placed at roughly \SI{400}{\kilo\meter} altitude, two CubeSats placed at \SI{3,500}{\kilo\meter}, rovers, and a lander is analyzed. In the study, CDMA is utilized as the multiple access technique because the areostationary relay satellite is considered to receive data from numerous users simultaneously. In contrast to more traditional methods like OFDM, which may need guard bands and frequency allocations, the CDMA concept is reported to be more resistant to Doppler effects and did not need frequency/time synchronization among users. Nonetheless, multiuser interference is said to be the CDMA system's major drawback. As a solution, the initial state of spreading pseudo-random noise (\acrshort{pn}) codes given to every user are unique. Therefore, it is claimed to have resulted in minimal cross-correlation when all users simultaneously connect with the areostationary relay satellite on a specific band, thereby reducing multiuser interference. The performance of a CDMA system for CubeSats operating close to the Moon is proposed in \cite{Babuscia_2016}, and the impact of Doppler effects on this system's performance is examined. The end-to-end optical CubeSat CDMA system architecture is designed in \cite{Divsalar_Opt_2018}, along with the characterization of the signal processing components needed for efficient operation and a performance evaluation for the uncoded scenario for multiple CubeSats.
	\subsubsection{NOMA}
	NOMA has recently been identified as a potential multiple access technique for gradually improving the spectral efficiency of mobile networks. NOMA's core principle is to use the power domain for multiple access, while the earlier generations of mobile networks relied on the time, frequency, and code domain. As an example, consider 3GPP-LTE's standard orthogonal frequency-division multiple access (\acrshort{ofdma}). One major problem with this orthogonal multiple access (\acrshort{oma}) technique is that when various bandwidth sources, like subcarrier channels, are given to users with weak channel conditions, their spectral efficiency is low. On the other hand, using NOMA allows each user to gain access to all of the subcarrier channels. As a result, the bandwidth resources assigned to users with weak channel conditions may still be utilized by users having strong channel conditions, which considerably increases spectral efficiency \cite{Zhiguo_2017_application_of_NOMA}. The primary benefits of NOMA can be summarized as follows \cite{Liu_2017_NOMA_for_5G}:
	\begin{enumerate}
		\item Because NOMA enables each time or frequency resource to be used by several users, it has a high bandwidth efficiency and therefore enhances system throughput.
		\item The number of connected devices can be substantially increased because the resources allocated per device decrease
		\item Because NOMA takes advantage of a new dimension, the power-domain, it can theoretically be used as an "add-on" approach for every present OMA technique, such as time division multiple access (\acrshort{tdma}), frequency division multiple access (\acrshort{fdma}), CDMA, and OFDMA.
		\item NOMA offers a low-complexity design. The underlying ideas of existing multiple access schemes and NOMA are highly comparable, as they both depend on assigning multiple users to a single resource block, such as time and frequency.
	\end{enumerate}
	NOMA can make a difference at this point by using very limited time and frequency resources in future Mars missions. For example, a rover is on an exploration mission far away from Mars habitat, and another rover is considerably close to the habitat. In that case, NOMA enables both rovers to use the same frequency and time resources in both downlink and uplink. 
	NOMA can also be useful when the number of UEs and surface operators substantially increase. Building a livable habitat and exploring an extraterrestrial planet in terms of various disciplines would require sophisticated WSNs containing large number of devices. These devices should last for long in terms of battery and provide enough throughput in a short time. NOMA can play a key role by providing efficient use of these resources. In \cite{Xiaojuan_2018}, the outage probability of hybrid satellite-terrestrial relay networks that use an amplify-and-forward protocol is examined. Following the similar architecture, \cite{Jiao_2020} demonstrated a NOMA uplink relay system, in which a relay satellite amplifies and forwards the signal from the BS on Earth to a lander and a rover operating on Moon's surface using NOMA. In a NOMA-based GEO and LEO satellite network, \cite{Ruixing_2021} examined a joint user pairing and power allocation system, which can be helpfull in areostationary relay-orbiter networks on Mars. Additionaly, \cite{Zhixiang_2020} investigated the performance of NOMA with LEO satellites taking into account the Doppler shift. \cite{Xinwei_2022} takes into account the utilization of RIS in NOMA, where a base station sends superposed signals to several users with the help of a RIS. The NOMA technique in a downlink land mobile satellite (\acrshort{lms}) network is presented in \cite{Yan_Perf_2018}, along with a thorough performance analysis of the system under consideration. An effective and reliable backhaul network is necessary to provide point-to-point optical wireless communication with high-quality service. \cite{Le_Tran_2021} suggested an optical backhauled NOMA in optical communication networks to achieve this goal. NASA is working on a new scheme for optical communications called wavelength division multiplexing (\acrshort{wdm}) to achiave higher data rates. An analysis on WDM, optical wavelength division multiple access (\acrshort{wdma}), and integrated optical WDMA and CDMA is published in \cite{Divsalar_Wave_2020}.
	
	\subsection{LTE on Moon}
	Nokia Bell Laboratories announced that the first LTE cellular network on the Moon is going to be employed. They stated that Lunar Outpost and its M1 mobile autonomous prospecting platform (\acrshort{mapp}) rover are chosen to integrate the LTE user equipment. After landing on the Moon, the MAPP rover is planned to move away from the Nova-C lander, communicating with it and being remotely controlled via Nokia's LTE/4G technology \cite{klein_2021_nokia_LTE}. It is claimed that LTE technology has the potential to transform lunar surface communications by offering consistent, high data speeds while reducing power, size, and cost. Communication is critical for NASA's Artemis mission, which aims to maintain a long-term existence on the Moon by the end of the decade. The network is envisioned to offer significant communication features for a wide range of data transmission operations, particularly critical command and control tasks, real-time navigation, remote control of lunar rovers, and high-definition video streaming \cite{nokia_2020_moon}. This development can only be just the beginning of new cellular networks built for the exploration habitats on other planets, including Mars. 
	
	\subsection{Flexible 3D Networks}
	A flexible 3D network structure based on 6G for mobile connection on Mars is studied in \cite{Bonafini_2021_E2E_performance}. As stated in the study, it is a unique, beneficial, and adaptable technique for establishing a new generation communications infrastructure in an extraterrestrial context with no need for a stable ground power source. The study mostly examined the potential altitudes that CubeSat orbits could reach and potential trade-offs between various parameters. The significance of such a study was to observe the inversely proportional behavior of mechanical and aerospace parameters versus communication and networking parameters. The lifetime and coverage of an orbiting node are noted to be expanded as its altitude rises, and the amount of propulsion power required to achieve station keeping is decreased. On the other side, it is expressed that it may also worsen service time, as well as delays and attenuation, which will have a negative effect on the quality of service (\acrshort{qos}) performance.
	
	\subsection{Environment-Aware Communications}
	Channel knowledge is an essential element of wireless communications that need to be tracked and acquired for building a reliable wireless network. One of the ways to obtain channel knowledge is through environment-aware communication, which has been gaining attention in the last couple of years. The storage and use of physical environment data, such as a 3D city or terrain map, is one of the simplest methods for acquiring environment awareness. Still, storing a realistic environment map with small details is expensive. Additionally, ray tracing is required because a physical environment map by itself can not indicate the radio propagation characteristics straight away. On the other hand, ray tracing is dependent not only on a correct physical characterization of the surrounding environment but also on their dielectric properties, which are not always accessible. Furthermore, ray tracing methods typically have high computing costs, particularly in complex metropolitan areas, which makes them more appropriate for offline numerical computer simulations rather than real-time use. 
	The idea of a channel knowledge map (\acrshort{ckm}) is put forward in \cite{Zeng_toward_env_aware_2021} as a tool for wireless communications that are aware of their environment. The CKM is defined as a site-specific database, which is marked with the locations of the transmitters and receivers. It is stated that it provides channel-related data that can be used to improve environmental awareness and make more advanced real-time channel state information (\acrshort{csi}) evaluation possible or even unnecessary. In the study, various examples of CKM are given. For instance, the channel gains of wireless links inside a specific area are predicted using a channel gain map (\acrshort{cgm}). Also, the channel shadowing map (\acrshort{csm}), which forecasts the shadowing loss, and the channel path map (\acrshort{cpm}), which forecasts the channel path data, including the number of dominant paths and their delays, phases, powers, etc., are other examples given to CKM.
	CKM can also play an essential role in Mars communication because the constraints of various channel types that CKM is appealing are mostly comparable with Mars's physical constraints. Channels of areas on out-of-reach locations are good examples. Communication devices are required to reach those areas where CSI knowledge is mainly acquired by pilot-based channel training. Another example is channels with huge dimensions. The cost of channel training and feedback increases dramatically with channel dimension in terms of communication resources such as pilot sequence length, training energy, and response delay. Channels with heavy hardware limitations are also good examples of how CKM could make a difference. Low-complexity hardware and signal processing are required to use the energy efficiently over large dimensions. All these three types of challenging channels combine the common features of Martian channels. As the Mars terrain is rough, the transmission of command data to Mars takes a long time, and the amount of hardware that can be transferred to Mars is limited, real-time solutions like CKM may potentially become vital in the exploration of Mars. CKM has also come to light as a promising new method for facilitating UAV communications placement and trajectory optimization. A CKM-assisted multi-UAV wireless network is studied in \cite{Haoyun_2022} to optimize the placement of multi-UAVs via designing and using a suitable CKM. By combining the CKM with the user's location information, a novel environment-aware hybrid beamforming approach is developed in \cite{Wu_Env_2022} to annihilate the need for real-time channel training for hybrid beamforming designs implemented on MIMO systems. A novel, CKM-based joint active and passive beam selection system for RIS-empowered wireless communication is suggested in \cite{Ding_2021}.
	
	\subsection{Artificial Intelligence}
	The history of artificial intelligence (\acrshort{ai}) goes back to 1940, when neural networks (\acrshort{nn}) are first employed. Over the decades, machine learning (\acrshort{ml}) and deep learning (\acrshort{dl}) techniques have been demonstrated to be effective architectures for complex tasks such as robotics, computer vision, and natural language processing, where developing a solid mathematical model is quite challenging. Communication technologies, on the other side, rely on a wide range of mathematical models and theories that call for in-depth expertise in information theory and channel modeling. In addition, the capability of secure transmission and the elimination of numerous hardware flaws make communications a complex and advanced field. Therefore, ML and DL methods can perform better than current technologies and offer observable benefits by optimizing current sophisticated problems. DL methods can be used to individually optimize the building blocks, including channel estimation algorithms, symbol detection, modulation, and coding schemes \cite{Basar_deep_learning_2022}. Mars communication can also benefit from ML and DL methods as the whole process of Mars colonization, and exploration experience is envisioned to be automated and self-adapting for large periods. Channel estimation is one of the biggest challenges in wireless communications, especially for Mars, because of heavy atmospheric conditions, rough terrains, and large distances. In such conditions, it is essential to have a dataset to train the ML and DL models. A Mars dataset may be generated in the future so that DL and ML algorithms can benefit. The use of both empirical datasets and statistical channel models is available in the literature. However, no matter what, it is almost impossible to perform channel estimation with high efficiency and low loss. At this point, RF chains and pilot-based estimations can be utilized. Also, it might be better to support optimization algorithms that are used in conventional channel estimation methods with ML and DL. Potential techniques for improving data return, lowering operational costs, and managing the complexity of automated systems in Mars missions include the use of automation augmented by cognition and machine learning. In \cite{Ferreira_2019}, authors went through how machine learning could be used in the link-to-link part of communication systems. A broad overview of the most significant issues in the subject of space explorations handled by AI applications and their possible solutions are given in \cite{app12105106} and \cite{Oche_2021}. In \cite{Hackett_Imp_2018}, a radio-resource-allocation controller for space communications is implemented utilizing a multi-objective reinforcement-learning method and deep artificial neural networks. The design, development, and implementations of ML for resource management in multibeam GEO satellite systems are covered in \cite{Gomez_2022}. The use of ML-based techniques in practical satellite communication networks is investigated in \cite{Miguel_2021} and \cite{Zhou_Machine_2021}. A multi-period-clustering (\acrshort{mpc}) technique, which introduces the clustering concept in machine learning, is suggested in \cite{Zhang_Multi_2022} to handle the energy dispersion problem for the carrier Doppler-shift acquisition in space communications. In \cite{Kothari_2020}, deep learning in space is highlighted as one of the research directions for embedded and mobile machine learning. In the study, different ways that machine learning has been used to space data, such satellite imaging, are gathered and how on-device deep learning might significantly enhance the functioning of a spacecraft by, for example, lowering communication costs or making navigation easier is discussed. Reinforcement learning-based techniques to overcoming spacecraft control issues are presented and examined in \cite{TIPALDI20221}. Various application areas are taken into consideration in the study, including maneuver planning in orbit transfers, constellation orbital control, and guidance, navigation, and control mechanisms for spacecraft landing on planetary bodies.
	
	\section{Conclusions} \label{sec:Conclusions}
	In this paper, we have reviewed the past, present, and future of Mars communications in terms of various aspects such as the past and ongoing Mars missions and their capabilities, Mars's characteristic features, atmospheric effects on radio propagation, channel modeling for near-Earth, deep space, and near-Mars links, 3D channel modeling computer simulations on Mars environments, and emerging technologies for Mars communication. Particularly, we have provided a very broad engineering view for Mars communications. We have also presented our own path loss exponent, PDP, and RMS delay spread calculations of various communication scenarios built upon 3D Martian terrains via computer simulations. These results could possibly help to modify future waveform design and standardization efforts for intra-planetary Mars communication. We conclude that there is significant room for growth in interplanetary communications, especially on Mars. First, more comprehensive channel modeling and wireless communication standard development can be carried out for future Mars missions. Furthermore, the potential of emerging communication techniques such as optical communication, RIS, \si{\tera\hertz} waves, NOMA, environment-aware communication, and AI should be investigated in a more comprehensive way for communications on the red planet.

	\ifCLASSOPTIONcaptionsoff
	\newpage
	\fi

	\bibliographystyle{IEEEtran}
	\bibliography{mars_ref}
	
\end{document}